\documentclass[final]{siamart190516}
\usepackage{graphicx, epsfig}
\usepackage{amsmath,amsfonts,ulem}

\newcommand{\mb}[1]{ \mbox{\boldmath$#1$} }

\newcommand{\ds}{\displaystyle}
\newcommand{\beq}{\begin{eqnarray}}
\newcommand{\eeq}{\end{eqnarray}}
\newcommand{\beqq}{\begin{eqnarray*}}
\newcommand{\eeqq}{\end{eqnarray*}}
\newcommand{\p}{\partial}

\newcommand{\eps}{\varepsilon}
\newcommand{\Eta}{\mbox{\boldmath$\eta$}}
\newcommand{\x}{\mbox{\boldmath$x$}}

\newcommand{\y}{\mbox{\boldmath$y$}}

\newcommand{\n}{\mbox{\boldmath$n$}}

\newcommand{\w}{\mbox{\boldmath$w$}}

\newcommand{\sm}{\setminus}
\newcommand{\PP}{\mbox{\boldmath$P$}}
\newcommand{\FF}{\mbox{\boldmath$F$}}
\newcommand{\GG}{\mbox{\boldmath$G$}}
\font\bb=msbm10 at 12pt
\def\rR{\hbox{\bb R}}

\def\eE{\hbox{\bb E}}
\def\v{\mathbf{v}}

\title{Computational methods and fast hybrid stochastic simulations for triangulation sensing and identifying principles of cell navigation in the brain}
\author{%
U. Dobramysl\thanks{Department of Applied Mathematics and Theoretical Physics, University of Cambridge, Cambridge, UK}%
\and D. Holcman\thanks{Group of data modeling and computational biology, IBENS-PSL Ecole Normale Superieure, Paris, France.}%
}
\headers{}{}
\begin{document}
\maketitle
\begin{abstract}
Brownian simulations can be used to generate statistics relevant for studying molecular interactions or trafficking. However, the concurrent simulation of many Brownian trajectories at can become computationally intractable. Replacing detailed Brownian simulations by a rate model was the basis of Gillespie's algorithm, but requires one to disregard spatial information. However, this information is crucial in molecular and cellular biology. Alternatively one can use a hybrid approach, generating Brownian paths only in a small region where the spatial organization is relevant and avoiding it in the remainder of the domain. Here we review the recent progress of hybrid methods and simulations in the context of cell sensing and guidance via external chemical gradients. Specifically, we highlight the reconstruction of the location of a point source in 2D and 3D from diffusion fluxes arriving at narrow windows located on the cell. We discuss cases in which these windows are located on the boundary of the 2D or 3D half-space, on a disk in free space, inside a 2D corridor, or a 3D ball. The hybrid method in question performs Brownian simulations only inside a region of interest. It uses the Neumann-Green's function for the mentioned geometries to generate exact mappings exit and entry points when the trajectory leaves the region, thus avoiding the explicit computation of Brownian paths in an infinite domain. Matched asymptotics is used to compute the probability fluxes to small windows and we review how such an approach can be used to reconstruct the location of a point source and estimating the uncertainty in the source reconstruction due to an additive perturbation present in the fluxes. We also review the influence of various window configurations on the source position recovery. Finally, we discuss potential applications in developmental cell biology and possible computational principles.
\end{abstract}

\section{Introduction}
Asymptotic analysis has been a powerful tool to obtain approximate solutions for various types of  partial differential equations, such as the wave or diffusion equations \cite{Schuss:Book,o1991singular,bender2013advanced}, but also nonlinear equations such as reaction-diffusion equations for pattern formation, the Poisson-Nernst-Planck equation for the charge distribution in semi-conductors and electrolytes, and many more. These methods include the classical matched asymptotics \cite{Ward1993_2,Ward2005,Ward3,Ward1993_1,Ward1992,Pillay2010} that glues a solution found inside a domain with a solution in a small layer near the boundary or the Green's function \cite{Holcmanschuss2018}. Other methods for singular perturbation include the well known WKB equation to account for exponentially small terms \cite{Schuss1}. In the past 20 years, asymptotic analysis has also played a key role in the analysis of numerical schemes associated with diffusion processes \cite{Schuss4} to relate microscopic statistical quantities to parameters used in stochastic simulations, but also to analyse boundary layers in to finite elements discretization. Discrete simulations are based on jump processes, while the microscopic world is described by continuum equations. Recently asymptotic analysis has been used to design fast hybrid simulations, where the continuum description is used to avoid excessive computational cost \cite{dobramysl2018reconstructing,erban2019stochastic,yates2020}: Trajectories are explicitly simulated only in a small fraction of the domain, and then coupled to a more efficient continuum or compartment model outside of this region of interest. \\
The goal of this review is to present recent efforts in developing hybrid simulations to understand the developmental biology of the brain. A pressing question in this field is to find the principles by which neuronal projections navigate the developing brain from the cell body to their target (to form synapses), a distance that could measure up to tens of centimeters.  Below, we briefly expand on this background, where finding navigation principles for a cell requires the formulation of physical models and the associated numerical simulations for a single cell guidance in a complex environment. In this, the fluxes associated with the chemical gradients of molecular guidance cues emerge as the primary coordinates. It would be impossible to simulate the detailed position of each molecule, hence proper coarse-graining remains fundamental to reveal the spatial organization of molecular guidance cues and to derive statistical laws for cell guidance. \\
For more than 30 years, one of the main challenges of neuronal development was to identify the molecules involved in neuronal migration in the brain. The field of neuronal development was mostly driven by discovering these molecular cues, and the associated receptors, located on the cell membrane, which convey spatial signaling information \cite{chedotal2010wiring,kolodkin2011mechanisms,blockus2014multifaceted}. Yet no consensus has emerged about the mechanism that converts a molecular gradient of cues into positional information used by a cell to navigate. \\
In parallel to these experimental effort, the physical literature \cite{BergBook,Berg1977,bialek2008cooperativity} has focused on developing scenarios and mechanisms for estimating how a local concentration gradient at a small test surface (such as a ball) can be sampled when counting the number of diffusing molecules. The mathematical problem consists in computing the flux of Brownian particles arriving to a small target when the condition is absorbing, partially absorbing or simply transparent. Using the well-known dipole expansion, this approach was used to extract the direction of a gradient \cite{aquino2016know,endres2008accuracy}. \\
However, the computational principles that characterize how neurons in the brain migrate and stop at their final location remain unclear. In particular, to navigate long distances (mm to cm), a cell embedded in a tissue has to determine its position relative to guidance points. For example, bacteria are finding a local gradient source of diffusing molecules via temporal sampling and are basing their movement decisions on this information. Interestingly neurons need to pass through narrow corridors to avoid invading other regions, they need to spread over an entire domains or sometimes compete equally for space: this is the case in the visual cortex where neurons project from the retinas of both eyes to equally innervate the lateral geniculate nucleus (part of the thalamus that relays audiovisual impulses to the cortex) and the cortical region V1 \cite{o2007area,espinosa2012development}.  These processes involve multiple navigational mechanisms: positive attraction to cues, negative repulsion by cues or a combination thereof.\\
Recently, we focused on the problem of neuronal guidance in the brain by answering the question: Is it possible to triangulate the point source of a molecular gradient using the arrival rate of Brownian particles in a domain containing many absorbing windows on its boundary (both in two and three dimensions)~\cite{dobramysl2018mixed,dobramysl2018reconstructing,dobramysl2020PRLTria,dobramysl2020threed}? \\
Here, we review recent computational approaches (modeling, asymptotics and hybrid simulations) designed to estimate the position of a point source that emits stochastic particles from the steady-state fluxes collected at narrow windows, located on a bounded surface. This review includes: resolving a triangulation problem (inverse problem) using the theory of diffusion at steady-state, solving a mixed boundary Laplace's equation using external Green's function. In the second part, we focus on efficient algorithms to simulate Brownian trajectories explicitly only close to the window targets. Finally, in the last section, we present the computations used for estimating the precision of the source recovery from various empirical estimators. To conclude, if the mathematical analysis reveals that triangulating the source location is quite noisy and imprecise, it could be sufficient to direct the cells, probably explaining why multiple sources are redundantly distributed along the way. To start, we describe the analysis of the biological question and its computational formulation before proceeding with the mathematical analysis.
\subsection{Sensing a gradient in biophysics and computational biology}
\begin{figure}
  \centering
\includegraphics[scale=1]{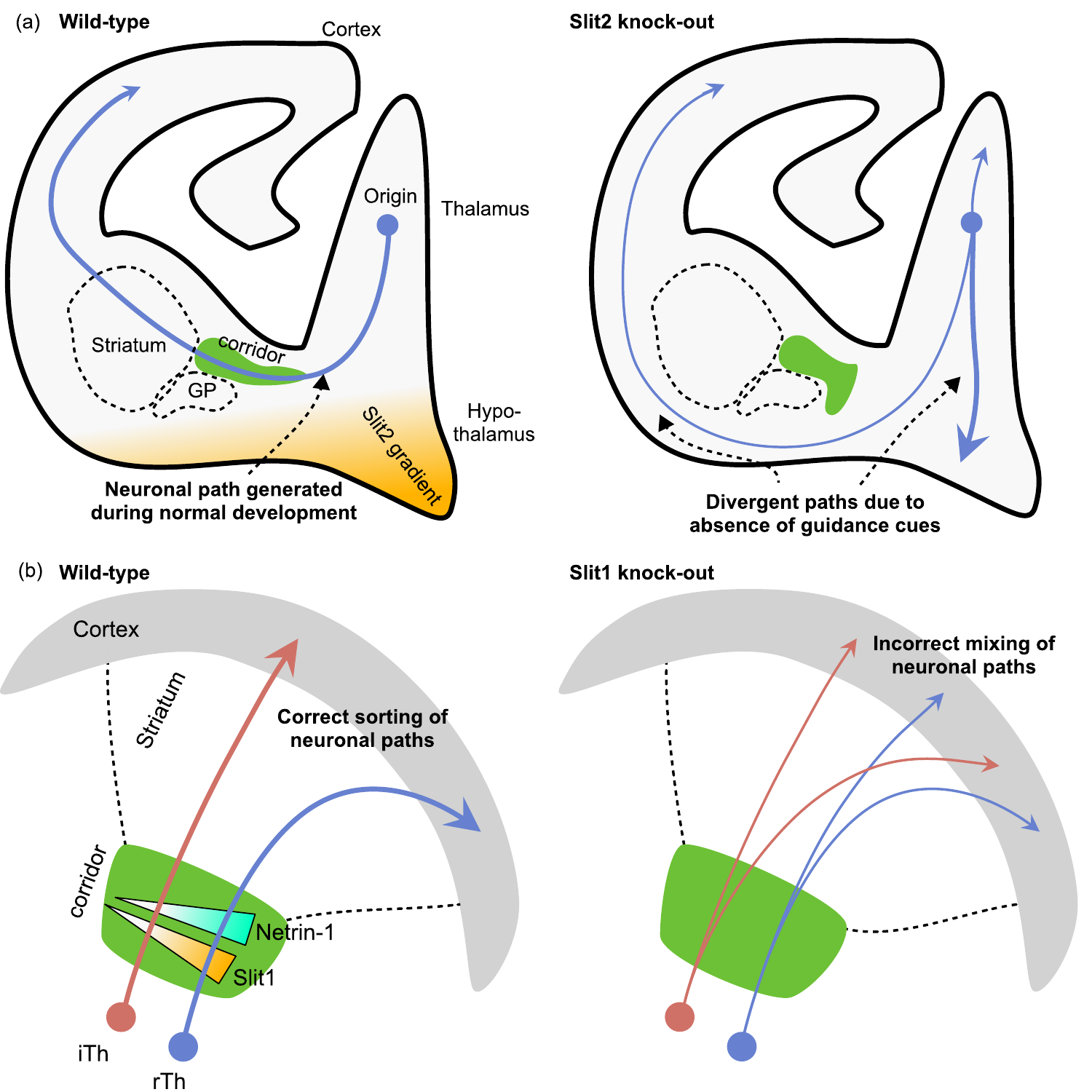}
\caption{ {\bf Neuronal navigation in chemical gradients.}
Role of two cues (Slits and Robos) in the development of thalamocortical axons. (a) When one cue species (Slits) is at regular expression levels (i.e. in wild-type mice) in the ventral telencephalon and diencephalon, it repells thalamocortical axons (TCAs, purple) away from these domains. This repulsion defines the positioning of corridor cells which constitute a permissive bridge for TCAs in the ventral telencephalon. When the expression of these gradients is altered, some TCAs are redirected to the dorsal and ventral diencephalon. Corridor cell migration is also perturbed preventing many TCAs from growing across the striatum. (b) Two parallel expression gradients of Slit1 and Netrin-1 within the corridor control the sorting of TCAs along the cortex rostro-caudal axis. Abbreviations: E - embryonic day; GP - globus pallidus; iTh - intermediate thalamus; Str - striatum; rTh - rostral thalamus.}\label{fig:navigation_ch}
\end{figure}

\begin{figure}[http!]
    \centering
    \includegraphics[scale=0.3]{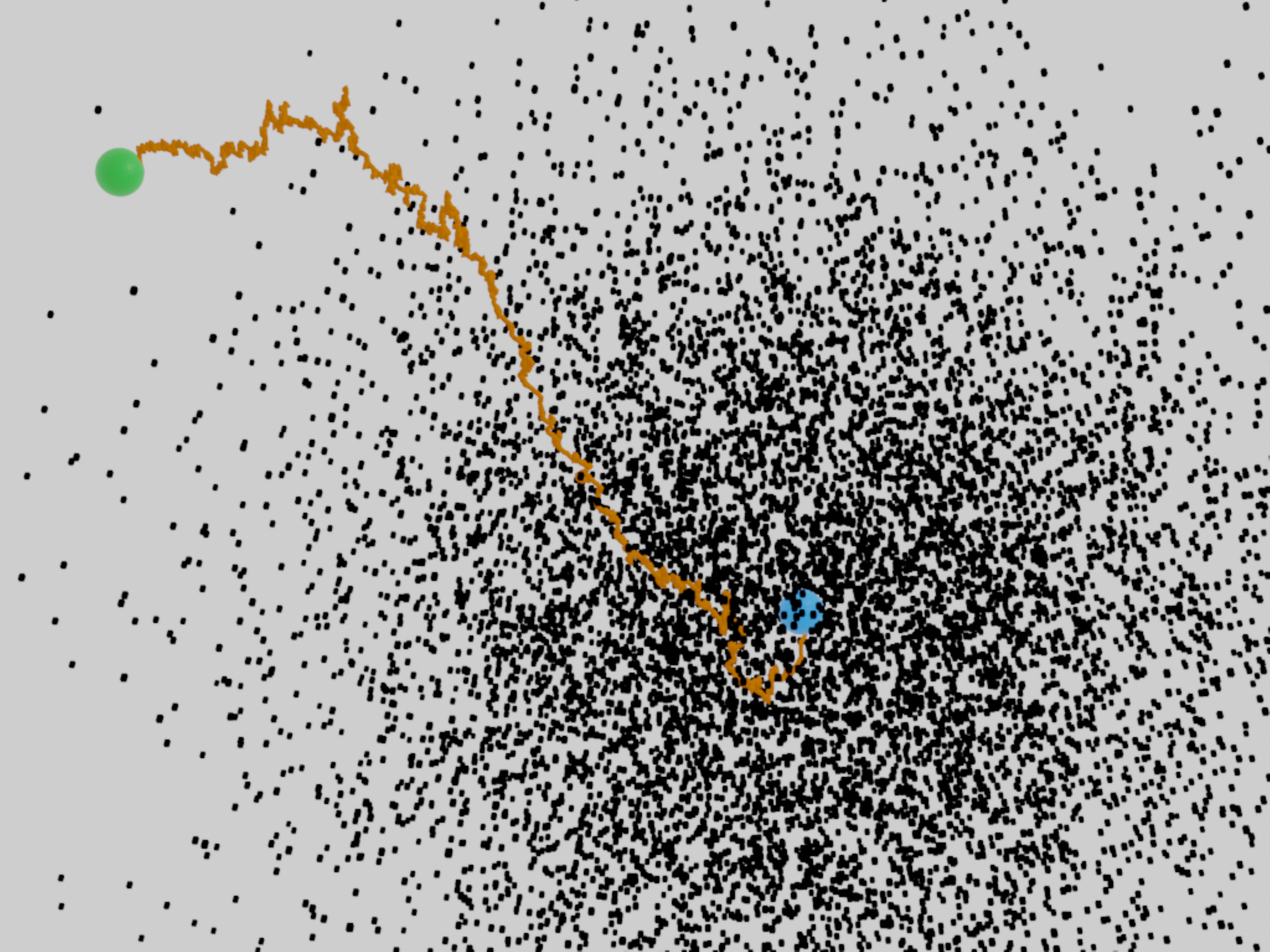}
    \caption{{\bf Schematic  modeling of a cell generating its path in a gradient}. A cell (green) is embedded in a gradient of cues (black) and has to find a target source (blue), generating a path, that depends on the gradient concentration.}
    \label{fig:figure0}
\end{figure}

Sensing a molecular gradient is a key process in cell biology and crucial for the detection of a concentration that can transform positional information into cell specialization and differentiation \cite{kasatkin2008morphogenetic,Malherbe,wolpert1996one}. During neuronal development, the tip of an axon (a neuronal projection) - the growth cone - uses the concentration of cue molecules \cite{reingruber2014computational} to decide whether to continue moving or to stop, to turn right or left (Fig. \ref{fig:figuremodel}A). Bacteria and spermatozoa are able to orient themselves in a molecular gradient \cite{kaupp2017signaling,Heinrich3} (Fig. \ref{fig:figure0}).\\
A large class of models have focused on the relation between the local fluctuations of the concentration near a test target and the sampling time to recover the concentration at the target. In this case, the test ball can be fully absorbing (uniform boundary conditions)~\cite{BergBook,endres2008accuracy} which was shown to be sufficient to detect a gradient direction, but does not allow the extraction of the source position as we will discuss later on. \\
The spatial distribution of cell surface receptors that bind cue molecules is a key element; measuring the fluxes at receptors is the first step for a cell to determine outside concentration gradients. Assuming binding at the receptors is fast compared to the timescales of diffusion or cell movement, the receptors report their binding state - i.e. the diffusive flux to the receptor - to the interior of the cell via biochemical signalling cascades. At this stage, we think that it is not enough for a cell to identify a local gradient, but it has to be able to estimate the position of the gradient source in two or three-dimensional space for a correct navigation. We refer the reader to a recent review \cite{aquino2016know} that discusses estimating the fluctuation at the test volume when a uniform condition is used at the boundary.\\
In some scenarios (including axonal growth cones) receptors are able to rearrange their positions in response to external gradients \cite{bouzigues2010mechanism}. On top of that, a cell can make complex computations regarding its navigation decisions by comparing chemical gradients with an gradients of obstacles \cite{wondergem2019chemotaxis}, however the mechanisms behind this remains unclear.\\
For neuronal cells, it is not clear whether they find an exact target position or stop at a well-defined position within a gradient field. But in both cases, this process requires the cell to estimate the gradient source position, hence information about the distance in addition to the direction is crucial. Experimental evidence for this comes from simply knocking down (eliminating) receptors: in that case, neurons are not able to find their correct target place and thus the brain patterning is disrupted, leading, for example, to vision impairment \cite{chedotal2010wiring}.\\
To give another example, the guidance molecule ephrin displays a gradient along the axes of the retina to control the temporal–nasal mapping of the retina in the optic tectum/superior colliculus by regulating the topographically-specific interstitial branching of axons (Fig. \ref{fig:navigation_ch}). Hence, ganglion cells project from the retina to the tectum and stop at very specific locations. Sometimes the combination of multiple gradients is required \cite{reingruber2014computational}.\\
To conclude, a large class of brain cells such as neurons and astrocytes need to migrate long distances and stop at a highly specific locations. Therefore, for the cell to stop at the correct point requires not only the estimation of the direction of the source but having good estimation of the gradient source position.

\subsection{Limits to concentration estimation via local sampling}
The standard limit for measuring a gradient concentration is due to the particle number fluctuations during a finite measurement time. This concept has been introduced in the context of concentration or gradient sensing \cite{BergBook}\cite{Berg1977,MuglerPNAS2015,aquino2016know} quantified by an ensemble of formulae to relate the local change of concentration at a test ball to the duration of the measurement $T$: for a concentration $c_0$, the variation of concentration $\delta c$ is related to the size $a$ of the test ball, the diffusion coefficient $D$ and the measurement time $T$ by
\beq \label{berg1}
\frac{\delta c}{c_0} \sim \frac{1}{D a c_0 T}.
\eeq
The exact coefficient depends whether the boundary of the ball is absorbing, partially reflecting or transparent. Formula \ref{berg1} has been extended to the case where the test ball behaves like a giant receptor where ligands can bind, unbind and rebind \cite{bialek2008cooperativity,aquino2016know}.  The sampling time $T$ is assumed to be much larger than the diffusion time $a^2/D$. However this framework does not provide much information about the source of the diffusing molecules. Furthermore, the noise measurement is only connected to the target geometry not the source. Improving on this, computing the ratio $R$ of the fluxes at the front versus the back of a ball \cite{endres2008accuracy} allows the recovery of the direction of a gradient but not the position of the source.\\
To conclude, a uniformly absorbing boundary condition at a test volume limits the possibility to recover additional properties about the gradient source location. In the remainder of this review, we will focus on a test ball surface hosting many discrete receptors and summarize the recent efforts to infer the position of the source from combining the values of steady-state fluxes. Developing numerical simulations allows an exploration of the range of validity of the analytical results. Sensitivity analysis reveals the accuracy of the source position recovery. When narrow windows are considered, the distribution of arrival time of Brownian particles is well approximated by a Poisson process, and the rate is precisely the steady-state flux \cite{Holcmanschuss2018,Holcman2015}.
\section{Modeling receptor absorption fluxes}\label{sec:analysis_fluxes}
\begin{figure}
  \centering
  \includegraphics[scale=0.18]{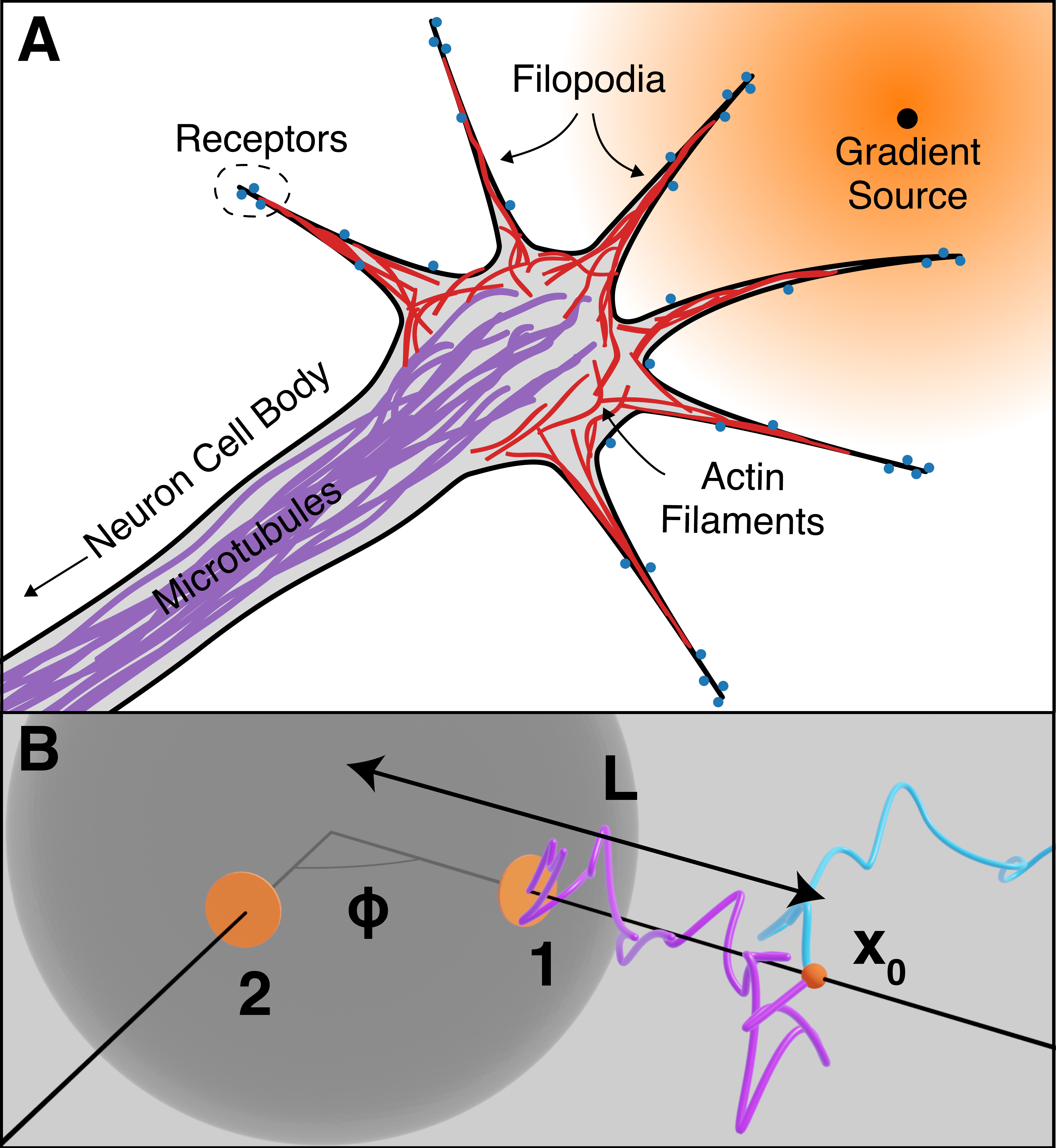}
  \caption{ (A) Neuronal growth cone in an external chemical gradient. Receptors located on the cell surface are able to sense the gradient, thereby defining the first step of navigation in the brain. (B) Sensing model - a ball containing small windows (receptors), two in this case. The source is located at $\x_0$, where Brownian particles are released (blue and purple).}
  \label{fig:figuremodel}
\end{figure}

The physical modeling framework is the following: a point source located at position $x_0$ releases independent identically distributed particles, called cues or ligands (Fig. \ref{fig:figuremodel}B). In the Smoluchowski limit, the Langevin equation for the cue position $X(t)$ is given as:
\beq \label{stochlocal0}
\dot{X}=\frac{F(X(t),t)}{\gamma} + \sqrt{2D}\, \dot{W},
\eeq
where $W$ represents Gaussian white noise with $\langle W(t)W(s)\rangle=\delta(t-s)$, $\gamma$ is the dynamical viscosity \cite{Schuss2} and $F(X,t)$ is a force (or equivalently a velocity) field. The underlying source of the noise is thermal agitation. We disregard any fluctuations in the rate of release, which is usually assumed to follow a Poisson distribution. We will also neglect possible cue killing mechanisms that could destroy particles, which would lead to exponential gradients \cite{lander2002morphogen,nahmad2011spatiotemporal}. Finally, we only consider the pure diffusion case without external forces $F(X(t),t)=0$.\\
At steady-state, the independent Brownian particles move in free space (we assume that there are no obstacles for now), but cannot penetrate a domain $\Omega$. This domain can either be a ball (in three dimensions) or a disk (in two dimensions) of radius $R$). Other shapes such as ellipses could be possible, but would lead to more involved analytical computations. \\
This domain $\Omega$ models a biological cell that can count the arrival of cues to the boundary $\p\Omega$. This boundary is divided into $N$ small and disjoint absorbing windows, $\partial\Omega_{\varepsilon_j}\ (j=1,\ldots,N)$ each of area $|\partial\Omega_{\eps_j}|=O(\eps^2)$ (in three dimensions), where the radius $\eps$ is small. We assume that the windows are sufficiently far apart to avoid non-linear effects~\cite{Holcman2008_1}. The absorbing boundary is
\beq
\p\Omega_a= \cup_1^{N} \partial\Omega_{\varepsilon_j}.
\eeq
The remaining boundary surface is reflective $\p\Omega_r=\p\Omega\setminus\p\Omega_a$ for the diffusing particles. Note that other models are possible, for example instead of purely absorbing boundary conditions one can consider partially absorbing (Robin) boundary conditions \cite{grebenkov2020single}.\\
To compute the particle flux to the absorbing boundary, we first use the transition probability density $p(\x,t\,|\,\x_0)$ to find a particle at position $\x$ at time $t$, when it started at position $\x_0$. It is the solution of the Diffusion equation
\begin{align}\label{IBVP}
\frac{\p p(\x,t\,|\,\x_0)}{\p t} =&D \Delta p(\x,t\,|\,\x_0)\quad\hbox{\rm for } \x,\x_0 \in \rR^d\setminus\Omega,\\
p(\x,0\,|\,\y)=& Q\delta(\x-\x_0)\quad \hbox{\rm for } \x,\x_0 \in \rR^d\setminus\Omega\nonumber\\
\frac{\p p(\x,t\,|\,\x_0)}{\p \n} =&0\quad \hbox{\rm for } \x \in\p\Omega_r, \x_0\in \rR^d\setminus\Omega,\nonumber\\
p(\x,t\,|\,\x_0)=&0\quad \hbox{\rm for } \x \in \p\Omega_a,  \x_0\in \rR^d\setminus\Omega,\nonumber
\end{align}
where $D$ is the diffusion coefficient, $Q$ is the rate of particle emission and $d$ is the number of dimensions (two or three). The steady-state gradient $P_0$ is obtained by integrating equation \eqref{IBVP} from $t=0$ to infinity, which is equivalent to resetting a particle after it disappears through a window \cite{Schuss:Book}. It is given by the solution of the mixed-boundary value problem
\beq
D\Delta P_0(\x) & = & -Q \delta(\x-\x_0) \hbox{ for } \x\, \in \,\rR^d\setminus\Omega \label{eqDP3b}\\
\frac{\p P_0}{\p \n}(\x) & = & 0\hbox{ for } \x \,   \in\, \p\Omega_r  \label{eqDP1} \nonumber\\
P_0(\x) & = & 0 \hbox{ for } \x \,   \in \p\Omega_a.\nonumber
\eeq
The probability fluxes associated with $P_0$ at each individual window $\Omega_{\eps_j}$ can be computed from the fluxes and depend on the specific window arrangement and the domain $\Omega$. The parameter $Q>0$ can be calibrated so that there are a fixed number of particles located in a volume. At infinity, the density $P_0(\x)$ has to tend to zero in three dimensions. More complex domains can be studied if their associated Green's function can be found explicitly. We focus below on several cases such as a ball in the entire space or half-space and also in narrow band.
\subsection{Fluxes of Brownian particles to small targets in \texorpdfstring{$\rR^2$}{2D}}
The flux received by the disk containing the $N$-small absorbing windows $\p\Omega_{1}\cup\hdots\cup \p\Omega_{N}$ on the disk boundary $\p \Omega$ is computed by solving the mixed boundary value problem, given as \eqref{eqDP3b} with $d=2$ and $Q=1$ without loss of generality:
\beq
-D\Delta P_0(\x) & = & \delta(\x-\x_0) \;\;\text{ for }\;\; \x\, \in \,\rR^2\sm \Omega \label{eqDP2}\\
 \nonumber \ds \frac{\p P_0}{\p n}(\x) & = & 0 \;\;\text{ for }\;\; \x\, \in\, \p\Omega \sm (\p \Omega_{1}\cup\hdots \cup\p\Omega_{N})\\
 \nonumber P_0(\x) & = & 0 \;\;\text{for}\;\; \x\, \in\, \p\Omega_{1}\cup\hdots\cup \p\Omega_{N}
 \eeq
Although the probability density $P_0(\x)$ diverges when $|\x|\rightarrow\infty$ in two dimensions, the splitting probability between windows is finite because it is the ratio of the steady-state flux at each hole divided by the total flux through all windows:
\beq
J_k= \ds \frac{\ds \int_{\p\Omega_{k}} \ds \frac{\p P_0(\x)}{\p \n} dS_{\x}}{\ds \sum_{q} \int_{\p\Omega_{q}} \ds \frac{\p P_0(\x)}{\p \n} dS_{\x}}.
\eeq
Additionally, in two dimensions and due to the recurrence property of the Brownian motion, the probability to hit a window before going to infinity is one. Thus the total flux is one:
\beq
\sum_{q} \int_{\p\Omega_{q}} \ds \frac{\p P_0(\x)}{\p \n} dS_{\x}=1.
\eeq
\subsection{The two-dimensional problem}
In this section, we present the computation of the fluxes in three different configurations: (1) when the windows are distributed on a line on the boundary of the infinite half-plane, (2) located on the boundary of a disk embedded in the entire 2D plane, and (3) when the disk is located in a narrow band with reflecting boundaries. We conclude this section by showing the case of arbitrarily many windows. We use the Neumann-Green's function and the method of matched asymptotics throughout this section~\cite{Ward1993_1,Ward3}.
\subsubsection{Fluxes to two small absorbers on a half-plane} \label{sec:half-plane}
\begin{figure}[http!]
\centering
\includegraphics[scale=0.99]{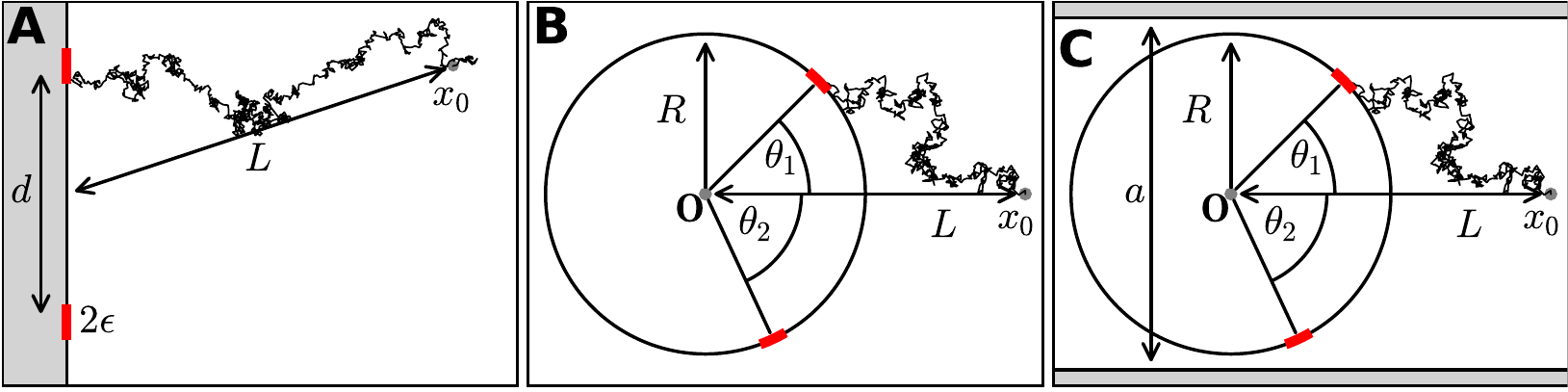}
\caption{\textbf{Brownian fluxes to small windows in different geometries}. (A) Two windows of size $2\epsilon$ are placed on the boundary of half-space a distance $d$ apart. Diffusing particles are released from a source at $\x_0$ at a distance $L=|\x_0|$ and are absorbed by one of the windows. (B) Two absorbing windows of size $2\epsilon$ are placed on the circumference of a disk with radius $R$ at angles $\theta_1$ and $\theta_2$ with the $x$-axes. As before, particles are released at the source position $\x_0$. (C) Two windows are placed on a disk as in (B), inside an infinitely long strip with reflecting walls at $y=\pm a$.}
\label{Fig1}
\end{figure}

We now start with case (1), where two small, absorbing holes windows, $\p \Omega_1 = \lbrace x=0,\;z=z_1+s| s \in [-\eps_1/2, \eps_1/2]\rbrace$ and $\p \Omega_2= \lbrace x=0,\;z=z_2+s | s \in [-\eps_2/2, \eps_2/2]\rbrace$, are positioned on the boundary of the two-dimensional half-plane, $\Omega = \lbrace  (x,z) \in \mathbb{R}^2, x>0\rbrace$ (Fig.~\ref{Fig1}A). The source is located at $\x_0\in\Omega$, and diffusing particles are reflected everywhere on the boundary, except at the two small targets.
The boundary value problem in equation \ref{eqDP2} for two windows reduces to
\beq
-D\Delta P_0(\x) & = & \delta(\x-\x_0) \;\;\text{ for }\;\; \x\, \in \,\rR_{+}^2 \label{eqDP3}\\
 \nonumber \ds \frac{\p P_0}{\p n}(\x) & = & 0 \;\;\text{for}\;\; \x\, \in\, \p\rR_{+}^2 \sm (\p \Omega_{1}\cup\p\Omega_{2})\\
 \nonumber P_0(\x) & = & 0 \;\;\text{for}\;\; \x\, \in\, \p\Omega_{1}\cup \Omega_{2}.
\eeq
We set $D=1$ and derive a solution of equation \ref{eqDP3} in the small window limit. The derivation uses an inner and outer solution. The inner solution is constructed near each small window~\cite{Pillay2010} by scaling the arclength $s$ and the distance to the boundary $\eta$ by $\bar \eta=\frac{\eta}{\eps}$ and $\bar s=\frac{s}{\eps}$ (we use the same size for both windows $\eps_1=\eps_2=\eps$), so that the inner problem reduces to the classical two-dimensional Laplace equation
\beq
\Delta w=0 \hbox{ in } \rR_+^{2}\\
\frac{\p w}{\p n}=0 \hbox{ for } |\bar s|>\frac{1}{2},  \bar \eta=0\\
w(\bar s, \bar \eta)=0 \hbox{ for } |\bar s|<\frac{1}{2},  \bar \eta=0.
\eeq
The far field behavior for $|\x|\rightarrow \infty$  and for each hole $i=1, 2$ is
\beq
w_i(\x) \approx  A_i \{\log|\x-\x_i|-\log \eps +o(1)\},
\eeq
where $A_i$ is the flux
\beq\label{flux2Dhalfspace}
A_i=\frac{2}{\pi} \int_0^{1/2} \frac{\p w(0,\bar s)}{\p \bar \eta}d\bar s.
\eeq
The outer solution is the external Neumann-Green's function of the two-dimensional half-plane, i.e. the solution of
\beq
-\Delta_{\x} G(\x,\y) & = & \delta(\x-\x_0) \hbox{ for } \x\, \in \,\rR_{+}^2, \label{eqDP2b}\\
\frac{\p G}{\p n_{\x}}(\x,\x_0) & = & 0\hbox{ for } \x \,   \in\, {\partial\rR_{+}^2}.
\eeq
given for $\x, \x_0 \in \rR_{+}^2$ by
\beq\label{eqDP2bs}
G(\x,\x_0)=\frac{-1}{2\pi }\left(\ln |\x-\x_0| + \ln \left|\x-\bar{\x_0}\right| \right),
\eeq
where $\bar{\x_0}$ is the symmetric image of $\x_0$ through the boundary axis $0z$. The uniform solution is the sum of inner and outer solution
\beq\label{gexp}
P(\x,\x_0)=G(\x,\x_0)  +A_1 \{\log|\x-\x_1|-\log \eps\}+A_2 \{\log|\x-\x_2|-\log \eps\} +C,
\eeq
where $A_1,A_2,C$ are constants to be determined. To that end, we study the behavior of the solution near each window location $\x_i$. In the boundary layer, we require the uniform solution to approach the inner solution, i.e.
\beq
P(\x,\y)\approx A_i \{\log|\x-\x_i|-\log \eps\}.
\eeq
Using this condition on each window, we obtain the two equations:
\beq\label{cond1}
G(\x_1,\x_0)  +A_2 \{\log|\x_1-\x_2|-\log \eps\} +C=0\\
\nonumber
G(\x_2,\x_0)  +A_1 \{\log|\x_2-\x_1|-\log \eps\} +C=0.
\eeq
Due to the recursion property of the Brownian motion in two dimensions there are no fluxes at infinity, therefore the conservation of flux yields
\beq \label{compatibility}
\int_{\p\Omega_{1}} \frac{\p P(\x,\y))}{\p \n}dS_{\x} + \int_{\p\Omega_{2}} \frac{\p P(\x,\y))}{\p \n}dS_{\x} =-1.
\eeq
In the limit of two well separated windows ($|\x_1-\x_2| \gg 1$) and using the condition for the flux in equation \ref{flux2Dhalfspace} we get for each window $i=1,2$
\beq
\int_{\p\Omega_{i}} \frac{\p P(\x,\y))}{\p \n}dS_{\x}= -\pi A_i
\eeq
(the minus sign is due to the outer normal orientation), thus
\beq \label{compatibility2}
\pi A_1+\pi A_2=1.
\eeq
Using relations \ref{compatibility2} and \ref{cond1}, we finally obtain the system
\beq
\frac{G(\x_1,\x_0)-G(\x_2,\x_0)}{\{\log|\x_1-\x_2|-\log \eps\}} +(A_2-A_1) =0\\
A_1+ A_2=\frac{1}{\pi}.
\eeq
Finally, the absorbing probabilities are given by
\beq \label{eq:p2-halfplane}
P_2=\pi A_2&=&\frac12+ \frac{\pi}{2}\frac{G(\x_1,\x_0)-G(\x_2,\x_0)}{\{\log|\x_1-\x_2|-\log \eps\}} \\
\label{eq:p2-halfplane2}
&=&\frac12- \frac{1}{4}\frac{\ln \frac{|\x_1-\x_0|\left|\x_1-\bar{\x_0}\right|}{|\x_2-\x_0|\left|\x_2-\bar{\x_0}\right|}}{\{\log|\x_1-\x_2|-\log \eps\}}.
\eeq
and
\beq
P_1=\frac12+\frac{1}{4}\frac{\ln \frac{|\x_1-\x_0|\left|\x_1-\bar{\x_0}\right|}{|\x_2-\x_0|\left|\x_2-\bar{\x_0}\right|}}{\{\log|\x_1-\x_2|-\log \eps\}}.
\eeq
These probabilities precisely depend on the source position $x_0$ and the relative position of the two windows. When one of the splitting probabilities (either $P_1$ or $P_2$) is known and fixed in $[0,1]$, recovering the position of the source requires inverting equation \ref{eq:p2-halfplane2}. For $P_2=\alpha \in [0,1]$, the position $\x_0$ lies on the curve
\beq\label{source}
S_{source}=\{ \x_0 \hbox{ such that} \, \frac{|\x_1-\x_0|\left|\x_1-\bar{\x_0}\right|}{|\x_2-\x_0|\left|\x_2-\bar{\x_0}\right|}=\exp \left(  (4\alpha-2)\{\log|\x_1-\x_2|-\log \eps\} \right) \}.
\eeq
Therefore, knowing the splitting probability between two windows is insufficient to recover the exact source position $\x_0$. Rather, it can only be narrowed down to a one dimensional curve solution. However the direction can be obtained by simply checking which one of the two probability is the highest.
\subsubsection{Fluxes to small windows on a disk} \label{sec:disk-free-space}
\begin{figure}[http!]
  \centering
  \includegraphics{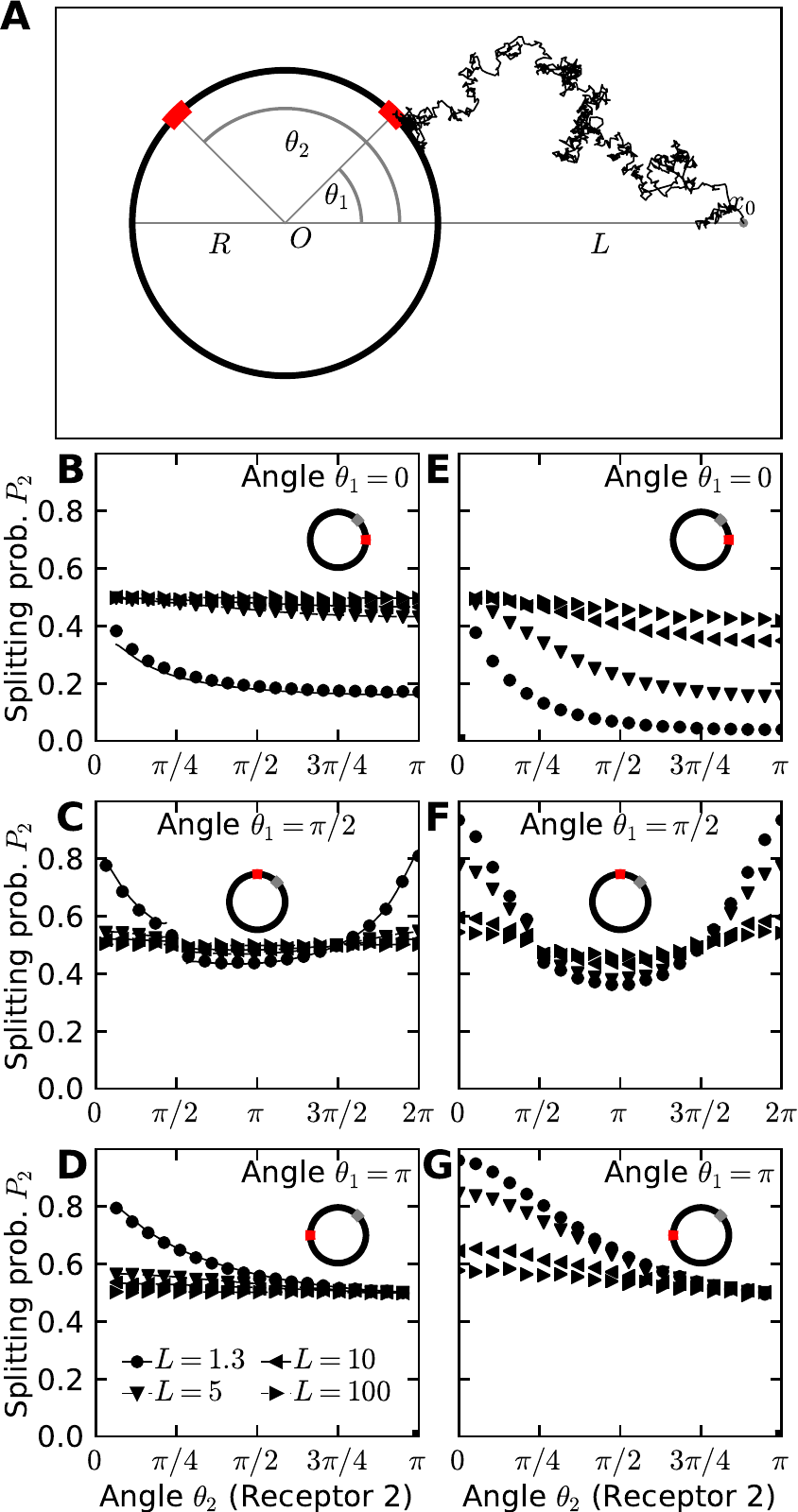}
  \caption{\textit{Diffusion fluxes to small windows on the disk surface.} (A) Schematic representation of a mixed stochastic simulation of Brownian particles released at position $x_0$ at a distance $L=|x_0|$ from the origin $O$. Two windows of size $2\epsilon$ are placed on the circumference of the disk of radius $R$ in two dimensions or the equator of a sphere in three dimensions at angles $\theta_1$ and $\theta_2$ with the $x$-axis. Brownian particles are injected at a distance $R_{e}$. (B) Splitting probability (normalized flux) at window 2 in two dimensions with angle $\theta_1=0$, (C) $\theta_1=\pi/2$ (the jump in the analytical solution at $\pi/2$ emerges due to divergence when windows overlap), and (D) $\theta_1=\pi$. Simulations (markers) are compared to analytical solutions (solid lines). (E) Splitting probability at window 2 in three dimensions (flux normalized to the total flux absorbed by any of the two windows) with angle $\theta_1=0$, (F) $\theta_1=\pi/2$, and (G) $\theta_1=\pi$.}\label{fig:disk}
\end{figure}

We now turn to case (2), where a similar asymptotic solution can be derived for the splitting probability when the domain containing the windows is a disk of radius R. The boundary conditions are similar: there are no particle fluxes except on $\p \Omega \sm \left(\p\Omega_1\cup \p\Omega_2\right)$ and the two windows $\p\Omega_1\cup \p\Omega_2$ remain absorbing (Fig. \ref{Fig1}B). The external Neumann-Green's function of a disk $D(R)$ of radius $R$ is the solution of the boundary value problem
\beq
-\Delta_{\x} G(\x,\y) & = & \delta(\x-\y) \hbox{ for } \x\, \in \,\rR^2 \setminus D(R), \label{eqDP2b_disk}\\
\frac{\p G}{\p n_{\y}}(\x,\y) & = & 0\hbox{ for } \x \,   \in\, \p D(R).
\eeq
and when $\x, \y \in \rR^2 \setminus D(R)$ it is given by
\beq
G_B(\x,\y)=\frac{-1}{2\pi }\left(\ln |\x-\y| + \ln \left|\frac{R^2}{|\x|^2}\x-\y\right| \right).\label{greens_function_disk}
\eeq
This is the sum of two harmonic functions with a singularity at  $\y  \in \,\rR^2 \setminus D(R)$  and an image singularity at $\frac{R^2}{|\y|^2}\y \in B(R)$.  A direct computation shows that $\ds \frac{\p G(\x,\y)}{\p r}|_{r=R}=0$, where $\x=r e^{i\theta}$. Following the derivation given for the half-plane above, we can use \eqref{eqDP2b_disk} directly in expression \eqref{eq:p2-halfplane} and obtain the probability to be absorbed by any of the windows. For window 2, the splitting probability is
\beq \label{eq:p2-disk}
P_2=\pi A_2&=&\frac12+ \frac{\pi}{2}\frac{G_B(\x_1,\x_0)-G_B(\x_2,\x_0)}{\{\log|\x_1-\x_2|-\log \eps\}} \\
&=&\frac12- \frac{1}{4}\frac{\ln\frac{\ds |\x_0-\x_1| \left|\ds \frac{R^2}{\ds |\x_0|^2}\x_0-\x_1\right|}{\ds |\x_0-\x_2| \left|\ds \frac{R^2}{\ds |\x_0|^2} \ds \x_0-\x_2\right|}}{\{\log|\x_1-\x_2|-\log \eps\}},
\eeq
and for window 1
\beq \label{eq:p2-disk2}
P_1=\pi A_1 =\ds \frac12+ \frac{1}{4}\frac{\ln\frac{\ds |\x_0-\x_1| \left|\frac{R^2}{|\x_0|^2}\x_0-\x_1\right|}{\ds|\x_0-\x_2| \left|\ds\frac{R^2}{\ds |\x_0|^2}\x_0-\x_2\right|}}{\{\log|\x_1-\x_2|-\log \eps\}}.
\eeq
To evaluate how the probability $P_2$ changes with the distance of the source $\x_0$ and the relative position of the windows, a comparison between Brownian simulations (see section~\ref{sec:hybrid_simulations}) versus expression~\eqref{eq:p2-disk2} is shown in Fig.~\ref{fig:disk}.
Interestingly, already at a distance of $L=10R$, the absolute difference between the fluxes $\Delta P=|P_1-P_2|$ is within $5\%$. This rapid decay renders a determination of the source direction (or equivalently the concentration difference) difficult to impossible in noisy environments. This result is independent of the window positions and $\Delta P\to 0$ as $L$ increases, see Fig. \ref{fig:disk}B-D.
\subsubsection{Diffusion in a narrow strip}
\begin{figure}[http!]
  \centering
  \includegraphics{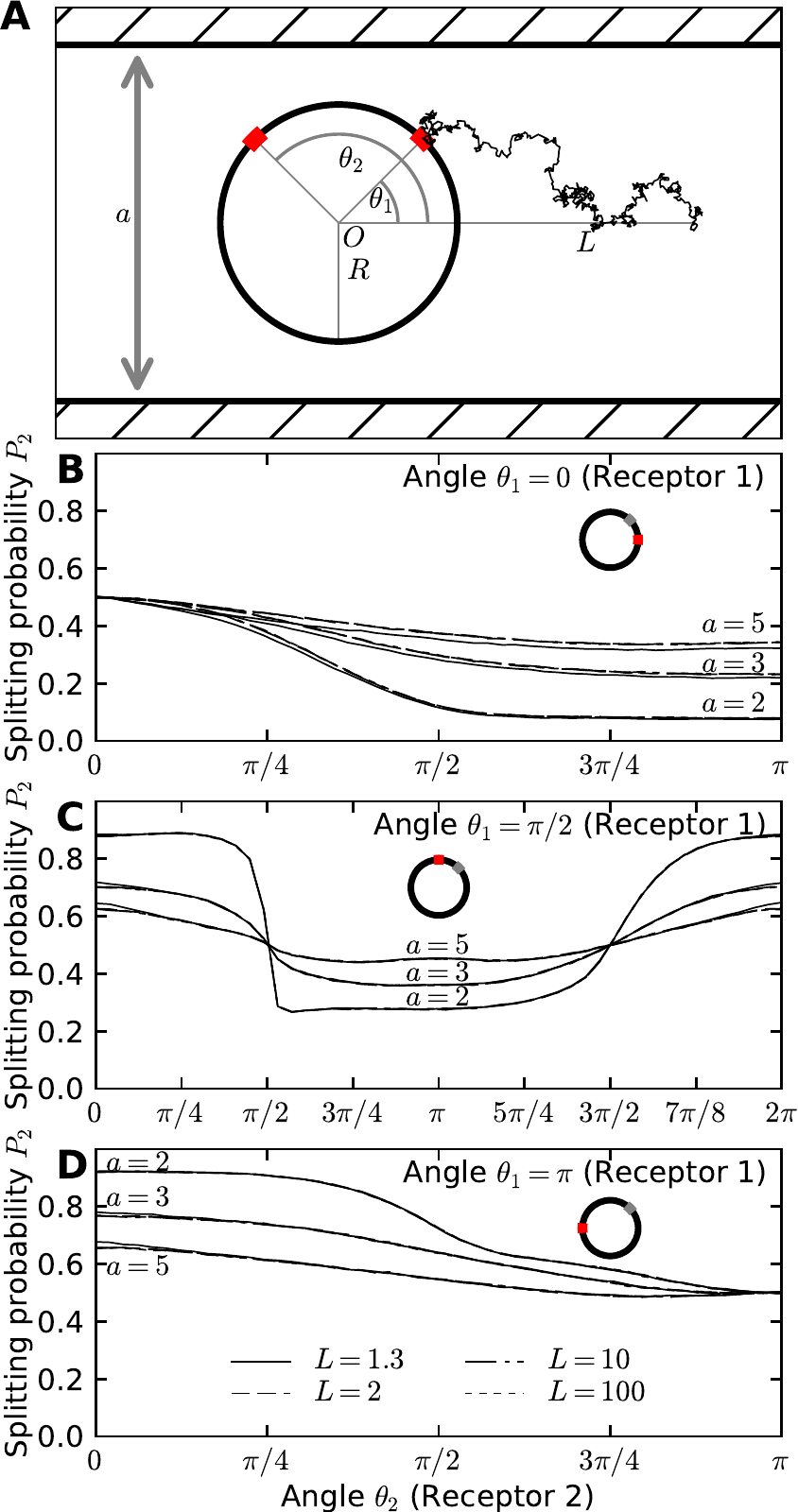}
\caption{\textbf{Diffusion fluxes to small windows for a disk in a narrow strip of width $a$.} (A) Scheme of the mixed stochastic simulations of Brownian particles confined in the strip and released at position $x_0$ at a distance $L=|x_0|$ from the origin $O$. Two windows of size $2\epsilon$ are placed on the circumference of the disk of radius $R$ at angles $\theta_1$ and $\theta_2$ with the $x$-axis. Brownian particles are injected at a distance $d$ on both sides of the disk (dashed vertical lines) and reflected from the strip walls at $y=\pm a/2$. (B) Splitting probability (normalized flux) at window 2 with angle $\theta_1=0$, (C) $\theta_1=\pi/2$, and (D) $\theta_1=\pi$.}
  \label{fig:strip}
\end{figure}

In case (3), when the test disk is located in a narrow strip (Fig. \ref{fig:strip}A), the difference of fluxes between the two windows converges asymptotically to a finite difference $\Delta P$ depending on the strip width $a$, even for large source distances $L\geq 100R$ (see Fig. \ref{fig:strip}B-D.). Indeed, the fluxes hardly show any dependence on the source distance $L$. The narrow funnel~\cite{PPR2013} between the strip and the disk prevents Brownian particles to reach a window located on the opposite side of the disk, leading to the observed effects.

\subsubsection{Splitting fluxes with many windows} \label{manywindows}
To conclude this section, we now present the general solution of equation \ref{eqDP2} in the form
\beq\label{gexp_many}
P(\x,\x_0)=G(\x,\x_0)  +\sum_{k}A_k \{\log|\x-\x_k|-\log \eps\} +C,
\eeq
where $A_1,..,A_N,C$ are $N+1$ constants to be determined.  Using the behavior near the center of each window $\x_i$,
\beq
P(\x,\y)\approx A_i \{\log|\x-\x_i|-\log \eps\},
\eeq
we obtain the ensemble of conditions for $i=1..N$
\beq\label{cond_many}
G(\x_i,\x_0)  +\sum_{k\neq i} A_k \log\frac{|\x_i-\x_k|}{\eps} +C=0.
\eeq
The last equation is given by total flux condition:
\beq
\sum_k \int_{\p\Omega_{k}} \frac{\p P(\x,\y))}{\p \n}dS_{\x} =-1.
\eeq
When the absorbing windows are well separated compared to the distance $|\x_i-\x_j| \gg 1$, a direct computation using \ref{gexp_many} gives
\beq\label{condn}
\sum_i \pi A_i =1.
\eeq
Conditions \ref{cond_many} and \ref{condn} can be rewritten in a matrix form
\beq\label{sys1}
[a] \mathcal{A} =\mathcal{B},
\eeq
{where for $i\neq j$, $i,j\le N$, $a_{ij}=\log\frac{|\x_i-\x_j|}{\eps}$, $a_{i,N+1}=a_{N+1,i}=1$ for $i\le N$, and $a_{ii}=0$ for $i=1..N+1$,
\beq
\mathcal{A}&=& (A_1,..,A_n,C)^T.\\
\mathcal{B}&=& (-G(\x_1,\x_0) ,..,-G(\x_n,\x_0),1/\pi)^T
\eeq
The matrix $[a]$ is symmetric and invertible, but does not have a specific structure, rendering it difficult to compute an explicit solution for a large number of windows in general. However, system \ref{sys1} can directly be solved numerically to find the unique solution $A_1,..,A_n$ and the constant $C$.}
\subsection{Computing the fluxes in three dimensions} \label{s:fluxes3D}
Next, we switch to the three-dimensional scenario, where general procedure is similar to the two-dimensional cases discussed in the previous section. As we shall see, the crucial difference is that here, Brownian particles can escape to infinity, while in 2D particles are guaranteed to re-visit every point on the plane. We again start with the case in which windows are located on the boundary of three-dimensional half-space and proceed by showing the explicit expressions for $n=1,2$ and $3$ windows. We conclude with showing calculations for the case in which windows are located on the surface of a ball.
\subsubsection{Computing the fluxes of Brownian particles to small windows in half--space} \label{s:HP}
To compute the fluxes to narrow windows located on the plane $\rR^2$, when the Brownian particles can evolve in $\rR_{+}^3$, we will again employ the method of matched asymptotics. The general solution of equation~\eqref{eqDP3b} is constructed using the Green's function:
\beq
\Delta G_0(\x) & = & -\delta(\x-\x_0) \hbox{ for } \x\, \in \,\rR^3_+  \label{eqDP3plus}\\
\frac{\p G_0}{\p \n}(\x) & = & 0\hbox{ for } \x \,   \in\, \p\rR_{+}^3.
\eeq
In three dimensions, the solution that tends to zero at infinity is given by
\beq\label{eqDP33}
G_0(\x,\x_0) = \frac{1}{4\pi}\left(\frac1{|\x-\x_0|}  + \frac1{|\x - \overline{\x}_0|}\right),
\eeq
where $\overline{\x}_0$ is the mirror image of $\x_0$ with respect to the plane at $z=0$. The function $w= P_0-G_0$ is the solution of
\beq
\Delta w & = & 0 \hbox{ for } \x\, \in \,\rR^3_+ \\
\frac{\p w}{\p \n}(\x) & = & 0\hbox{ for } \x \,   \in\, \p\rR_{+}^3\setminus\cup_{i=1}^N\Omega_{\eps_i}\\
w(\x) & = & \alpha_i \hbox{ for } \x \,   \in \Omega_{\eps_i}, \, i=1..N,
\eeq
where we assume that the windows $\Omega_{\eps_i}$ are circular and centered around the points $\x_i$, respectively. We also assume that $\eps_i$ is small enough such that we can approximate the Green's function as being constant over the window extent:
\beq
\alpha_i= -G_0(\x_i,\x_0).
\eeq
We construct the solution $w_j$ for each window $j$ by mapping to the electrified disk problem from electrostatics \cite{jackson2007classical}, which is equivalent to the boundary layer equation for $w_j$ near the window $\Omega_{\eps_j}$~\cite{Ward2005,Pillay2010,Ward1992}
\beq
{\cal L} w_j &\equiv&\, w_{j,\eta\eta} + w_{j,s_1s_1} + w_{c,s_2s_2}
=0\hspace{0.5em}\mbox{for}\ \eta\geq 0, \ -\infty<s_1,s_2<\infty\label{5:wc_1}\\
\partial_\eta w_j &=& \,0\hspace{0.5em}\mbox{for}\
\eta=0,\ s_1^2+s_2^2\geq \eps_j^2, \quad   w_j = 1 \hspace{0.5em}
\mbox{for} \ \eta=0, \ s_1^2+s_2^2 \leq \eps_j^2\label{5:wc_2}\\
w_j&\to&\,0\hspace{0.5em}\mbox{for}\ \rho=\eps^{-1}|\x-\x_j|\to\infty,
\label{mcdn}
\eeq
with the notations: $\eta=z$, $s_1=x$ and $s_2=y$.
The result is the classical Weber solution \cite{Ward2005,Pillay2010,Ward1992}
 \beq
 w_{j}(\sigma,\eta) = \frac{2}{\pi} \int\limits_{0}^{\infty} \frac{\sin\mu}{\mu} \, e^{-\mu
\eta/\eps_j} \, J_{0}\left( \frac{\mu \sigma}{\eps_j} \right) \, d\mu = \frac{2}{\pi}
\sin^{-1}\left(\frac{\eps_j}{L} \right), \label{5:wcsol_1}
 \eeq
where $\sigma \equiv (s_1^2 + s_2^2)^{1/2}$, $J_{0}(z)$ is the
Bessel function of the first kind of order zero. The distance $L=L(\eta,\sigma)$ is
defined by
 \beq
 L(\eta,\sigma) \equiv  \frac{1}{2} \left(  \left[ (\sigma + \eps_j)^2 + \eta^2
\right]^{1/2}  +   \left[ (\sigma - \eps_j)^2 + \eta^2 \right]^{1/2} \right) .
  \label{5:wcsol_2}
\eeq
The far-field behavior of $w_j$ in \eqref{5:wcsol_1} is given by
\beq
w_{j} \sim \frac{2\eps_j}{\pi} \left[ \frac{1}{\rho} + \frac{\eps_j^2}{6}
 \left( \frac{1}{\rho^3} - \frac{3\eta^2}{\rho^5} \right) + \cdots
  \right]\hspace{0.5em}\mbox{as} \quad \rho \to \infty , \label{5:wcff}
\eeq
which is uniformly valid in $\eta$, $s_1$, and $s_2$. Thus \eqref{5:wcff} gives the far-field expansion of $w_j$ as
\beq\label{5:w0ff}
w_j \sim \left( 1 - \frac{c_j}{\rho} + O(\rho^{-3}) \right)
\hspace{0.5em} \mbox{for} \ \rho \to \infty, \ c_j=\frac{2\eps_j}{\pi},
\eeq
where $c_j$ is the expression for the electrostatic capacitance of the circular disk of radius
$\eps_j$. The total flux is given by
\beq \label{flux}
\int_{D(\eps_i)}\frac{\p w_j}{\p \n}_{\eta=0} ds= 4\eps_i.
\eeq
Due to the linearity of the Laplace equation, we can write the general solution as a linear combination of the solutions $w_{i}$ for window $\Omega_{\eps,i}$
\beq \label{solution}
w(\x) =\sum_{i=1}^{n} a_i \alpha_i w_{i}(\x).
\eeq
The coefficients $a_i$ have to be determined using the absorbing boundary conditions
\beq
\alpha_j=\sum_{i=1}^{n} a_i \alpha_i w_{i}(\x_j), \hbox{ for } j=1...N .
\eeq
By definition, $w_{i}(x_i)=1$. Therefore, defining the matrix
\beq\mb{M}= \left( \begin{array}{cccc}
\alpha_1 & \alpha_2 w_{2}(\x_1) &\cdots & \alpha_n w_{n}(\x_1)  \\
&&&\\
\alpha_1 w_{1}(\x_2) & \ddots&&\vdots\\
&&&\\
\vdots& &\ddots&\vdots\\
&&&\\ \alpha_1 w_{1}(\x_n)& \cdots&\cdots& \alpha_n
\end{array}\right)\label{M}
 \eeq
and the approximation that for windows sufficiently far apart with the same radius $\eps$,
\beq
w_i(x_j)\approx \frac{2 \eps \alpha_i}{\pi |\x_0-\x_j|}
\eeq
we can now derive a Matrix equation. To this end, we separate $\mb{M}$ as
\beqq
\mb{M}=\mb{\Delta}_{\alpha}+ \frac{2\eps}{\pi} \mb{A},
 \eeqq
where $\mb{\Delta}_{\alpha}$ is the diagonal matrix
\beq
\mb{\Delta}_{\alpha}= \left( \begin{array}{cccc}
\alpha_1 & 0 &\cdots & 0  \\
&&&\\
0 & \ddots&&\vdots\\
&&&\\
\vdots&&\ddots&\vdots\\
&&&\\ 0& \cdots&\cdots& \alpha_n
\end{array}\right)\label{M2}
 \eeq
and $\mb{A}$ contains the off-diagonal terms:
\beq\frac{2\eps}{\pi}\mb{A}= \left( \begin{array}{cccc}
0 & \alpha_2 w_{2}(\x_1) &\cdots & \alpha_n w_{n}(\x_1)  \\
&&&\\
\alpha_1 w_{1}(\x_2) & \ddots&&\vdots\\
&&&\\
\vdots& &\ddots&\vdots\\
&&&\\ \alpha_1 w_{1}(\x_n)& \cdots&\cdots& 0
\end{array}\right).\label{M3}
\eeq
Writing $\tilde{\mb{\alpha}}$ and $\tilde{\mb{a}}$ for the vectors containing the $\alpha_j$ and the $a_j$ respectively, equation \eqref{solution} becomes
\beq \label{sysMatrix}
\left(\mb{\Delta}_{\alpha}+ \frac{2\eps}{\pi} \mb{A}\right) \tilde{\mb{a}}=\tilde{\mb{\alpha}}.
 \eeq
This can be inverted as the following convergent series
\beq \label{formal1s}
\tilde{\mb{a}}= \left( \mb{1}_M+ \frac{2\eps}{\pi} \mb{\Delta}_{\alpha}^{-1} \mb{A}\right)^{-1} \mb{\Delta}_{\alpha}^{-1} (\tilde{\mb{\alpha}}) =-\sum_{k=0}^{\infty} \ds(- \frac{2\eps}{\pi} \mb{\Delta}_{\alpha}^{-1}\mb{A})^k
\mb{\Delta}_{\alpha}^{-1} (\tilde{\mb{\alpha}}).
 \eeq
Relation \eqref{formal1s} is the formal solution for the coefficients $a_i$ in the asymptotic solution \eqref{solution}. Finally, we recall that the flux through each window is given by
\beq\label{totalflux}
\Phi_i=\int_{\Omega_{i}}\frac{\p w}{\p \n}(\y) dS_{\y} =4 \eps \pi a_i \alpha_i.
\eeq
\subsubsection{Explicit expression in the cases of \(N=1,2\) and \(3\) windows in the \(z=0\) plane} \label{ss:explicit}
Proceeding, we now show the expressions for the fluxes for one, two and three windows. In the case of one window only, solution \eqref{solution} becomes
\beq
P_0(x)=G_0(x)-G_0(x_1)\frac{2}{\pi}
\sin^{-1}\left(\frac{a_j}{L(x)} \right),
\eeq
where $L(x)$ is given in \eqref{5:wcsol_2}. Therefore, we get
\beq
\int_{\Omega_{i}}\frac{\p P_0}{\p \n}(\y) dS_{\y} =\int_{\Omega_{i}}\frac{\p G_0}{\p \n}(\y) dS_{\y} -G_0(x_1) \int_{\Omega_{i}}\frac{\p w_1(\y)}{\p \n} dS_{\y}.
\eeq
By definition $\int_{\Omega_{\eps}}\frac{\p G_0}{\p \n}(\y) dS_{\y} =0$, since $x_1$ is on the $z=0$ plane. Using \ref{eqDP33}, we retrieve the probability flux
\beq
\Phi_{\eps}=\int_{\Omega_{\eps}}\frac{\p P_0}{\p \n}(\y) dS_{\y} =\frac{2}{\pi}\frac{\eps}{|x_0-x_1|}.
\eeq
Hence given the flux $\Phi_{\eps}$, the ensemble of possible positions $\x_0$ is a sphere centered around $x_1$ with radius $\frac{2}{\pi}\frac{\eps}{\Phi_{\eps}}$.

With two windows centered at $x_1$ and $x_2$, the solution \eqref{solution} is
\beq \label{solution2}
w(x) =a_1 \alpha_1 w_{x_1}(x)+a_2 \alpha_2 w_{x_2}(x).
\eeq
System \eqref{sysMatrix} then becomes an elementary 2 by 2 matrix equation
\beq
a_1&=&\frac{1-d_{\eps}\alpha_b/\alpha_a}{1-d_{\eps}^2}\\
a_2&=&\frac{1-d_{\eps}\alpha_a/\alpha_b}{1-d_{\eps}^2},
\eeq
where $d_{\eps}=\frac{2 \eps}{\pi |x_1-x_2|}$. The flux at each window can be computed from relation \eqref{totalflux}
\beq
\label{eq:halfspacefluxn=2}
\Phi_1=\int_{x_1+\Omega_{\eps}}\frac{\p P}{\p \n}(\y) dS_{\y} =4 \eps \pi a_1 \alpha_a,\\
\Phi_2=\int_{x_2+\Omega_{\eps}}\frac{\p P}{\p \n}(\y) dS_{\y} =4 \eps \pi a_2 \alpha_b,
\eeq
with the explicit expressions
\beq \label{twowindows}
\Phi_1=\frac{2\eps}{\pi|x_1-x_0|} \ds \left( \frac{1- \ds\frac{2 \eps |x_2-x_0| }{\pi |x_1-x_2||x_1-x_0|}}{1-\ds(\frac{2 \eps}{\pi |x_1-x_2|})^2}\right),\\
\Phi_2=\frac{2\eps}{\pi|x_2-x_0|} \ds \left( \frac{1- \ds \frac{2 \eps |x_1-x_0| }{\pi |x_1-x_2||x_2-x_0|}}{1-\ds (\frac{2 \eps}{\pi |x_1-x_2|})^2}\right).
\eeq
Given the two fluxes $\Phi_1$ and $\Phi_2$ the source $x_0$ lies on the one dimensional curve described by the intersection of two surfaces in three-dimensional space.

Finally, in the case of three windows located at $x_1$, $x_2$ and $x_3$ the solution \eqref{solution} is
\beq \label{solution2_1}
w(x) =a_1 \alpha_1 w_{x_1}(x)+a_2 \alpha_2 w_{x_2}(x)+a_3 \alpha_3 w_{x_3}(x)
\eeq
Inverting the $3\times3$ matrix \eqref{sysMatrix} yields
\beq\label{exactN=3a}
a_1&=&\frac{1-d_{23}^2+\frac{\alpha_2}{\alpha_1}(d_{13}d_{23}-d_{12})+\frac{\alpha_3}{\alpha_1}(d_{12}d_{23}-d_{13})}{1-\Delta^2}\\
\label{exactN=3b}
a_2&=&\frac{1-d_{13}^2+\frac{\alpha_1}{\alpha_2}(d_{13}d_{23}-d_{12})+\frac{\alpha_3}{\alpha_2}(d_{12}d_{13}-d_{23})}{1-\Delta^2}\\
\label{exactN=3c}
a_3&=&\frac{1-d_{12}^2+\frac{\alpha_1}{\alpha_3}(d_{12}d_{23}-d_{13})+\frac{\alpha_2}{\alpha_1}(d_{12}d_{13}-d_{23})}{1-\Delta^2},
\eeq
where $d_{ij}=\frac{2 \eps}{\pi |x_i-x_j|}$ and $\Delta^2=d_{12}^2+d_{13}^2+d_{23}^2+2d_{12}d_{13}d_{23}$. Inserting (\ref{exactN=3a}-\ref{exactN=3c}) into \eqref{totalflux} and performing a second order expansion in $\eps$ leads to
\beq
\label{fluxesN=3a}
\Phi_1&=&\frac{2\eps}{\pi}\frac{1}{|x_1-x_0|}-4\eps^2\left[\frac{1}{|x_2-x_0|}\frac{1}{|x_1-x_2|}+\frac{1}{|x_3-x_0|}\frac{1}{|x_1-x_3|}\right]+O(\eps^3)\\
\label{fluxesN=3b}
\Phi_2&=&\frac{2\eps}{\pi}\frac{1}{|x_2-x_0|}-4\eps^2\left[\frac{1}{|x_1-x_0|}\frac{1}{|x_1-x_2|}+\frac{1}{|x_3-x_0|}\frac{1}{|x_2-x_3|}\right]+O(\eps^3)\\
\label{fluxesN=3c}
\Phi_3&=&\frac{2\eps}{\pi}\frac{1}{|x_3-x_0|}-4\eps^2\left[\frac{1}{|x_1-x_0|}\frac{1}{|x_1-x_3|}+\frac{1}{|x_2-x_0|}\frac{1}{|x_2-x_3|}\right]+O(\eps^3).
\eeq
When the three fluxes $(\Phi_1,\Phi_2,\Phi_3)$ are given, the source can finally be narrowed down to the single intersection point of the three surfaces described by Eqns. (\ref{exactN=3a}-\ref{exactN=3c}) and \eqref{totalflux}. To order $\eps$, these surfaces are spheres centered on their respective windows, as shown by equations~(\ref{fluxesN=3a}-\ref{fluxesN=3c}). The solution is then uniquely determined from the flux coordinates $\x_0=\x_0(\Phi_1,\Phi_2,\Phi_3)$.
\subsubsection{Computing the fluxes of Brownian particles to small targets on the surface of a ball} \label{s:ball}
To conclude this section, we present the derivation of the fluxes to an arbitrary number of small windows located on the surface of a ball $B_a$ with radius $a$ in three-dimensional space. The Brownian particles which are released at a source at position $\x_0$ outside $B_a$. The fluxes are computed from the solution of the associated Laplace's equation
\beq
D\Delta P_0(\x) & = & -\delta(\x-\x_0) \hbox{ for } \x\, \in \,\rR^3 -B_a \label{eqDP3bb}\\
\frac{\p P_0}{\p \n}(\x) & = & 0\hbox{ for } \x \,   \in\, \p B_a\setminus \Sigma_a \\
P_0(\x) & = & 0 \hbox{ for } \x \,   \in \Sigma_a
\eeq
where $\Sigma_a=S_1(\eps) \cup...\cup S_n(\eps)$ and $S_k(\epsilon)$ are non-overlapping windows of radius $\eps$ located on the surface of the ball and centered around the point $\x_k$. Contrary to the 2D case, the probability of finding particles is required to decay to zero at infinity
\beq
\lim_{|\x| \rightarrow \infty}  P_0(\x) =0.
\eeq
As in subsection \ref{s:HP}, we compute the difference $w= P_0-\tilde N$, where $\tilde N$ is the Neumann-Green's function for the external Ball defined in~\eqref{Neumann} (see appendix~\ref{appendix1}). It can be found as the solution of
\beq \label{eqfdt}
\Delta w & = & 0 \hbox{ for } \x\, \in \,\rR^3 -B_a \\
\frac{\p w}{\p \n}(\x) & = & 0\hbox{ for } \x \,   \in\, \p B_a\setminus \Sigma_a\\
w(\x) & = & \alpha_i \hbox{ for } \x \,   \in \Sigma_a,
\eeq
where we again assume that the windows $\Omega_{\eps_i}$ are small enough such that we can approximate the Neumann-Green's function as a constant over their extent
\beq
\alpha_i= -N(\x_i,\x_0).
\eeq
To solve \eqref{eqfdt}, we use Green's identity over the large domain $\Omega$,
\beq
\int_{\Omega}\left( \mathcal{N}(\x,\x_0)  \Delta w(\mathbf{x})-\tilde{p}(\mathbf{x}) \Delta \mathcal{N}(\x,\x_0) \right)d\mathbf{x}= \int_{\p \Omega} \left( \mathcal{N}(\x,\x_0) \frac{\p w(\mathbf{x})}{\p n} -w(\x) \frac{\p \mathcal{N}(\x,\x_0)}{\p n}\right).
\eeq
With expressions \eqref{eqfdt} and \eqref{Neumann}, we obtain
\beq
w(\x)=\sum_{k} \int_{S_k(\eps)} \mathcal{N}(\x,\x_k) \frac{\p w(\x)}{\p n} dS_{\x}.
\eeq
Note that the unbounded part of the surface integral in $\Omega$ converges to zero at infinity due to the decay condition \eqref{decay}. The flux to an absorbing hole is given by~\cite{Holcman2015}
\beq
\frac{\p P_0}{\p \n}(\y) =\frac{A_i}{\sqrt{\eps^2-r^2}}, \hbox{ for } \y \in S_k(\eps).
\eeq
To compute the unknown constants $A_i$, we use the Dirichlet condition at each window
\beq
\alpha_q&=&\int_{S_q(\eps)} \mathcal{N}(\x_q,\x) \frac{\p w(\x)}{\p n} dS_{\x}+\sum_{k\neq q} \int_{S_k(\eps)} \mathcal{N}(\x_q,\x_0) \frac{\p w(\x)}{\p n} dS_{\x},\\
&=&\mathcal{N}(\x_q,\x_k) \int_{S_q(\eps)} \frac{\p w(\x)}{\p n} dS_{\x}+\sum_{k\neq q} \int_{S_k(\eps)} \mathcal{N}(\x_q,\x) \frac{\p w(\x)}{\p n} dS_{\x}. \label{matrixball}
\eeq
Using Neumann's representation \eqref{Neumann2} for the singularity located on the surface of the disk, the first integral term in expression \eqref{matrixball} yields~\cite{lagache2017extended}:
\beqq
\int_{S_q(\eps)} \mathcal{N}(\x_q,\x) \frac{\p w(\x)}{\p n} dS_{\x}
&\approx &\int_{0}^{\epsilon} \left(\frac{g_0^i}{\sqrt{\epsilon^2-s^2}}+f_i(s)\right)\left(\frac{1}{2\pi Ds}+ \frac{1}{4\pi a }\log\left(\frac{s}{2a+s}\right)+O(1)\right) 2\pi s ds \nonumber \\
&=& A_k\left(\frac{\pi}{2}+\frac{\epsilon}{2a} \log\left(\frac{\epsilon}{a}\right)+B_k\epsilon \right),\nonumber
\eeqq
where $B_k$ is a constant term appearing in the third order expansion of the Green's function~\cite{lagache2017extended}. For the second term, we recall that
\beq
\int_{0}^{\epsilon}\frac{A_k}{\sqrt{\epsilon^2-s^2}} 2\pi s ds=2\pi \eps A_k,
\eeq
and for $k=1...N$ obtain the relations
\beq \label{matrixball2}
\alpha_q &=& 2\pi \eps \sum_{k\neq q} A_k \mathcal{N}(\x_q,\x_k)+A_q\left(\frac{\pi}{2}+\frac{\epsilon}{2a} \log\left(\frac{\epsilon}{a}\right)+B_k\epsilon \right).
\eeq
This can be written in Matrix form as before
\beq \label{sysMatrix2}
[\mb{\tilde M}]{\mb{ \tilde A}}=\tilde{\mb{\alpha}}.
\eeq
We separate $[\mb{\tilde M}]$ as
\beqq
[\mb{\tilde M}]=\mb{\Delta}+ \frac{2\eps}{\pi} \mb{N},
 \eeqq
where
\beq\mb{N}= \left( \begin{array}{cccc}
0 & \mathcal{N}(\x_1,\x_2) &... & \mathcal{N}(\x_1,\x_n)  \\
&&&\\
\mathcal{N}(\x_1,\x_2) & .&.&.\\
&&&\\
.&. &.&.\\
&&&\\ \mathcal{N}(\x_1,\x_n)& ...&.& 0
\end{array}\right),\label{NMat}
 \eeq
and
\beqq
\mb{\Delta}=\theta_{\eps}\mb{I}.
\eeqq
Here,
$\theta_{\eps}=\left(\frac{\pi}{2}+\frac{\epsilon}{2a} \log\left(\frac{\epsilon}{a}\right)+B\epsilon \right)$
and
\beq
\tilde{\mb{\alpha}}=
\left( \begin{array}{cc}
\alpha_1 & \\
.&\\
.&\\
\alpha_n
\end{array}\right),\quad \label{alpha2}\,
{\mb{ \tilde A}}= \left( \begin{array}{cc}
A_1 & \\
.&\\
.&\\
A_n
\end{array}\right).
\label{alpha2_1}
\eeq
Inverting the matrix yields the solution for the flux constants
\beq \label{formals}
\tilde{\mb{A}}= \left( \theta_{\eps}I+ \frac{2\eps}{\pi} \mb{\Delta}_{\alpha}^{-1} \mb{A}\right)^{-1} \mb{\Delta}_{\alpha}^{-1} (\tilde{\mb{\alpha}}) =-\sum_{k=0}^{\infty} \ds(- \frac{2\eps}{\pi} \mb{\Delta}_{\alpha}^{-1}\mb{A})^k \mb{\Delta}_{\alpha}^{-1} (\tilde{\mb{\alpha}}).
\eeq
Finally, to first order, the flux to each window is
\beq \label{flux3dd}
\begin{aligned}
  \Phi_k&=\int_{S_q(\eps)} \frac{\p P(\x)}{\p n} dS_{\x}=\int_{S_q(\eps)} \frac{\p w(\x)}{\p n} dS_{\x}=2\pi A_k\\
  &=2\pi \theta_{\eps}^{-1} (\alpha_k -\frac{4 \eps}{\theta_{\eps}}\sum_{q\neq k}\mathcal{N}(\x_q,\x_k)\alpha_k )+O((\frac{2 \pi \eps}{\theta_{\eps}})^2).
\end{aligned}
\eeq
The system of equations~\eqref{flux3dd} can be solved numerically for the flux ($A_k$ and $\Phi_k$) as a function of the source position $\x_0$ and the window locations $\x_1,\x_2,...,\x_n$. We refer to the appendix for the explicit expression in the case of a ball for the Neumann's function $\mathcal{N}$ ~\cite{lagache2017extended} computed for the exterior in three dimensions.
\section{Hybrid stochastic simulations}\label{sec:hybrid_simulations}
In the previous sections, we presented the asymptotic analysis allowing the recovery of a source position from fluxes. The last step, inverting the matrix, cannot be done explicitly and requires a numerical approach. In parallel, the result of the analysis should be compared to stochastic simulations. Therefore, in this section, we present an approach for the efficient simulation of Brownian trajectories in large (or infinite) domains with a small region of interest. It simulates detailed particle paths only in this region of interest, and avoids the explicit calculation of the trajectory elsewhere.
\subsection{General hybrid stochastic algorithms}
Historically, spatial components in a chemical reaction model were introduced to account for heterogeneous distributions of interacting molecules. There are two traditional approaches to estimate the number of synthesized molecules from these interactions: one consists of the classical reaction-diffusion equations \cite{turing1990chemical} and the other is to use stochastic simulations. Note that the simplified Gillespie's algorithm \cite{gillespie1977exact,gillespie2013perspective} summarizes the geometry into rate constants. In the past decade, several algorithms have been developed to accelerate the computation time by mixing the two methods, resulting in stochastic reaction–diffusion simulations~\cite{flegg2012two,erban2019stochastic}. Deterministic reaction-diffusion occurs in the bulk of the domain, while in a region of interest the description is purely stochastic. Coupling the two regimes at their interface is then proscribed by the particular method and specific to the problem at hand \cite{franz2013multiscale}.\\
Another type of hybrid simulation concerns the scale between discrete binding of few molecular events, modeled by Markov chains and the continuum modeled by Mass-action laws \cite{guerrier2017multiscale}.\\
In this section, we will focus on hybrid simulations that concern single trajectories and avoiding simulating the entire path of a Brownian particle. The path is only simulated close to targets of interest. These targets could represent cooperative or competitive interaction events. Therefore it is crucial to restore the stochastic behavior to collect the statistics of interest such as the probability of activating a threshold ~\cite{holcman2005stochastic,duc2010threshold,duc2012using,holcman2015post}
\subsection{Hybrid simulations based on Green's function mapping}
With Brownian motion it is always possible to naively run trajectories starting from an initial point and estimate any statistic of interest. This could for example be the arrival time at a small target. However, in open space, i.e. when the domain is infinite in at least one direction, the mean arrival time of a Brownian particle to a target is infinite in two dimensions~\cite{ito2012diffusion}, even though the probability of arrival is one. In three dimensions, the situation is even worse as the probability to hit the target is strictly smaller than one (because escape to infinity is possible). These facts render the computational effort quite prohibitive: Naive simulations become inefficient due to the very large excursions of Brownian trajectories before hitting targets of interest. We note however, that although the mean time is infinite, the splitting probability of hitting one of several windows is finite and thus estimating it can provide relevant information for many applications~\cite{dobramysl2020threed,dobramysl2020PRLTria}. In this context, hybrid stochastic simulations circumnavigates these issues of naive Brownian simulations and avoids to simulate explicitly long trajectories with large excursions and thus it circumvents the need for an arbitrary cutoff distance for our infinite domain. \\
The procedure consists of using the analytical structure when possible so that we know the spatial arrival distribution of Brownian trajectories at an arbitrary boundary encompassing target positions. Then these trajectories can be simulated only inside this region of interest. Note that the spatial structure inside the region of interest can be arbitrarily complex. As we shall see below, this procedure allows to efficiently create large ensembles of trajectories at steady-state.  Note that the achieving similar results for the transient regime is still under investigation. In the next subsections, we will consider the cases the two- and three-dimensional half-space, the 2D disk and the 3D ball. We will also discuss the case of a narrow band in two dimensions.
\subsection{Hybrid algorithm in two dimensions: half and full space} \label{s:algo}
\begin{figure}[http!]
\centering
\includegraphics[scale=0.99]{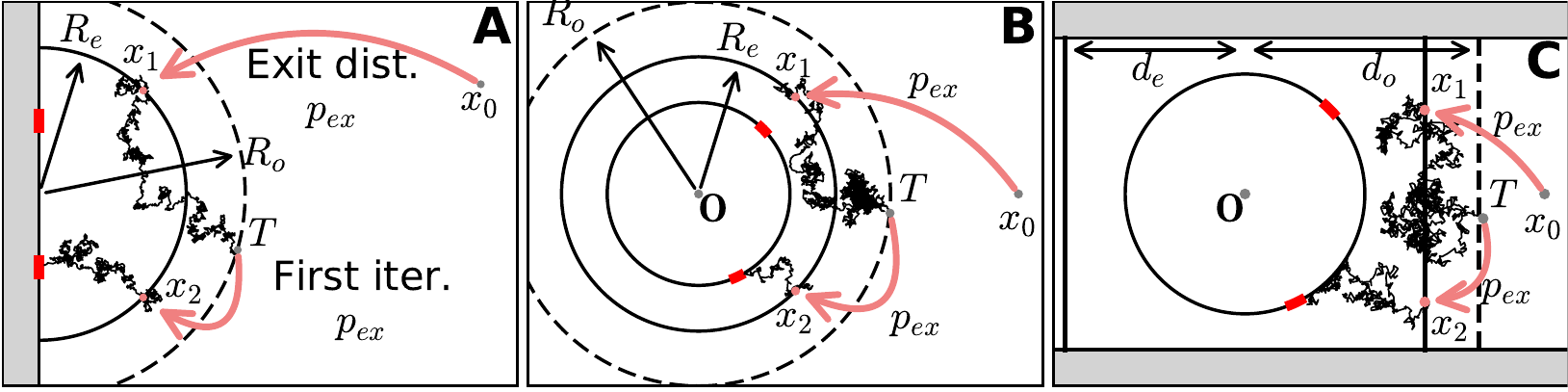}
\caption{(A) hybrid stochastic simulation procedure for two windows on the boundary of half-space. Brownian particles injected at $\x_0$ are directly place on semi-circle with radius $R_e$ according to the exit pdf $p_{HS}$ (red arrow). Inside the disk, Trajectories are generated by the Euler's scheme \ref{euler} until it passe outside the radius $R_o > R_e$, where the trajectory is terminated at point $T$ and restarted at a new position determined by the pdf $p_{HS}$ . (B) same as in (A) but for a ball. (C) hybrid simulation  scheme for windows on a disk in a strip. Brownian particles are injected at the boundary $x=d_e$ based on the exit probability distribution $p_S$. Trajectories with $x>d_o$ or $x<-d_o$ are re-injected at $x=\pm d_e$ according to $p_S$ (same procedure as in (A)).}
\label{fig:simulation}
\end{figure}

To illustrate the algorithm, we present the case where windows are located on the boundary of two-dimensional half-space, i.e. the line with $x=0$ (figure~\ref{fig:simulation}):
\begin{enumerate}
    \item The source releases a particle at position $x_0$.
    \item If $|\x(t)|>R_0$, map the particle's position to the circumference of the disk $D(R_e)$, using the mapping $p_{ex}$~\eqref{flux-halfspace}, i.e. choose a new position $\x_j$ according to $p_{ex}$ and set $\x(t)=\x_j$. As the algorithm progresses, this generates a new entry point $\x_j$ whenever a particle leaves the disk $D(R_0)$ at a crossing point $\mathbf{T}$ (see figure~\ref{fig:simulation})A.
    \item Simulate the a step of the particle trajectory inside $D(R_0)$ via the Euler-Maruyama scheme
        \beq \label{euler2D}
        \x(t)=\x(t-\Delta t)+\sqrt{2D\Delta t}\mathbf{Z}(t),
        \eeq
        where $\mathbf{Z}(t)$ is a vector of standard normal random variables and $D$ is the diffusion coefficient.
    \item When $\x(t)\cdot \mathbf{e}_y \le 0$ and $|\x(t)-\x_i|<\eps$ for any window $i$, the particle is absorbed by window $i$ the trajectory is terminated. Note that this is specific to receptor binding only, but any reaction is possible here.
    \item If the particle crossed any reflective boundary, we go back to step 3 to generate a new position. Otherwise we return to step 2.
\end{enumerate}
The disk $D(R_e)$ with radius $R_e$ corresponds to our region of interest. The larger concentric disk $D(R_0)$, $R_0>R_e$ is chosen such that frequent re-crossings of the boundary are unlikely, to enhance efficiency. Note that the absolute time of the particle trajectory is meaningless because the duration of jumps is not taken into account.\\
The situation for a disk in free space is shown in figure~\ref{fig:simulation}B. The algorithm is unchanged in this case, except that $p_{ex}$ is given by equation~\eqref{flux-disk1} and the condition for absorption at window $i$ is $|\x(t)|\le R$ and $|\x(t)-\x_i|<\eps$. Similarly, when the region of interest is inside an infinite strip with reflective walls as shown in figure~\ref{fig:simulation}C, the exit distribution $p_{ex}$ is given by equation~\eqref{eq:strip-y-distribution}. In this case, the situation when the particle is to the right and to the left of the region of interest have to be treated separately.
\subsection{Hybrid algorithm in three dimensions: half and full space} \label{s:algo3D}
\begin{figure}[http!]
\centering
\includegraphics[trim=0 275 0 0,clip,scale=0.8]{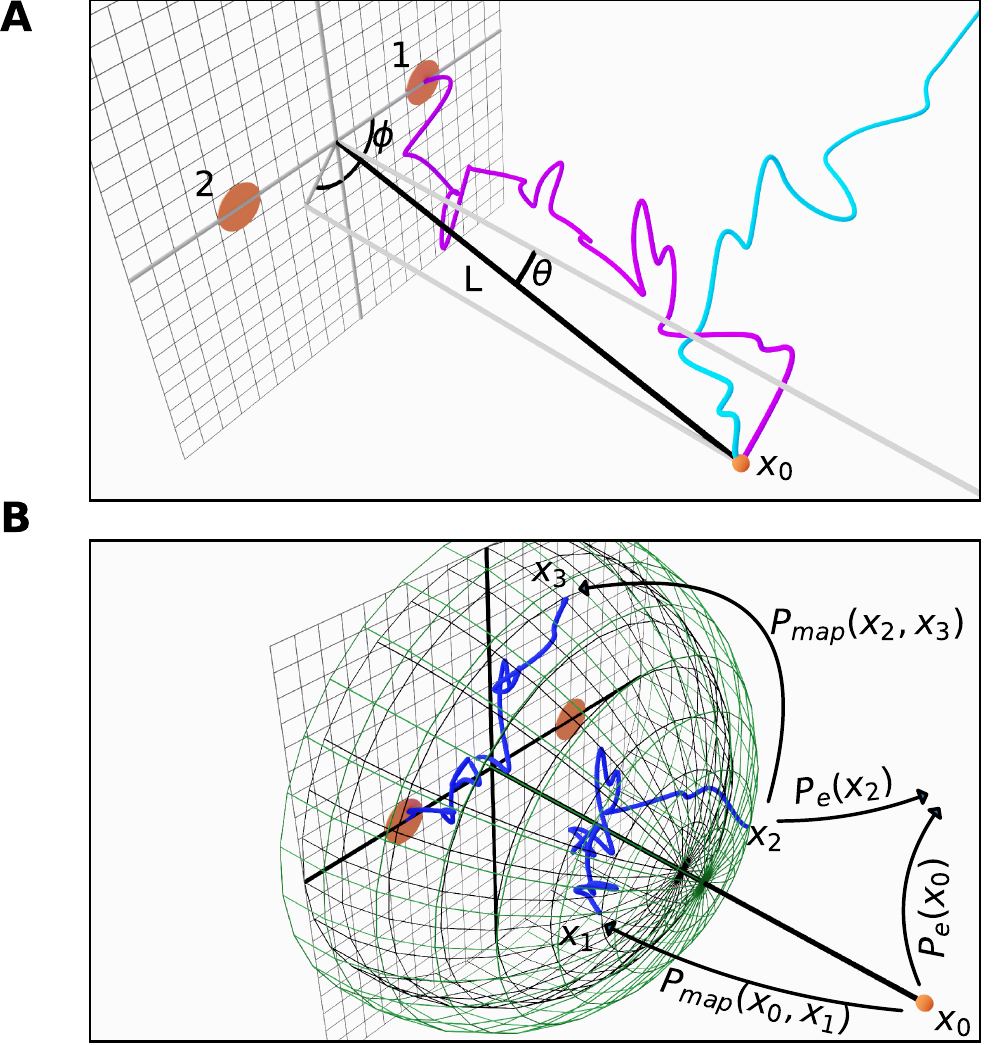}
\includegraphics[trim=0 0 0 300,clip,scale=0.8]{figure1-last2019.pdf}
\caption{Fluxes to two windows located on a plane. (A) Simulation scheme: Brownian particles are released from the source $\x_0$ at a distance $L$ from the origin located on the plane (measured in units of the distance between the windows). A particle is either absorbed by window 1 or 2 (magenta trajectory), or escapes to infinity (trajectory in cyan). (B) The position of a particle released by the source at $\x_0$ is mapped to the boundary of an imaginary half-sphere of radius $R$ enclosing the windows (black mesh) via the mapping probability distribution $P_{map}(x, y)$ given in Eq.~\eqref{eq:MappingHalfspace}. Particles perform Brownian motion inside the half-sphere until they are absorbed by a window or they leave the half-sphere with radius $R'>R$ (green mesh) upon which they are mapped back again (see algorithm listing below). (C) Flux through window 2 vs. the source zenith angle $\theta$, the azimuthal angle $\phi$ and the distance $L$. The analytical solution \eqref{eq:halfspacefluxn=2} (solid lines) is compared to simulation results (cross markers). (D) Splitting probability for a particle to hit window 2 conditional on hitting either one of the two windows.}
    \label{fig:figure3d}
\end{figure}

\begin{figure}
\centering
\includegraphics[scale=1]{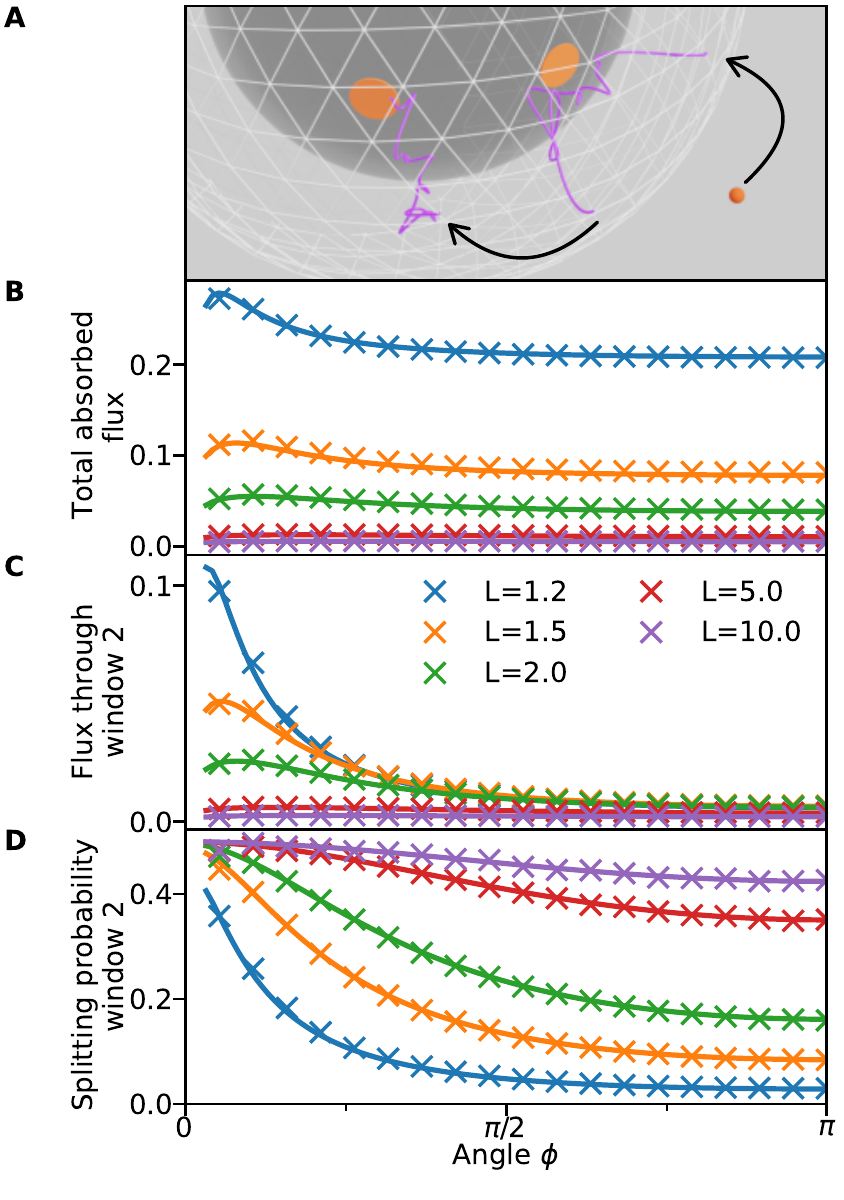}
\vspace{-25pt}
\caption{ {\bf Diffusion fluxes to two windows} (A) Reflecting ball with window one (orange) facing the source (small orange sphere) and a second window at an angle $\phi$. Brownian particles are released by the source (hybrid simulation) either absorbed by one of the windows (magenta) or can escape to infinity. (B) Total flux through both windows vs the angle $\phi$ , (C) absolute flux through window two and (D) splitting probability for a particle to hit window two. Curves are for various distance $L$ to the source: analytical results Eq. \eqref{flux3dd} (solid lines) are compared to simulation data (crosses).}
\label{fig:figure2}
\end{figure}

In three dimensions, the algorithm is very similar to the two-dimensional case discussed above. After the initial mapping of the source position to the boundary of a region of interest, a particle performs Brownian motion (Euler-Maruyama scheme) until it is absorbed by a window. A particle can also leave the test ball of radius $R'>R$ (this larger radius exists to prevent frequent mappings), upon which it is mapped back to the surface $\p B_{R}$. In detail, it consists of the following steps (Fig. \ref{fig:figure3d}A-B):
\begin{enumerate}
    \item The source releases a particle at position $\x(t=0)=\x_0$.
    \item If $|\x(t)|>R'$, allow the particle to escape to infinity with probability $P_e(x(t))$ and terminate the trajectory. Otherwise, map the particle's position to the surface of the sphere $S(R)$, using the mapping $P_{map}$~(Eq. \ref{eq:MappingHalfspace}. Again, the mapping leads to a sequence of positions $\x_1,..\x_n$ until the particle is absorbed or escapes.
    \item Time stepping is again achieved via the Euler-Maruyama where a Brownian step is given by
        \beq \label{euler}
        \x(t)=\x(t-\Delta t)+\sqrt{2D\Delta t}\mathbf{R}(t),
        \eeq
        where $\mathbf{R}(t)$ is a vector of standard normal random variables.
    \item When either $\x(t)\cdot \mathbf{e}_z \le 0$ and $|\x(t)-\x_i|<\eps$ for any $i$, the particle is absorbed by window $i$.
    \item If the particle crossed any reflective boundary, we go back to step 3 to generate a new position. Otherwise we return to step 2.
\end{enumerate}
The choice of $R$ is arbitrary as long as $S(R)$ encloses all windows with a safety margin of at least $\epsilon$. Again, as above, the radius $R'$ is chosen such that frequent re-crossings are avoided, e.g. $R'\leq R+10\sqrt{2D\Delta t}$. When the region of interest is a ball in free space, the mapping distribution is given by equation~\ref{probas}.\\
This algorithm can be used to simulate trajectories of Brownian particles at steady-state close to a region of interest of any shape. Two illustrations of the algorithm are shown for a half-space in Fig. \ref{fig:figure3d} (the test domain is half a sphere) and for a full space where the test domain is an entire ball in Fig. \ref{fig:figure2}.
\subsection{Construction of mapping distributions from explicit Green's functions} \label{sec:greens-funct-halfspace}
The stochas\-tic-analytic hybrid algorithm requires a probabilistic mapping of a point outside the region of interest to a point on its boundary $C_i$. This is given by the explicit exit distribution $p_{ex}$ computed by solving the Laplace equation with an absorbing boundary condition on $C_i$. When $p_{ex}$ is found, the initial point $x_1$ is then chosen randomly according to this distribution on $C_i$. In the subsections below we derive the exit distributions for each of the cases above.
\subsubsection{Mapping for 2D full space} \label{sec:full}
The explicit external Neumann-Green's function in $\rR^{2}$ with zero absorbing boundary condition on a disk $D(R)$ of radius R is the solution of
\beq
\begin{aligned}
\label{green3_full}
-\Delta_{\y} G(\x,\y) & = \delta(\x-\y), &\quad &\text{for}\;\; \x, \y\, \in \,\rR^{2}, \\
 G(\x,\y) & =0 &\quad &\text{for}\;\; \y\, \in\, \p B \cap \rR^{2},\ \x \in \,\rR^{2}.
\end{aligned}
\eeq
The solution is constructed via the method of images~\cite{melnikov2011green} and given by
\beq
\label{greenfct-disk}
G(\x,\y)=-\frac{1}{2\pi}\left(\ln|\x-\y|-\ln\left|\x-\frac{R^2}{|\y|^2}\y\right|-\ln\frac{|\y|}{R}\right).
\eeq
Thus the probability distribution of exit points $p_{ex}$ on the boundary $\p D(R)$ when the source is located at position $\x_0$ is computed by normalizing the flux \cite{DSP},
\beq
p_a(\y|\x_0)=\ds \frac{\ds \frac{\p G }{\p \n_{y}}(\y,\x_0)}{\ds \oint_{\p D(R)}\frac{\p G }{\p \n_{y}}(\y,\x_0)dS_{\y}},
\eeq
Due to symmetry, the flux can be straightforwardly expressed in polar coordinates $r=|\x|$, $\rho=|\y|$ and the angles $\theta$ and $\theta'$ of the vectors $\x$ and $\y$ respectively:
\beq
\label{flux-disk1}
p_{ex}(r,\theta;\theta')=R\frac{\p G}{\p
  \rho}\Bigl|_{\rho=R}=\ds \frac{1}{2\pi}\frac{\ds \frac{r^2}{R^2}-1}{\ds \frac{r^2}{R^2}-2\frac{r}{R}\cos(\theta-\theta')+1}.
\eeq
Due to the recurrence property of Brownian motion in two dimensions $\int_{\p D(R)}\frac{\p G }{\p \n_{y}}(\x,\y)dS_{\y}=1$. The probability density \ref{flux-disk1} is then used to determine the position of the sequence of points $\x_1, \x_2,..$, until the particle is finally absorbed at one of the target windows. 
\subsubsection{Mapping for 2D half-space} \label{sec:halfspace}
The Neumann-Green's function $G_{HS}$ for the half-space $\rR^{2}_+$ with zero absorbing boundary condition on a half a disk of radius R is the solution of the boundary value problem
\beq \begin{aligned}
\label{green3_halfspace}
-\Delta_{\y} G_{HS}(\x,\y) & = \delta(\x-\y), &\quad &\text{for}\;\; \x, \y\, \in \,\rR^{2}_+, \\
 \frac{\p G_{HS}}{\p n_{\y}}(\x,\y) & =  0, &\quad &\text{for}\;\; \y\, \in\, \p\rR^{2}_+,\ \x \in \,\rR^{2}_+,\\
 G_{HS}(\x,\y) & =0 &\quad &\text{for}\;\; \y\, \in\, \p B \cap \rR^{2}_+,\ \x \in \,\rR^{2}_+,
\end{aligned}
\eeq
Using the method of image charges, the solution is obtained from the Green's function $G(\x,\y)$ with an absorbing disk in the free space. The Green's function for the half-space $\rR^{2}_+$ is then constructed by symmetrizing with respect to the reflecting z-axis:
\beq
\label{greenfct-halfspace}
G_{HS}(\x,\y) & =& \frac{1}{2}[G(\x,\y)+G(x,\tilde \y)] -\frac{1}{4\pi} \Bigl( \left(\ln|\x-\y|-\ln\left|\x-\frac{R^2}{|\y|^2}\y\right|-\ln\frac{|\y|}{R}\right)+\\
&&\left(\ln|\x-\tilde\y|-\ln\left|\x-\frac{R^2}{|\tilde\y|^2}\tilde\y\right|-\ln\frac{|\tilde\y|}{R}\right)\Bigr), \nonumber
\eeq
where $\tilde \y$ is the mirror reflection of $\y$ on the vertical axis. The exit probability distribution is the flux through the absorbing half disk boundary
\beq \label{flux-halfspace}
\begin{aligned}
p_{ex}(r,\theta;\theta')=2R\ds\frac{\partial G}{\partial \rho}\Bigl|_{\rho=R}
                         =\ds\frac{\ds\frac{r^2}{R^2}-1}{2\pi}&\left[\frac{1}{\ds 1-2\frac{r}{R}\cos(\theta-\theta')+\frac{r^2}{R^2}}\right.\\ &\;\;+\left.\frac{1}{\ds 1+2\frac{r}{R}\cos(\theta+\theta')+\frac{r^2}{R^2}}\right],
\end{aligned}
\eeq
where the length in polar coordinates are $r=|\x|$, $\rho=|\y|$ and
 the angles $\theta$ and $\theta'$ of $\x$ and $\y$ are given with respect to the horizontal axis respectively.
\subsubsection{Mapping for the 2D semi-strip} \label{sec:greens-function-semistrip}
Computing the mapping for the case of the region of interest inside a (semi-)infinite strip with reflective boundaries (figure~\ref{fig:simulation}C) requires the Neumann-Green's function $G_{Se}$ of the semi-strip of width $a>0$
\beq
\Omega_a=\{(x_1,x_2)\in \rR^2 |x_1>0,0<x_2<a\}.
\eeq
A zero absorbing boundary condition is imposed on the boundary $\p\Omega_1=\{(0,x_2)|0<x_2<a\}$ and a reflecting boundary condition on the rest of the strip $\p\Omega_2=\{(x_1,0)|x_1>0\}\cup\{(x_1,a)|x_1>0\}$ (Fig. \ref{fig:simulation}C). The function $G_{Se}$ is solution of the boundary value problem (see appendix~\ref{sec:green_band})
\beq
\begin{aligned}
\label{green-semistrip}
-\Delta_{\y} G_{Se}(\x,\y) & = \delta(\x-\y), &\quad &\text{for}\;\; \x, \y\, \in \,\Omega, \\
 \frac{\p G_{Se}}{\p n_{\y}}(\x,\y) & =  0, &\quad &\text{for}\;\; \y\, \in\, \p\Omega_2,\ \x \in \,\Omega,\\
 G_{Se}(\x,\y) & =0 &\quad &\text{for}\;\; \y\, \in\, \p\Omega_1,\ \x \in \,\Omega.
\end{aligned}
\eeq
As before, the normalized flux is the distribution of exit points \cite{DSP}. Hence, the exit probability distribution $p_{ex}(x_2;y_1,y_2)$ is given explicitly by the flux through the boundary $\p\Omega_1$
\begin{equation}
  \label{eq:strip-y-distribution}
  p_{ex}(x_2;y_1,y_2)=\frac{\p G_{Se}}{\p y_1}\Bigl|_{y_1=0}=\frac{\sinh\omega y_1}{2a}\Bigl[\frac{1}{\cosh\omega y_1-\cos\omega(x_2+y_2)}+\frac{1}{\cosh\omega y_1-\cos\omega(x_2-y_2)}\Bigr]\,.
\end{equation}
\subsubsection{Mapping for the 3D half-sphere on a reflecting plane}\label{a:mapping}
The mapping of a particle with original position $|\x|>R$ to a position on the surface of the half-sphere with radius $R$ is given by the diffusive flux through this surface with absorbing boundary conditions. Hence, we need to find the Green's function for the infinite domain $\mathbb{R}^3_+-B(R)$ with Dirichlet boundary conditions at $\p B(R)$:
\beq
    \begin{aligned}
        -\Delta G(\x, \y) &= \delta(\x-\y) \quad \text{for} \quad \x\in\mathbb{R}^3_+\backslash B(R) \\
        \frac{\p G}{\p \n}(\x,\y) &= 0 \quad \text{for} \quad \x\in\p\Omega\\
        G(\x,\y) &= 0 \quad \text{for} \quad \x\in\p B(R)
    \end{aligned}
\eeq
Again, given the symmetries of the reflective half-plane and the sphere, we apply the method of images, starting with the Green's function for the absorbing ball in free space~\eqref{eq:GreenFctMap3DBall}. The solution of this problem is
\beq\label{eq:GreenFctMap3DHalfspace}
G(\x, \y) = -\frac{1}{4\pi}\left[\frac{1}{|\x-\y|} - \frac{|\x|}{R}\frac{1}{|\x-\y|\x|^2/R^2|} + \frac{1}{|\x-\tilde{\y}|} - \frac{|\x|}{R}\frac{1}{|\x-\tilde{\y}|\x|^2/R^2|}\right],
\eeq
where $\tilde{\y}$ is the reflected image of $\y$ through the plane. The probability distribution for the mapped positions is therefore
\beq \label{eq:MappingHalfspace}
P(\x,\y)=\frac{1}{\sqrt{R^2+\rho^2-R\rho\kappa}^3} +\frac{1}{\sqrt{R^2+\rho^2-R\rho\tilde{\kappa}}^3}\,,
\eeq
where
\beq
\kappa=\cos(\theta-\theta')(\cos[\phi-\phi']+1)+\cos(\theta+\theta')(\cos[\phi-\phi']-1) \\ \tilde{\kappa}=\cos(\theta-\theta')(\cos[\phi-\phi']-1)+\cos(\theta+\theta')(\cos[\phi-\phi']+1),
\eeq
$\phi$ and $\phi'$ are the polar angles of $\x$ and $\y$ in the $x-y$ plane and $\theta$ and $\theta'$ are their respective angles with the $z$-axis.
\subsubsection{Mapping for the 3D ball in free space} \label{a:mappingball}
The mapping of a particle with original position $|\x|>R$ is given by the diffusive flux through an absorbing ball with radius $R$. Therefore, we require the Green's function for the infinite domain $\mathbb{R}^3/B(R)$ with Dirichlet boundary conditions at $\p B(R)$:
\beq
\begin{aligned}
-\Delta G(\x, \y) &= \delta(\x-\y) \quad \text{for} \quad \x\in\mathbb{R}^3\backslash B(R) \\
G(\x,\y) &= 0 \quad \text{for} \quad \x\in\p B(R)
\end{aligned}
\eeq
Again, this is straightforward to solve via the method of images and gives
\beq\label{eq:GreenFctMap3DBall}
 G(\x, \y) = -\frac{1}{4\pi}\left[\frac{1}{|\x-\y|} - \frac{|\x|}{R}\frac{1}{|\x-\y|\x|^2/R^2}\right].
\eeq
The boundary flux is then
\beq \label{eq:pmap_3dball}
\frac{\p G}{\p r}(r=R, \y)=\frac{1}{4\pi}\frac{\beta^2-1}{(1+\beta^2-2\beta\cos\gamma)^{3/2}},
\eeq
where $r=|\x|$, $\beta=|\y|/R$ and $|\x||\y|\cos\gamma=\x\cdot\y$.
Integration over the surface of the ball yields
\beq \label{intfl}
\int_{\p B(R)}P(\x,\y)dS_{\x}=\beta^{-1}=\frac{R}{|\y|},
\eeq
which is the probability for a particle hitting the ball before escaping to infinity.
The mapping probability distribution is then obtained by normalizing~\eqref{eq:pmap_3dball}
\beq\label{probas}
 P(\x,\y)=\frac{|\y|}{R}\frac{1}{4\pi}\frac{\beta^2-1}{(1+\beta^2-2\beta\cos\gamma)^{3/2}}
\eeq
with $|\x|=R$, $|\x||\y|\cos\gamma=\x\cdot\y$ and $\beta=|\y|/R$.
A random new location on the ball of radius $R$ can then be generated by using the probability \eqref{probas}.
\section{Applications of the hybrid algorithm}
We proceed by discussing applications of both the hybrid algorithm presented in the previous section and the analytical expressions obtained for the fluxes in section~\ref{sec:analysis_fluxes}. We first look at the splitting probabilities for two windows in two and three-dimensional half-space, followed by three windows in three-dimensional space. Subsequently, we discuss how the source position can be recovered in two or three dimensions from the fluxes to three or more windows.
\subsection{Computing the splitting probability in 2D}
\begin{figure}[http!]
\centering
\includegraphics[scale=0.98]{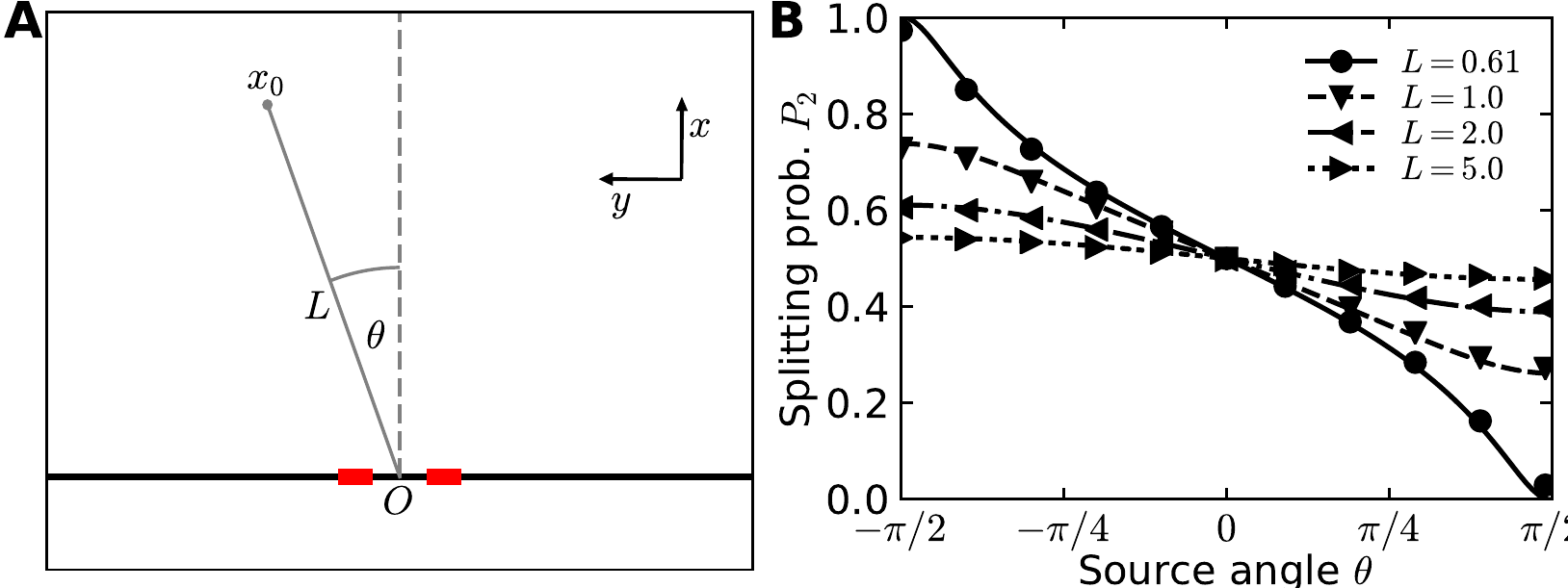}
\caption{Diffusion fluxes to small windows on the boundary of a half-plane. (A) Particles are released at the source $\x_0$ at a distance $L=|\x_0|$ from the origin forming an angle $\theta$ with the $x$-axis. (B) Splitting probability (normalized flux) at window 2 as a function of $\theta$ for different $L$. {The exact analytical solution given in equation~\eqref{eq:p2-halfplane2} (lines) is compared to hybrid simulations (markers).}}
\label{Fig-result}
\end{figure}

As discussed in section~\ref{sec:hybrid_simulations}, hybrid algorithms can be used to compute the splitting probability between small windows located on the surface of a domain. The biological background for this is binding of chemical ligands to receptors, where binding need to be collected. In two-dimensional half-space, for two windows located on the x-axis (i.e. the line $y=0$) and Brownian particles being released at position $\x_0$ (see figure \ref{Fig-result}A), the splitting probability can be computed analytically. The expression is given in equation~\eqref{eq:p2-halfplane}). Figure~\ref{Fig-result}B compares this with results obtained from the hybrid stochastic-analytical simulations as a function of the distance of the source distance $L=|\x_0|$ and the direction angle $\theta$.\\
Other examples for the splitting probability with two windows are shown for a disk in the full 2D plane (Figure \ref{fig:disk}B-G) or in a band with reflecting boundaries (Fig. \ref{fig:strip}B-D). In all these cases, the splitting probability quickly tends to $1/2$ when the distance to the source increases; i.e. the ability to detect the direction to the source decays. The exception is when the band is very narrow and one window is facing the source while the other is on the opposite side of the disk.
\subsection{Computing the fluxes to two windows in 3D}
As in the 2D case, the 3D hybrid algorithm agrees with the analytical formula given in equation~\eqref{twowindows}: Figure~\ref{fig:figure3d}C-D) shows the flux $\Phi_2$ through window 2 and the splitting probability $p_2=\frac{\Phi_2}{\Phi_1+\Phi_2}$ for a continuous zenith angle $\theta$, various source distances $L=d(0,\x_0)$ and the azimuthal angle $\phi$ either set to zero or to $\pi/4$. As $L$ increases, the splitting probability increases quickly and converges to $1/2$, hence the difference in the flux between the windows becomes small. Just as in two dimensions, this suggests that determining the direction of the source becomes increasingly difficult already for rather small distances $L$ - i.e. when $L$ on the order of 10 times the distance between the two windows.
\subsection{Computing the fluxes to three windows in 3D}
\begin{figure}[http!]
\centering
\includegraphics[scale=0.7]{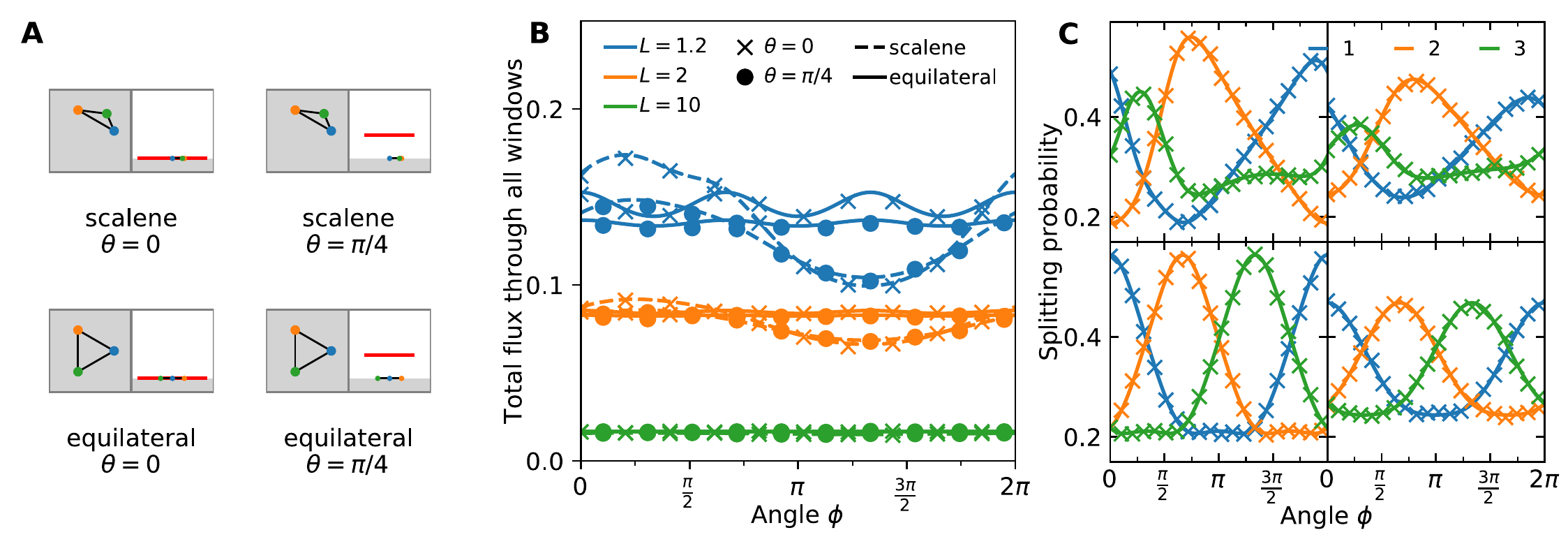}
\caption{Fluxes to three windows located on the boundary of half-space. (A) Window configurations in the plane. The windows are arranged either in an equilateral ($\alpha=2\pi/3$, $\beta=-2\pi/3$) or a scalene ($\alpha=2\pi/3$, $\beta=0.4$) triangle, with a circumcircle radius of one. The source is kept at a distance $L$ and we varied its azimuthal angle $\phi$ continuously between $0$ and $2\pi$ while the zenith angle $\theta$ equals $0$ or $\pi/4$. (B) Total summed flux through all windows for the four different configurations as a function of the azimuthal angle $\phi$ and the source distance $L$. (C) Splitting probability for particles to hit a given window.}
\label{fig:figure2b}
\end{figure}

In 3D, three windows can be arranged in a continuum of configurations in a plan. We show two representative example, an equilateral triangle and a scalene configuration, both forming a triangle inscribed into a circle with unit radius and centered on the origin. In the scalene case, the angle between window 1 and 2 is $\alpha=2\pi/3$ and the angle between window 1 and 3 is $\beta=0.4$. The source position varies on a circle in a plane above and parallel to the boundary plane containing the windows at $z=0$. It is at a distance $L$ to the origin and the distance perpendicular to the $z=0$ plane is $L\sin\theta$, with the zenith angle $\theta$. The radius of the circle is given by $L\cos\theta$. \\
These two configurations allow us to compare the symmetrical (equilateral) arrangement with a strongly anisotropical one. The total flux through all three windows over the in-plane source position angle $\phi$ for $\theta=0$ and $\pi/4$, reveals that for a source positioned very close to the windows, $L=1.2$, about $15\%$ of the flux is captured by the windows with the remainder escaping to infinity (Fig. \ref{fig:figure2b}B). \\
For a source far away from the windows ($L=10$), the captured flux decreases to $\sim 2\%$. Neither the window configuration nor the angle $\theta$ has much influence on the total flux except when the source is very close. There is good agreement between analytical and hybrid simulations for the total flux and the splitting probabilities $p_i=\frac{\Phi_i}{\Phi_1+\Phi_2+\Phi_3}$, $i=1...3$ (Fig. \ref{fig:figure2b}C). The peaks in the splitting probability indicate the in-plane angles $\phi$ at which the source is closest to the corresponding window.
\subsection{Recovering the source position in 2D} \label{sec:3receptod}
\begin{figure}[http!]
\centering
\includegraphics[scale=0.99]{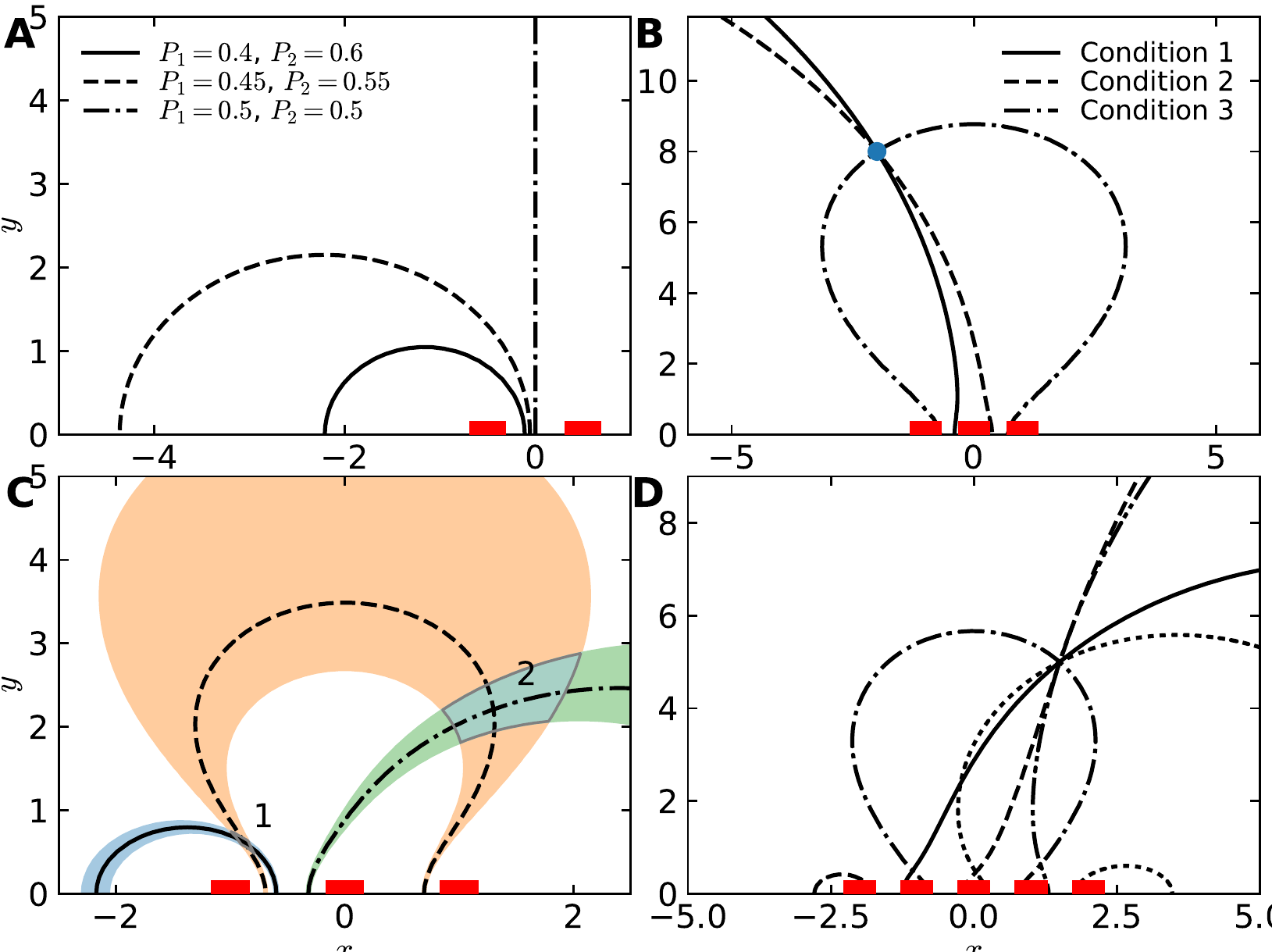}
\caption{Reconstruction of the source position in the half-plane. (A) Two windows placed a distance $d=1$ apart allow the recovery of the source position up to a curve. Three different flux configurations are shown. (B) Three windows positioned at $y=-1, 0, 1$ yield two independent curves, the intersection of which is the position of the source {at $\x_0=(-2, 8)$}. The redundant third condition is shown for completeness. (C) The shaded areas indicate the uncertainty resulting from fluctuations in the fluxes with an amplitude of $\eta=0.005$. The resulting sensitivity of the reconstructed source position (overlapping shaded areas 1 and 2) is highly inhomogeneous. (D) Five windows with nearest-neighbor distance $d=1$ yield four independent curves. The point of intersection of all curves is the recovered source position.}
\label{Fig3}
\end{figure}

\begin{figure}[http!]
  \centering
 \includegraphics{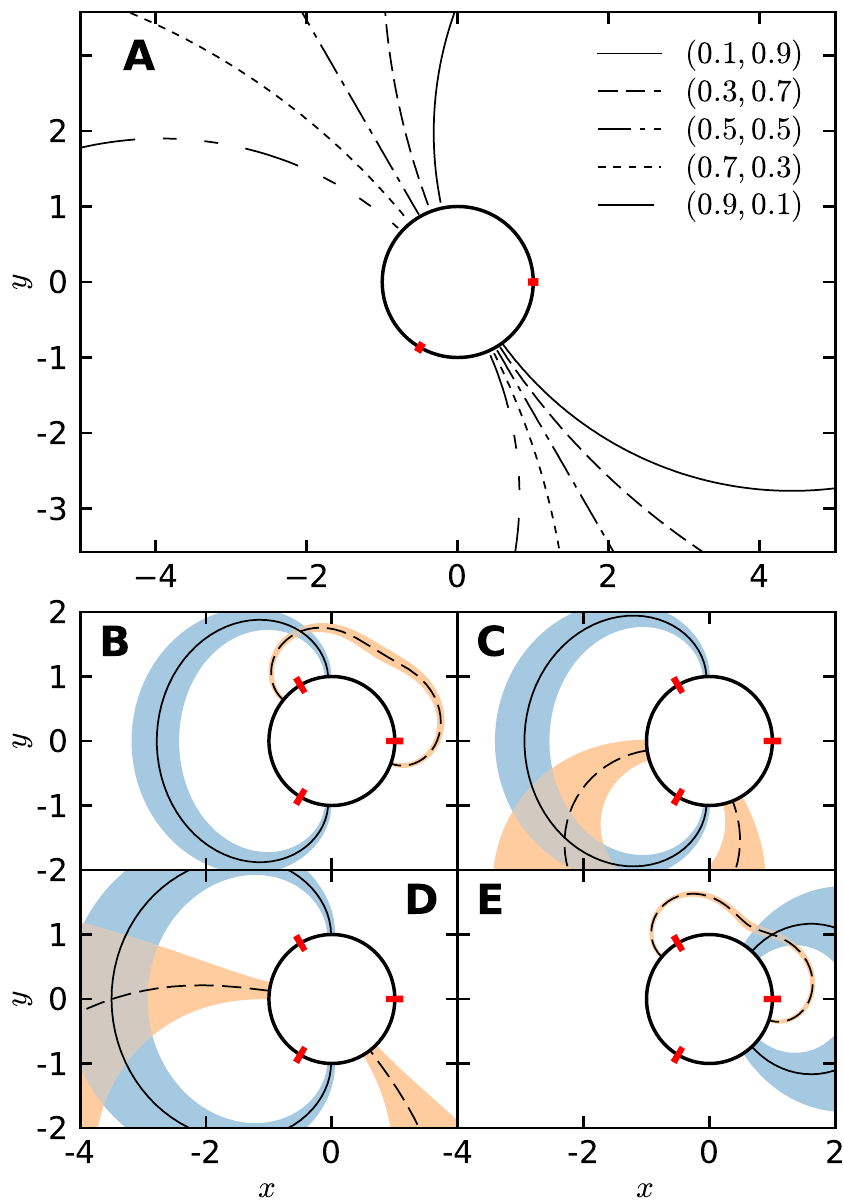}
  \caption{\textbf{Recovery or triangulation of the source position from fluxes to small absorbing windows}. (A) Two windows are positioned on a detecting disk with an angular spacing of $2\pi/3$. This arrangement allows recovery for the position $\x_0$ only a one dimensional curve. The curves are the ensemble of solutions for $\x_0$ computed from equation~(\ref{eq:p2-disk}) for the fluxes $(P_1, P_2)$ displayed in the figure legend. { (B-E) Intersection of the solid and dashed curves computed from equation~(\ref{eq:p2-disk}) for three absorbing windows. In that case, a unique position is recovered for the source position $\x_0$. We provide several examples: the original source positions are (B) $\x_0=(-0.5, 1.7)$, (C) $\x_0=(-2.3, -1.5)$, (D) $\x_0=(-3.5, 0) $ and (E) $\x_0=(1, 1)$. The amplitude of fluctuations for a fixed uncertainty $\eta=0.15$ added to the fluxes (chosen arbitrarily for illustration purposes) is represented in shaded areas.  The resulting uncertainty in the recovered source position  corresponds to overlap of the shaded regions .}}
 \label{fig:pos}
\end{figure}

To reconstruct the location of a source from the measured fluxes, at least three windows are needed. Due to recurrence of Brownian motion in 2D, particles are guaranteed to be absorbed by one of the windows. Hence the sum of the fluxes to all windows is one, and the fluxes are not independent quantities. With two windows, we therefore only have one piece of information where two are needed to recover the two coordinates of the source. The most we can do is to narrow the source location down to a curve in the 2D plane (see Fig.~\ref{Fig3}A for windows on the half-plane and Fig.~\ref{fig:pos}A for windows on the 2D disk). \\
Reconstructing the source location $\x_0$ from the splitting probabilities for three windows, requires the inversion of system \ref{cond_many}.  For three windows, the general solution is given by
\beqq
\begin{aligned}
    P(\x,\x_0)=&G(\x,\x_0)  +A_1 \{\log|\x-\x_1|-\log \eps\}+A_2 \{\log|\x-\x_2|-\log \eps\}\\
               &+A_3 \{\log|\x-\x_3|-\log \eps\} +C,
\end{aligned}
\eeqq
where $A_1,A_2,A_3,C$ are constants to be determined. Following the steps of section \ref{manywindows}, the three absorbing boundary conditions for $P(\x,\x_0)$ lead to the system of equations
\beq\label{cond2}
G(\x_1,\x_0)  +A_2 \{\log|\x_1-\x_2|-\log \eps\}+A_3 \{\log|\x_1-\x_3|-\log \eps\} +C=0 \nonumber\\
G(\x_2,\x_0)  +A_1 \{\log|\x_2-\x_1|-\log \eps\} +A_3 \{\log|\x_2-\x_3|-\log \eps\}+C=0\nonumber\\
G(\x_3,\x_0)  +A_1 \{\log|\x_1-\x_3|-\log \eps\}+A_2 \{\log|\x_2-\x_3|-\log \eps\} +C=0,
\eeq
with the normalization condition for the fluxes
\beq \label{compatibility3}
\pi A_1+\pi A_2+\pi A_3=1.
\eeq

With $ \Delta_{123}=\left(\log\frac{d_{13}d_{12}}{d_{32}\eps}\right)^2-4\log \frac{d_{12}}{\eps} \log \frac{d_{13}}{\eps}$, and
using the notations $d_{ij}=|\x_i-\x_j|$, we obtain the solution
\beq\label{a22}
A_2=\frac{  \log\frac{d_{13}d_{12}}{d_{32}\eps}(G_{30}-G_{10} +\frac{1}{\pi} \log\frac{d_{13}}{\eps})-( G_{1 0}-G_{2 0}+\frac{1}{\pi}\log\frac{d_{12}}{\eps})  \log\frac{d_{13}^2}{\eps^2  })}{\Delta_{123}}.
%
\eeq
\beq\label{a33}
A_3=\frac{  \log\frac{d_{13}d_{12}}{d_{32}\eps}(G_{20}-G_{10} +\frac{1}{\pi} \log\frac{d_{12}}{\eps})-( G_{1 0}-G_{3 0}+\frac{1}{\pi}\log\frac{d_{13}}{\eps})  \log\frac{d_{12}^2}{\eps^2  })}{\Delta_{123}}
%
\eeq
and
\beq\label{a11}
A_1=\frac{1}{\pi}-A_2 -A_3.
\eeq
Hence, we can specify the flux values $\alpha>0$ and $\beta>0$ such that $\alpha+\beta<1$,
\beq \label{compatibility1}
\alpha&=&\pi A_1=\int_{\p\Omega_{1}} \frac{\p P(\x,\y))}{\p \n}dS_{\x}\\
\nonumber
\beta&=&\pi A_2=\int_{\p\Omega_{2}} \frac{\p P(\x,\y))}{\p \n}dS_{\x}.
\eeq
Due to the normalization condition \ref{compatibility}, the flux condition on window 3 is redundant.   The source position can be recovered numerically by inverting system \ref{compatibility1} and solving for $\x0$ using expressions \ref{a22}-\ref{a33} (see appendix~\ref{sec:triang_algo_2D_halfplane}. This procedure is valid regardless of the spatial structure of the domain as long as the Green's function can be found. The resulting two curves intersect at $\x0$, as shown in figure \ref{Fig3}B for when the three windows are located on the boundary of 2D half-space~\cite{dobramysl2018mixed}. Figure~\ref{fig:pos}B-E shows the case of a disk on the full two-dimensional plane~\cite{dobramysl2018reconstructing}. \\
Measurement uncertainty in the steady-state fluxes will influence the accuracy of the recovered source location $\x_0$. A possible model is to add a small perturbation to the fluxes, such that $\alpha=\alpha_0(1+\eta),\beta=\beta_0(1+\eta)$ with $|\eta| \ll 1$ in relation \ref{compatibility1}. Direct Numerical computation reveals the resulting uncertainty in the position $\x_0$ (intersection of the colored regions) in Figs.~\ref{Fig3}C (for the half-space case) and~\ref{fig:pos}B-E (for the disk). It shows a highly non-linear spatial dependency, quantified by the relative sizes of the areas labelled 1 and 2.\\
Finally, adding more windows allows a refinement of the reconstruction of the source: For e.g. 5 windows (Fig.~\ref{Fig3}D), the source is located at the intersection of all curves for a given set of fluxes. There are other points at which two curves intersect, however, there is only one location where more than two curves (and all curves simultaneously) intersect, which corresponds to the source position. Therefore, having more than three windows could reduce the area of the uncertainty region when the fluxes contain steady-state measurement error, however an exact formula for reduction is yet to be found.
\subsection{Triangulation the source in 3D}\label{triangulating_source}
\begin{figure}
  \centering
  \includegraphics[scale=0.99]{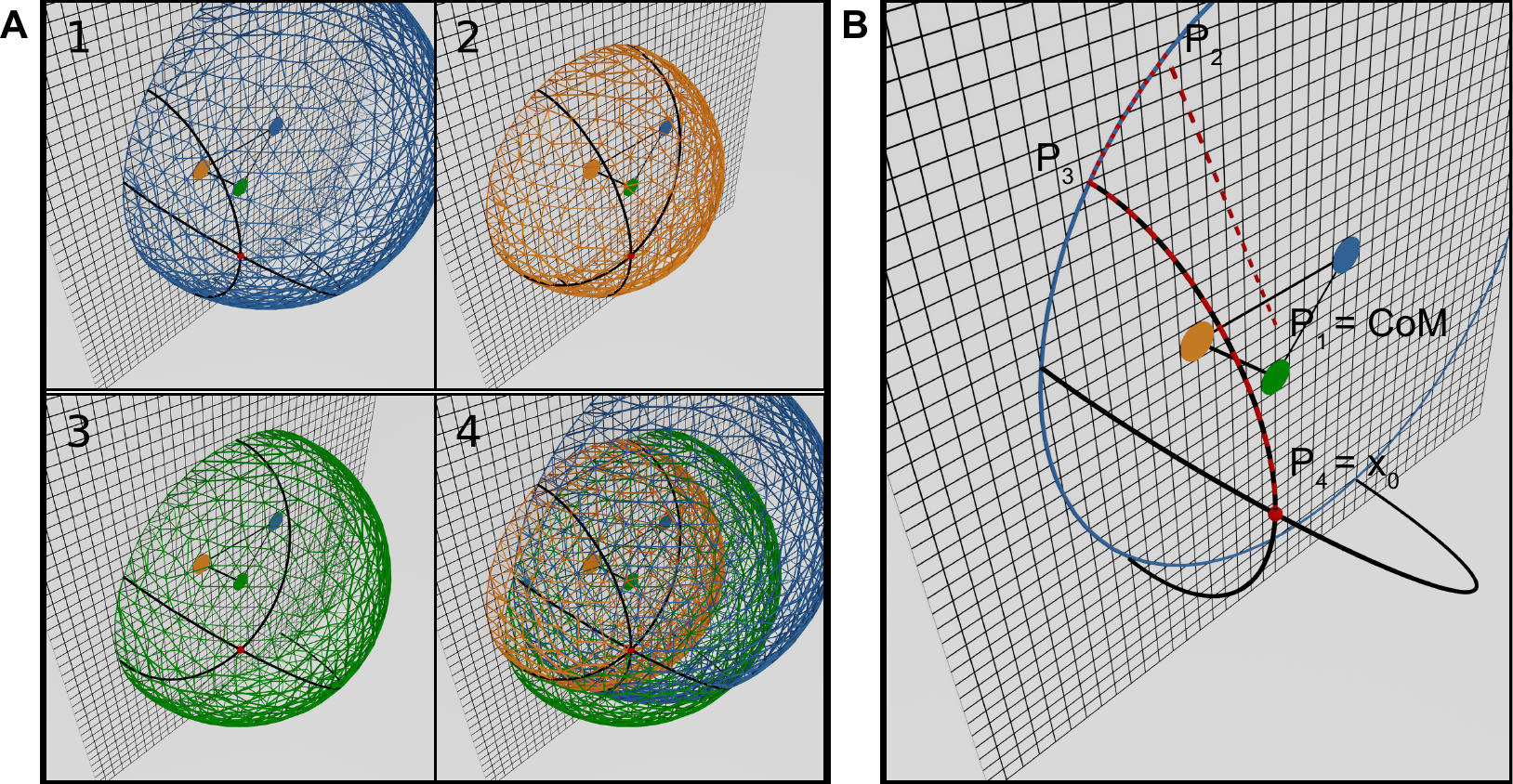}
  \caption{(A) Reconstruction of the source from the intersection of the surfaces defined by Eqs.~\eqref{exactN=3a} (color corresponding to the originating window's color). The source $\x0= (1, -1, 0)$(red dot) is recovered from the intersection of all three surfaces for the flux values $\phi_1\approx 0.03068, \phi_2\approx 0.04947$ and $\phi_3\approx 0.0358$. (B) Curve following algorithm schematic for windows located on the $x-y$ plane. The red dashed lines indicate the path the algorithm traces, starting close to the origin and ending at the source position. The blue circle is the intersection of the surface $S_k$ with the $x-y$ plane. The individual segments are labelled with the corresponding steps in the algorithm.}
  \label{curveTracing}
\end{figure}

The method to recover the source in three dimensions is similar to the two-dimensional case. However, the numerical procedure is more complex as we lay out in this section. When the three fluxes are given, as shown by equations (\ref{exactN=3a}-\ref{exactN=3c}), the source is located at the intersection point $\x_0$ of three overlapping closed surfaces. The position $\x_0$ only appears as the argument of the Neumann-Green's function $\mathcal{N}(\x,\y)$. Therefore, when the distance between the windows and the source, and the distances between the windows are large compared to the window size $\eps$, we can use the leading order approximation to recover $\x_0$ from equations~(\ref{fluxesN=3a}-\ref{fluxesN=3c}).\\
We start with the example of three windows in the x-y plane (Fig. \ref{curveTracing}): without loss of generality, we assume that the window positions are $\x_1=(0, 0, 0)$, $\x_2=(d, 0, 0)$ and $\x_3=(e, f, 0)$ (window 1 is at the origin and window 2 is on the x-axis). Then, using the leading order from the expansion of the fluxes in~(\ref{fluxesN=3a}-\ref{fluxesN=3c}), the location of the source is the solution of the three non-linear equations
\beq
\gamma_1^2 &= (x_0^{(1)})^2 + (x_0^{(2)})^2 + (x_0^{(3)})^2\\
\gamma_2^2 &= (d-x_0^{(1)})^2 + (x_0^{(2)})^2 + (x_0^{(3)})^2\\
\gamma_3^2 &= (e-x_0^{(1)})^2 + (f-x_0^{(2)})^2 + (x_0^{(3)})^2,
\eeq
where $\gamma_i=\frac{2\eps}{\pi\Phi_i}$. Solving for the coordinates of $\x_0$ and requiring that $x_0^{3}>0$, leads to the analytical solution
\beq
x_0^{(1)}&=&\ds \frac{d^2+\gamma_1-\gamma_2}{2d}\\
x_0^{(2)}&=&\ds \frac{1}{2df}\left[d(e^2+f^2+\gamma_1-\gamma_3)-e(d^2+\gamma_1-\gamma_2)\right]\\
x_0^{(3)}&=& \ds \frac{1}{2df}\left[(e^2+f^2)(\{\gamma_1-\gamma_2\}^2-d^4)+2de(e^2+f^2+\gamma_1-\gamma_3)(d^2+\gamma_1-\gamma_2)\right.\\
&& \left.-d^2(e^4+f^4+[\gamma_1-\gamma_3]^2+2e^2[f^2+2\gamma_1-\gamma_2-\gamma_3]-2f^2[\gamma_2+\gamma_3])\right]^{1/2}.
\eeq
In general, no explicit analytical inverse of $\mathcal{N}$ can be found, hence numerical procedures need to be used to find the position of the source $\x_0$ for any order in $\eps$. To this end, knowing the measured fluxes $\Phi_i$, $i=1...N$, we need to invert Eqs.~\eqref{sysMatrix} and~\eqref{totalflux} (or Eqs.~\eqref{sysMatrix2} in the case of a ball). Each of these equations describes a non-planar surface $S_{i}$, for window $i$, in three dimensions. Each surface intersects the half-plane (in the case of the windows located on the half-plane) or the unit ball (in the case of the windows located on the ball). Each pair of surfaces $S_i$ and $S_j$ intersect, forming three-dimensional curves $C_{ij}$ and all of these curves intersect at the location of the source $\x_0$. In the case of $N>3$ windows, the system is over-determined and we can simply choose any combination $k$, $l$ and $m$ of three fluxes from the $N$ available. Any combination will lead to the same source position.\\
To help the numerical procedure, we define the error function from equations~\eqref{sysMatrix} and~\eqref{totalflux}
\beq
F_i(\x_0)=\frac{\Phi_i}{\alpha_i}+\frac{2\eps}{\pi}\sum_{j\neq i}\Phi_jw_j(\x_j)-\alpha_i\,,
\eeq
in the case of half-space. For a ball in free space, the error function from equations~\eqref{sysMatrix2} is
\beq \label{sourceequation}
F_i(\x_0)=\theta_\eps\Phi_i+\sum_{j\neq i}\mathcal{N}(\x_i,\x_j)\Phi_j-2\pi\mathcal{N}(\x_i,\x_0)=0.
\eeq
In all cases, the global minimum of the squared sum
\beq
E_{klm}(\x)=|F_k(\x)|^2+|F_l(\x)|^2+|F_m(\x)|^2
\eeq
gives the source location $\x_0$. However, in contrast to the 2D case, there are many shallow local minima formed by the intersection curves $C_{ij}$, i.e. the collection of points where two of the three conditions are close to zero. These trap global minimization algorithms and thus render the determination of the global minimum difficult. Hence, this approach cannot be used directly.  Alternatively one can find and follow one curve $C_{ij}$ to the root of all three conditions $F_k=0$, $F_l=0$ and $F_m=0$. This is the approach described in the two algorithms we shall discuss now, one for windows in the $x-y$-plane and one for windows on the surface of the unit ball.\\
\subsubsection{Triangulation algorithm for windows on a plane} \label{ss:trianplan}
\begin{figure}
    \centering
    \includegraphics[scale=1]{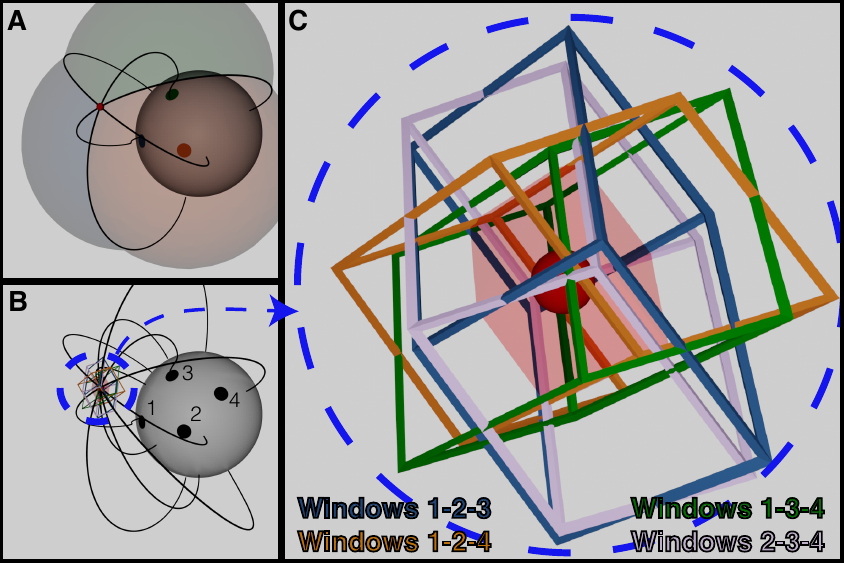}
    \caption{{\bf Triangulation of the source position with three out of four windows}. (A) Triangulation using three fluxes $F_1,F_2,F_3$. (B) Four windows yield redundant intersection lines. (C) Enlargement of the area around the source in (B). Each combination of three windows defines a volume around the source position {(here for a fixed perturbation $\eta=0.001$ of the fluxes)}, computed from combining 3 out of 4 windows (colors).}
    \label{fig:figure3bis}
\end{figure}

Each of the equations (\ref{sourceequation}) describes a closed surface in three dimensions, the intersection of which yields the source location. Therefore, to summarize the algorithm below, we search for the joint root of the $F_i(\x_0)$ by tracing the root contour of $F_k$ in the $x-y$ plane until we find its intersection with the root contour of $F_l$. We then plot the curve described by the joint root contour of $F_k$ and $F_l$ until $F_m=0$ is fulfilled (Fig.\ref{fig:figure3bis}A). This yields the source location $\x_0$ that depends on the measured fluxes $F_i$ and the window locations $\x_i$.
The detailed algorithm is as follows (see Fig.~\ref{curveTracing}B):
\begin{enumerate}
\item Define the initial step size $\Delta x$, the starting point $\mathbf{P}_1=(\Delta x, \Delta x, 0)$ and the error tolerance $\nu$ (typical values are $\Delta x=0.05$ and $\nu=10^{-6}$.  The origin is assumed to be the center of mass of the window positions.
\item Calculate the gradient vector $\v_1=[dF_k/d\x](\mathbf{P}_1)$ and its projection on the $x-y$ plane $\tilde{\v}_1=\v_1-(\v_1\cdot\mathbf{e}_z)\mathbf{e}_z$. Find the root $\mathbf{P}_2=\mathbf{P}_1+t\tilde{\v}_1$ where $t$ is such that $F_k(\mathbf{P}_1+t\tilde{\v}_1)=0$, using Newton's algorithm.
\item Calculate the gradient vector $\v_2=[dF_l/d\x](\mathbf{P}_2)$ and its projection to the $x-y$ plane $\tilde{\v}_2=\v-(\v_2\cdot\mathbf{e}_z)\mathbf{e}_z$. Find the root $\mathbf{P}_3=\mathbf{P}_2+t\tilde{\v}_2$ where $t$ is such that $F_l(\mathbf{P}_2+t\tilde{\v}_2)=0$ using Newton's algorithm.
\item Calculate the error on $F_k$ when we moved to $F_l$ by $e_{kl}=|F_k(\mathbf{P}_3)|$. If $e_{kl} > \nu$, go to step 2. Otherwise, we have now found the intersection $\mathbf{P}_3$ between the curve $C_{kl}$ and the $x-y$ plane within tolerance $\nu$ and can move on to tracing the curve $C_{kl}$.
\item Set $\y_0=\mathbf{P}_3$ and $\y_1=\mathbf{P}_3+\Delta x\mathbf{e}_z$.
\item Calculate the gradient vector $\v_1=[dF_k/d\x](\y_1)$. Find the root $y_2=\y_1+t\v_1$ where $t$ is such that $F_k(\y_1+t\v_1)=0$, using Newton's algorithm.
\item Calculate the gradient vector $\v_2=[dF_l/d\x](\y_2)$. Find the root $y_3=\y_2+t\v_2$ where $t$ is such that $F_l(\y_2+t\v_2)=0$, using Newton's algorithm.
\item Calculate the error on $F_k$ when we moved to $F_l$ by $E_{kl}=|F_k(\y_3)|$. If $E_{kl} > \nu$, go to step 6.
\item Set $\w=\y_3-\y_0$, $\y_0=\y_3$ and $\y_1=\y_0+\w$. Calculate the error on $F_m$ via $E_{m}=|F_m(\y_0)|$. If $E_{m} > \nu$, go to step 6. Otherwise, we have found the source location at $\mathbf{P}_4=y_0$ within tolerance $\nu$.
\end{enumerate}

\subsubsection{Triangulation algorithm for windows on a ball}
The case of windows on a ball is similar to the case of the windows on a plane. We again start with windows $k$, $l$ and $m$ arbitrarily chosen from the $N$ available windows. We proceed to trace the curves $C_{ij}$ formed by the intersections of the surfaces described by equations~\eqref{sysMatrix2} until the source position is recovered.
\begin{enumerate}
\item Define the initial step size $\Delta x$. Calculate the center of mass of the windows $x_m=\sum_i\x_i/N$ and its projection onto the unit ball $\tilde{\x}_m=\x_m/|\x_m|$. Define the starting point $\y_0=[\x_m+(\Delta x, \Delta x, 0)]/|\x_m+(\Delta x, \Delta x, 0)|$ and the error tolerance $\nu$.
\item Calculate the gradient vector $\v_1=[dF_k/d\x](\y_0)$. Define the geodesic $\mathbf{G}(t)=[\y_0+t\v_1]/|\y_0+t\v_1|$ and find the root $\y_1=\mathbf{G}(\tilde{t})$ where $\tilde{t}$ is such that $F_k(\tilde{t})=0$, using Newton's algorithm.
\item Calculate the gradient vector $\v_2=[dF_l/d\x](\y_1)$. Define the geodesic $\mathbf{G}(t)=[\y_1+t\v_2]/|\y_1+t\v_2|$ and find the root $\y_2=\mathbf{G}(\tilde{t})$ where $\tilde{t}$ is such that $F_l(\tilde{t})=0$ using Newton's algorithm.
\item Calculate the error on $F_k$ when we moved to $F_l$ by $e_{kl}=|F_k(\y_2)|$. If $e_{kl} > \nu$, go to step 2. Otherwise, we have now found the intersection between the curve $C_{kl}$ and the unit ball within tolerance $\nu$ and can move on to tracing the curve $C_{kl}$.
\item Set $\y_0=\y_2$ and $\y_1=(1+dx)\y_0$.
\item Calculate the gradient vector $\v_1=[dF_k/d\x](\y_1)$. Find the root $y_2=\y_1+t\v_1$ where $t$ is such that $F_k(\y_1+t\v_1)=0$ using Newton's algorithm.
\item Calculate the gradient vector $\v_2=[dF_l/d\x](\y_2)$. Find the root $y_3=\y_2+t\v_2$ where $t$ is such that $F_l(\y_2+t\v_2)=0$ using Newton's algorithm.
\item Calculate the error on $F_k$ when we moved to $F_l$ by $E_{kl}=|F_k(\y_3)|$. If $E_{kl} > \nu$, go to step 6.
\item Set $\w=\y_3-\y_0$, $\y_0=\y_3$ and $\y_1=\y_0+\w$. Calculate the error on $F_m$ via $E_{m}=|F_m(\y_0)|$. If $E_{m} > \nu$, go to step 6. Otherwise, we have found the source location at $y_0$ within tolerance $\nu$.
\end{enumerate}
\section{Sensitivity to recover the source position}
The accuracy at which a source can be reconstructed depends on the distance and the strength of fluctuations of the fluxes received at each receptor. In this section we present a perturbation analysis of the numerical reconstruction. Together with an assumed sensitivity threshold, this allows a characterization of the distance at which the direction to a source becomes impossible. We start with the two-dimensional case, showing how the sensitivity depends on distance in the case of the disk, followed by the case of windows on the boundary of half-space and finishing with the three-dimensional problem.
\subsection{Thresholded direction sensitivity in 2D}
\begin{figure}[http!]
  \centering
 \includegraphics{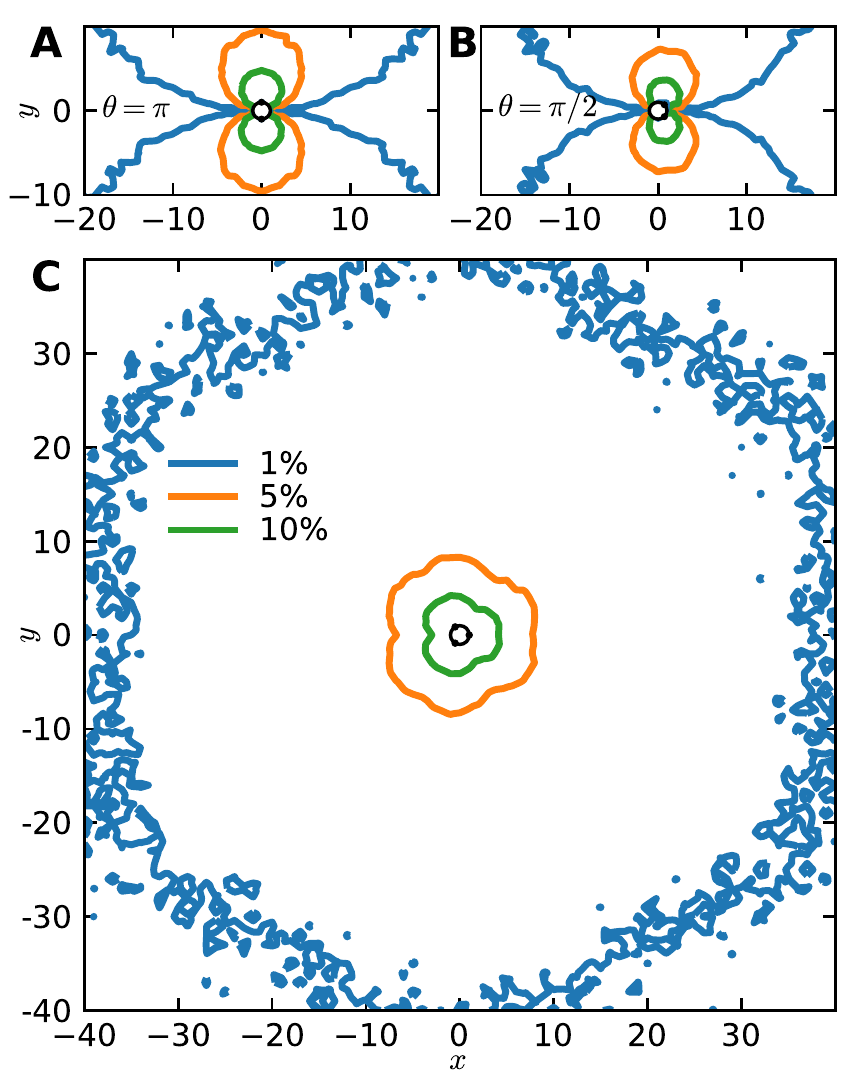}
  \caption{\textbf{Detectable region $D(S)$ and contours for small windows on a disk}  (A) Two windows are placed on a disk with angular spacing $\theta=\pi$. The contours indicate the position for a threshold  $T_h=1\%$, $5\%$ and $10\%$, given by the normalized flux difference or probability~(\ref{ratio1}). (B) Two windows are placed with an angle $\pi/2$ apart. (C) Three windows placed $2\pi/3$ apart. The contours of $D(S)$ are given by $r_3=T_h$ (relation~(\ref{ratio2})). 
  } \label{fig:sens}
\end{figure}

For two windows, directional sensitivity can be measured with the sensitivity ratio defined by~\cite{dobramysl2018reconstructing}
\beq \label{ratio1}
r(\x_1,\x_2,\x_0)=\frac{|P_1(\x_1,\x_2,\x_0)-P_2(\x_1,\x_2,\x_0)|}
{P_1(\x_1,\x_2,\x_0)+P_2(\x_1,\x_2,\x_0)}.
\eeq
This ratio measures the absolute difference in the measured fluxes at the two windows. As the source moves farther away, the splitting probability at both windows tends to $1/2$, hence $r\to 0$ as $|\x_0|\to\infty$. By setting a threshold value $T_{h}$, we can define the region where the source direction can be established as the interior of
\beq\label{dsregion}
D_{S}=\{\x_0 \hbox{ such that } r(\x_1,\x_2,\x_0)\geq T_{h} \}.
\eeq
The boundary of the region $D_S$ for two absorbing windows symmetrically positioned on a disk is shown in Fig.~\ref{fig:sens}A. Figure~\ref{fig:sens}B shows the same when the angle between the windows is $\theta_{12}=\pi/2$. In both cases the boundary of $D_{S}$ consists of two connected components and is shown for the sensitivity threshold values $T_{h}=1\%, 5\%$ and $10\%$. When $T_h=1\%$, the domain is around 20$\times$ the size of the detecting disk, which is a good measure of how far out cells would be able to be guided by an external chemical gradient. This is a surprisingly small distance and suggests that other mechanisms may be in play that allow guidance of biological cells with sources located much further away (see discussion below).
The optimal window placement that maximizes the detection sensitivity for a given source location $\x_0$ is defined by maximizing the ratio \ref{ratio1}
\beq \label{diff}
f(\x_0)=\max_{\x_1,\x_2} |P_1(\x_1,\x_2,\x_0)-P_2(\x_1,\x_2,\x_0)|.
\eeq
The maximum distance between the source and a disk containing two absorbing windows located at position $\x_1,\x_2$ that gives a minimal significant difference of probability flux is achieved for a window configuration aligned with the position of the source and symmetric with respect to the center of the disk centered at the origin. An explicit computation with $\x_2=-\x_1$, $|\x_1|=|\x_2|=R$ gives
 \beq
f(\x_0)=\ds \frac{1}{2}\frac{\ln \frac{|\x_1-\x_0|\left|\x_1-\bar{\x_0}\right|}{|\x_2-\x_0|\left|\x_2-\bar{\x_0}\right|}}{\{\log|\x_1-\x_2|-\log \eps\}}
       =\ds \frac{1}{2}\frac{\ds \ln \frac{\left|1-|\x_0|/R\right| \left|1-R/|\x_0|\right|}{\ds\left|1+|\x_0|/R\right| \left|1+R/|\x_0|\right|}}{\{\log|2R|-\log \eps\}}.
\eeq
In particular, a Taylor expansion of $f(\x_0)$ for large source position compared to the disk radius $L=|x_0|\gg R$, leads to the decay of the maximum detection threshold function
\beq \label{maxdetectionthreshold}
f(\x_0)= \ds\frac{2R}{L\log\frac{2R}{\eps}}+o\left(\frac{1}{L}\right).
\eeq
Hence, the direction sensitivity decreases as the reciprocal of the distance to the source. With three windows, direction sensing is possible if at least one of the difference between the splitting probability is higher than the threshold $T_h$. Therefore, we define the sensitivity ratio for three windows as
\beq \label{ratio2}
r_3(\x_1,\x_2,\x_3,\x_0)=\frac{\max\{|P_1-P_2|,|P_1-P_3|,|P_2-P_3|\}}{P_1+P_2+P_3},
\eeq
with $P_1$, $P_2$ and $P_3$ the splitting probabilities for particles to arrive at the respective windows, and depending on $\x_1$, $\x_2$, $\x_3$ and $\x_0$.
Numerical simulations reveal the region $D_{S,3}=\{\x_0\mathrm{such that }r_3(\x_1,\x_2,\x_3,\x_0)\ge T_h\}$ for the function $r_3$ with windows positioned at the corners of an equilateral triangle (Fig. \ref{fig:sens}C). The region is now connected, and for a threshold of $T_h=1\%$ extends up to 40 times the size of the detecting disk. Hence, adding a third window allowed a larger sensing distance and made the detection region more isotropic.\\
For two windows located on the boundary of half-space, the direction sensitivity can only be influenced by the spacing between the windows $d=|\x_1-\x_2|$. Therefore, we do not need to find the optimal arrangement and can directly compute the sensitivity ratio~\cite{dobramysl2018mixed}
\beq
r(d, L, \theta)= \left|\frac{1}{2}\frac{1}{\ln(d/\eps)}\ln\left[\frac{\frac{d^2}{4}+L^2-Ld\sin{\theta}}{\frac{d^2}{4}+L^2+Ld\sin{\theta}}\right]\right|\,,
\eeq
where $\theta$ is the angle between the $x$-axis and the vector from the origin $O$ to the source location $x_0$. A Taylor expansion for $L>>d$ of the logarithmic term yields to
\beq
r(d, L, \theta)=\frac{d}{L}\frac{|\sin\theta|}{\ln(d/\eps)}+o\left(\frac{d}{L}\right),
\eeq
where the maximum of the detection threshold is similar to the one of the disk in equation \ref{maxdetectionthreshold} with $d=2R$ and $\theta=\pm \pi/2$.\\
To conclude, in both the disk and the half-plane cases, the detection sensitivity decays algebraically with the distance. Cells often employ multiple different types of receptors for different ligands. Therefore, we can consider e.g. two different types of absorbing windows, each accepting only one of two types of Brownian particles. In this case, the splitting probabilities are independent and the sensitivity can be defined as the product of each window type's sensitivity function
\beq
f_{\text{2 classes}}(\x_0)= f(\x_0)^2 \propto \ds\left(\frac{d}{L\ln(d/\eps)}\right)^2+o\left(\frac{1}{L^4}\right).
\eeq
Interestingly, this formula predicts a decay of $1/L^2$ with respect to the source position.
\subsection{Sensitivity analysis in 3D}
\begin{figure}
\centering
\includegraphics[scale=0.8]{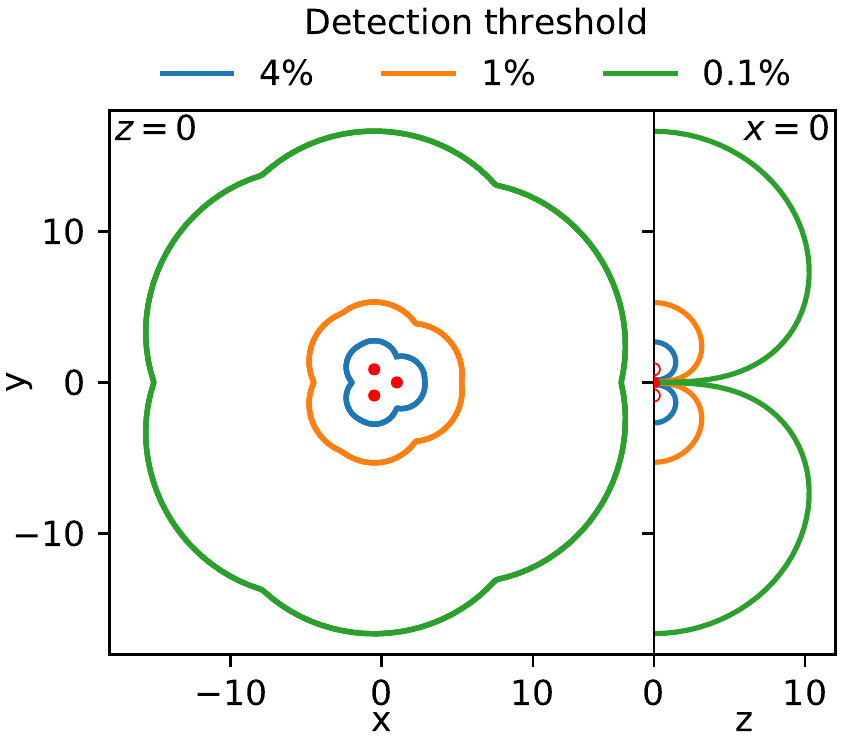}
\includegraphics[scale=0.6]{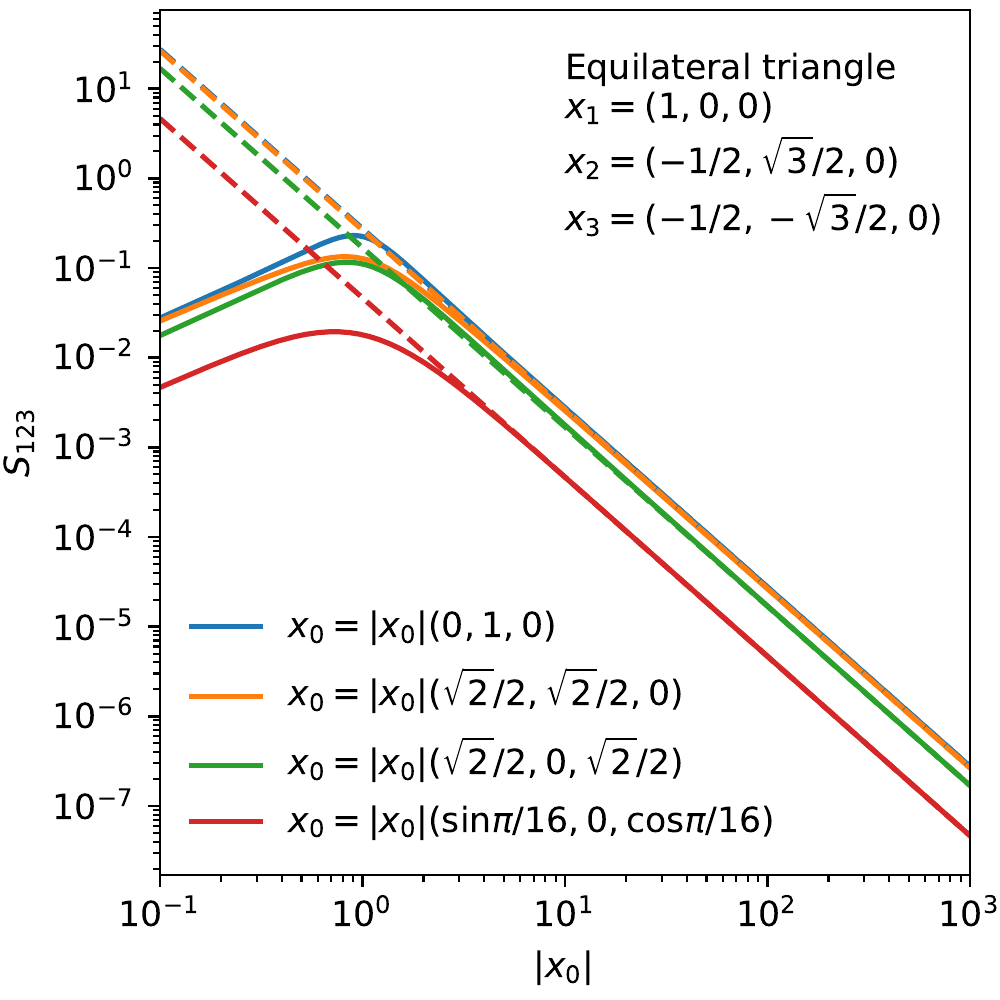}
\caption{{\bf Sensitivity of detecting the source position (from
Eq.~(\eqref{eq:S123}))}. (Left) a ball with three windows arranged as an
equilateral triangle on a geodesic. The detection contours is in
the plane that contains all 3 windows (left) and in plane perpendicular to the window plane (right), for three different detection thresholds (1\%, 0.1\% and 0.01\%). (Right) The sensitivity decays with $C/|x_0|^2$ for the source position $x_0$ (The constant C depend on the window organization and is adjusted for each case).}
\label{fig:figure3}
\end{figure}

In analogy to equation~\ref{ratio2}, the sensitivity function for three windows in three dimensions is expressed as the maximum of the differences between the splitting probabilities computed from the fluxes~\cite{dobramysl2020threed}
\beq\label{eq:S123}
S_{123}(\x_0;\x_1,\x_2,\x_3)=\max\{&|\Phi_1(\x_0)-\Phi_2(\x_0)|, \nonumber
  &|\Phi_2(\x_0)-\Phi_3(\x_0)|,\\&|\Phi_3(\x_0)-\Phi_1(\x_0)|\},
\eeq
where $\x_0$ is the position of the source and $\x_i$, $i=1, 2, 3$ are the positions of the three windows on the boundary $\p \Omega$ (this could be the surface of a ball in free space or the boundary of 3D half-space). The function $S_{123}$ describes the absolute imbalance between the fluxes through the windows. Fig.~\ref{fig:figure3}A shows the contours of this function for three windows arranged in an equilateral triangle on the $z=0$ plane in three-dimensional half-space. The distance at which directions can still be discerned is approximately an order of magnitude less for any given threshold compared to the equivalent situation in two dimensions, see the previous section. Indeed, using the dipole expansion for a source located far away $|\x_0|\gg 1$, we obtain that
\beq
\Phi_i(\x_0)-\Phi_j(\x_0)=\frac{2 \eps}{\pi} \frac{1}{|\x_0|} \left( \frac{(\x_i-\x_j.\x_0)}{|\x_0|^2}+
\frac{3(\x_0.\x_i)^2-x^2_i x_0^2 -3(\x_0.\x_j)^2+\x^2_j \x_0^2 }{|\x_0|^4}  \right)
\eeq
Thus, we use the approximation $S_{123}(\x_0;\x_1,\x_2,\x_3)\approx \frac{C}{|\x_0|^2}$, where $C>0$ is constant. Fig.~\ref{fig:figure3}B illustrates this decay for an equilateral triangle located on half-plan and for various source locations.
\section{Source location triangulation accuracy}
In this last section, we discuss the uncertainty introduced into the recovered source position due to fluctuations or measurement error in the diffusion fluxes at each window. In two dimensions, we discussed numerical results where the fluctuation level was held fixed. This led to large variability with strong spatial heterogeneity and anisotropy (see section~\ref{sec:3receptod}, figures~\ref{Fig3}C and~\ref{fig:pos}B-E).
\subsection{Region of uncertainty}\label{region_of_uncertainty}
\begin{figure}
    \centering
    \includegraphics[scale=0.9]{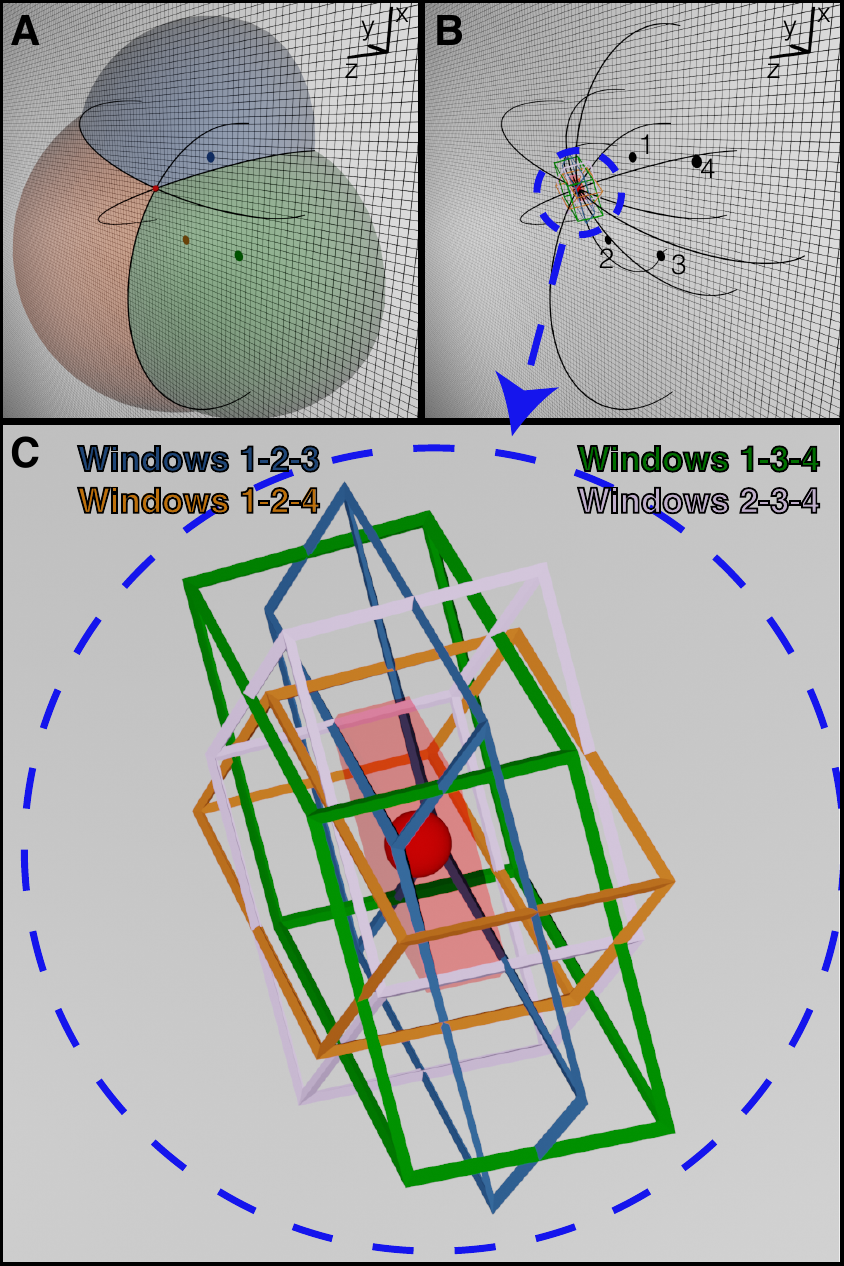}
    \caption{{\bf Triangulation of the source position with three out of four windows}. (A) Triangulation using fluxes from three windows only. (B) A further window yields additional redundant intersection lines. (C) Enlargement of the area around the source in (B). Each combination of three windows defines a volume (parallelepiped) around the source position, computed from combining 3 out of 4 windows (various colors). The intersection of these volumes defines the uncertainty volume $V_{unc}$ (shaded red).}
    \label{fig:figure3bisb}
\end{figure}

\begin{figure}[http!]
    \centering
    \includegraphics{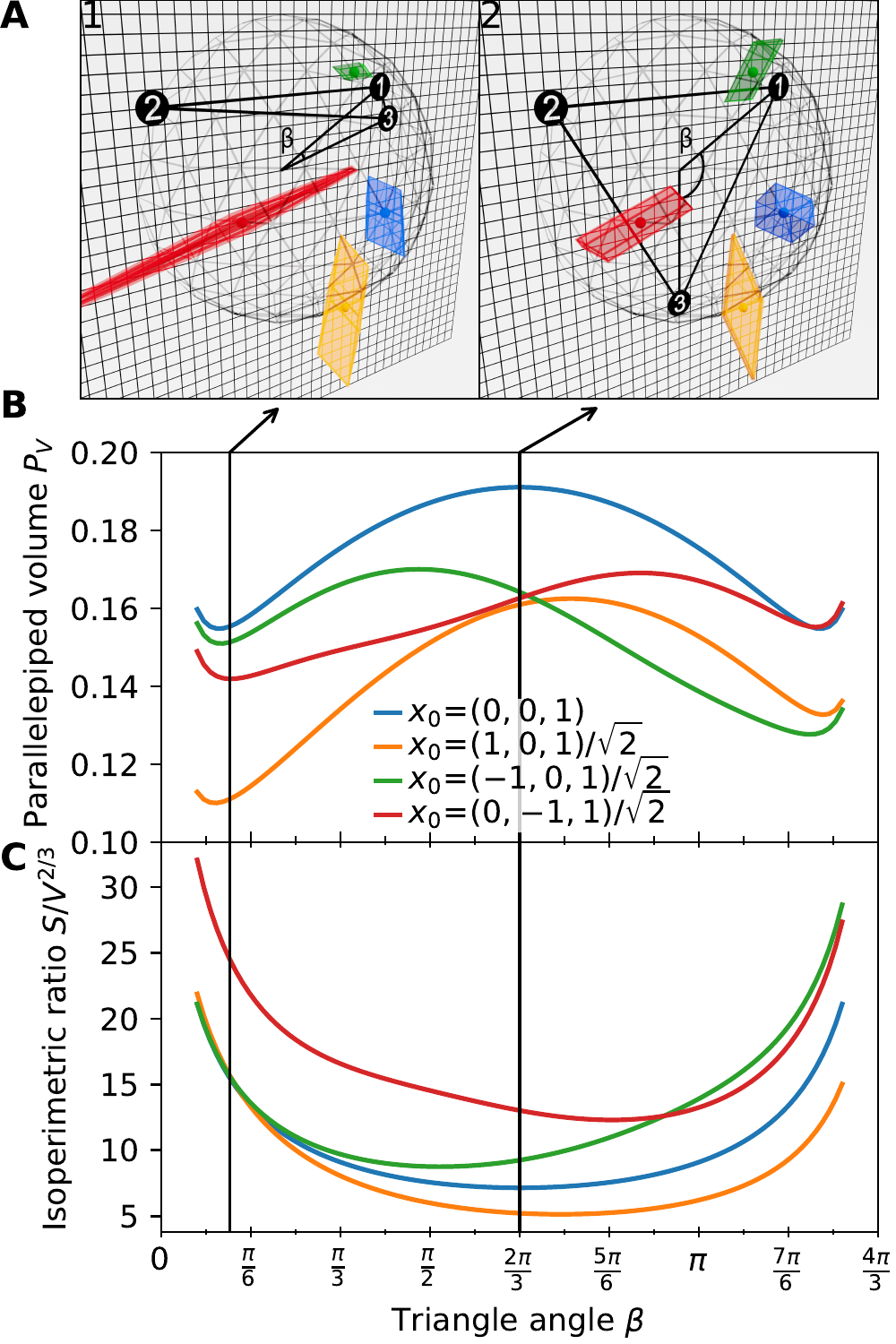}
    \caption{Uncertainty of the source recovery depends on the its location. (A) Volumes of uncertainty (colored regions) for four different position of the source. We show the volumes for two different configurations of three windows: (1) a scalene triangle and (2) an equilateral triangle. (B) Measured volume of uncertainty for different source positions vs the triangle angle $\beta$ ($\beta=0$ corresponds to window3 overlapping with window 1 while $\beta=2\pi/3$ corresponds to window 3 overlapping with window 2). (C) Uncertainty isotropy (isoperimetric ratio) as a function of the source position and the triangle angle $\beta$.}
    \label{fig:uncertainty_inhomog}
\end{figure}

\begin{figure}[http!]
\centering
\includegraphics{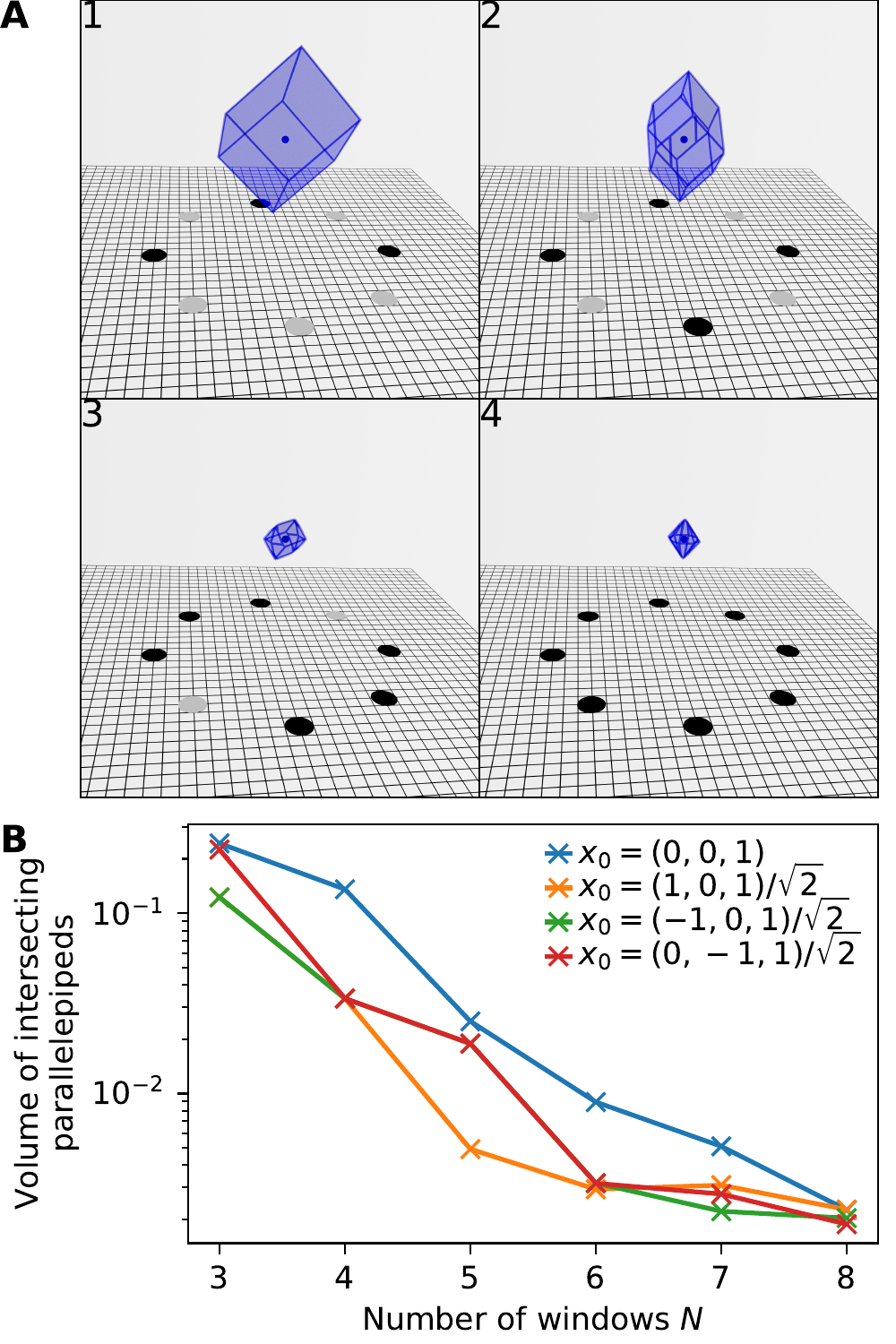}
\caption{Uncertainty is reduced by the number of windows. (A) Three-dimensional display of the total volume of uncertainty, defined as the intersection of all parallelepipeds from all combinations of three windows for (1) three, (2) four, (3) six and (4) eight windows. (B) The total uncertainty volume as a function of the number of windows for four different source positions.}
\label{fig:figure4}
\end{figure}

The volume of uncertainty $V_{unc}$ of the source location can be defined using a linear approximation of the fluxes when adding a small perturbation term $\eta$. This perturbation can either be deterministic or stochastic. When fluxes are large, Gaussian perturbation are possible; however when the perturbation size is comparable to the magnitude of the fluxes care must be taken (see below). The perturbed flux is defined as
\beq \label{pertub}
\tilde{\Phi}_i=\Phi_i+\eta,
\eeq
where $\eta\ll\Phi_i$ represents the size of the flux measurement error, and $\Phi_i$ is the original (unperturbed) flux.
To illustrate the volume of uncertainty to first order in $\eps$, we consider the 3D half-space case where the source is located on a sphere centered around the window $i$ and with a radius $\tilde{R}_i=2\eps/(\pi\Phi_1)$. Therefore, using equations~(\ref{fluxesN=3a}-\ref{fluxesN=3c}), the location of the source varies according to $-d\tilde{R}_i/d\Phi_i=2\eps/(\pi\Phi_i^2)$ along the radial vector $\x_0-\x_i$. The error vector associated with window $i$ is
\beq
\vec{e}_i=\eta_i\frac{2\eps}{\pi\Phi_i^2}\frac{\x_0-\x_i}{|\x_0-\x_i|}.
\eeq
For three windows, the principal directions $e_1$, $e_2$ and $e_3$ define a parallelepiped, the volume of which measures the uncertainty of the reconstruction. \\
With $N$ windows, the triangulation can be performed with any combination of three windows with indices $k$, $l$ and $m$. The $i-$th coordinate of the reconstructed source position is then given by a Taylor expansion in the flux coordinate system
\beq
\begin{aligned}
  \tilde{x}_{0}^i(\tilde{\Phi}_k, \tilde{\Phi}_l, \tilde{\Phi}_m) &= \tilde{x}_{0}^i(\Phi_k+\eta, \Phi_l+\eta, \Phi_m+\eta) \\
  &=\tilde{x}_{0}^i(\Phi_k, \Phi_l, \Phi_m)+\eta\left(\frac{\p x_{0}^i}{\p\Phi_k}+\frac{\p x_{0}^i}{\p \Phi_l}+\frac{\p x_{0}^i}{ \p \Phi_m}\right) + O(\eta^2)\\
  &=x^i_{0}+\eta(E^{(k,l,m)}_{1i}+E^{(k,l,m)}_{2i}+E^{(k,l,m)}_{3i}) + O(\eta^2),
\end{aligned}
\eeq
where we used that $\tilde{\x}_{0}(\Phi_k, \Phi_l, \Phi_m)=\x_0$ and the error matrix is defined as $E^{(k,l,m)}_{ij}= \ds \frac{\p x_0^j}{\p\Phi_i}$. Therefore, the uncertainty region can be computed from the Jacobian
\beq \label{eq:Jacob}
J_{ij}(\x_0)=\frac{\p \Phi_i}{\p x_0^j}(\x_0),
\eeq
for the three window fluxes $k$, $l$ and $m$ via equation~\eqref{sysMatrix}. This yields the linear uncertainty error vectors $E^{k,l,m}_{ij}$ as the column vectors of the inverse $[J_{ij}]^{-1}$. They span a parallelepiped $P_u$ at the location of the source $\x_0$. Therefore, similar to above, the volume of uncertainty for the source reconstruction is the volume of this parallelepiped
\beq
V_{\mathrm{unc}}^{(k,l,m)}(\eta)=\eta^3/|\det(J_{ij})|.
\eeq
The choice of the three windows $k$, $l$ and $m$ is arbitrary. The total volume of uncertainty $V_{unc}(\eta)$ can thus be defined as the volume of the geometric intersection of all parallelepipeds generated by the possible combinations of any three window fluxes from the $N$ available. The intersection of parallelepipeds is illustrated in Fig. \ref{fig:figure3bisb}: Using three windows only (Fig. \ref{fig:figure3bisb}A), the source location $\x_0$ is reconstructed using the algorithm presented in subsection \ref{ss:trianplan}. When adding a fourth window, there are four possible combinations of three from which the source can be reconstructed. Thus there are six curves arising from the intersection of the four surfaces (Fig. \ref{fig:figure3bisb}B). The resulting four parallelepipeds are displayed in Fig. \ref{fig:figure3bisb}C together with their geometric intersection (red volume). As we shall describe below, the volume $V_{unc}(\eta)$ is inhomogeneous, it depends both on the location of the source and the particular arrangement of the windows.\\
The region of uncertainty $R_{unc}$ and its volume $V_{unc}(\eta)$ strongly depend on the position of the source relative to the windows. This property is illustrated in Fig.\ref{fig:uncertainty_inhomog}A, where we plotted $R_{unc}$ for four different source positions and two different window configurations (a scalene and an equilateral triangle). Further properties of the uncertainty volume are obtained when varying the triangle angle $\beta$. In that case, the volume and the isoperimetric ratio $S/V^{2/3}$ are computed, where $S$ is the surface and $V$ the volume (Fig.~\ref{fig:uncertainty_inhomog}B,C). The parallelepipeds can be highly elongated (Fig.~\ref{fig:uncertainty_inhomog}C). Interestingly, the minimum of the isoperimetric ratio (i.e. the triangle angle $\beta$ at which $R_{unc}$ is most isotropic) strongly depends on the source position (Fig.\ref{fig:uncertainty_inhomog}C).\\
When the number of windows $N$ is larger than three, there are  $N!/(3![N-3]!)$ combinations of the $N$ error vectors $e_i$ (see Fig.~\ref{fig:figure4}A for illustrations of $N=3, 4, 6$ and $8$). Numerical simulations indicate that the volume of uncertainty decreases super-exponentially when the number of windows $N$ increases (see Fig.~\ref{fig:figure4}B).

\subsection{Estimation of the localization error}
Measuring the localization error for the source position $\x_{0}$ can also be achieved by using the empirical estimator
\beq
\hat{\x}_n=\frac{1}{n} \sum_{r=1}^n \tilde{x}_{0,r}(\tilde{\Phi}_j, \tilde{\Phi}_k, \tilde{\Phi}_l),
\eeq
where $r=(j,k,l)$ and $n=N!/(3![N-3]!)$ is the total number of window triplets (it also possible to chose a lower number of triplets). When noise is added to the fluxes (i.e. a new realization of the noisy fluxes $\tilde{\Phi}_i=\Phi_i+\eta_i$ is generated), each triplet provides a separate source localization $\tilde{x}_{0,r}(\tilde{\Phi}_j, \tilde{\Phi}_k, \tilde{\Phi}_l)$. By inverting the flux equation~\ref{sysMatrix} for a given noise realization, and by using a Taylor's expansion for the estimated position with respect to the fluxes
\beq
\tilde{x}_{0}(\tilde{\Phi}_j, \tilde{\Phi}_k, \tilde{\Phi}_l) = \tilde{x}_{0}(\Phi_j+\eta_j, \Phi_k+\eta_k, \Phi_l+\eta_l)=\x_{0}+J_{r}\Eta + O(\eta^2),
\eeq
the error for source location estimation is given by
\beq
\hat{\x}_n-\x_{0}= \frac{1}{n} \sum_{r=(j,k,l)} J_{r}(\x_{0}) \Eta_{j,k,l}.
\eeq
Here $\Eta=(\eta_j,\eta_k,\eta_l)$ could e.g. be a Gaussian perturbation with zero mean and variance $\sigma$. Averaging over q realizations of the noise
\beq \label{empsum}
\hat{\x}^q_n-\x_{0}= \frac{1}{q} \frac{1}{n} \sum_{i=1}^{q}\sum_{r} J_{r}(\x_0) \Eta_i,
\eeq
leads to the estimate of the deviation from the real source position
\beq
\sqrt{\eE(|\hat{\x}^q_r-\x_{0}|^2)}= \frac{\sigma}{\sqrt{n}} \sqrt{\frac{1}{n}\sum_{r}^n  \mathrm{tr}(J_r^T(\x_{0}) J_r(\x_{0}))}.
\eeq
This behaves like $ O(\frac{\sigma}{\sqrt{n}})$ because the trace is bounded from below and above. Here,
\beq
\mathrm{tr}(J_r^T(\x_{0}) J_r(\x_{0}))=\lambda^r_1+\lambda^r_2+\lambda^r_3
\eeq
and $(\lambda^r_1,\lambda^r_2,\lambda^r_3)$ are the eigenvalues of the matrix $J_r^T(\x_0) J_r(\x_0)$ at $x_0$, Thus the empirical sum \eqref{empsum} tends to zero as $1/\sqrt{n}$ by the central limit theorem.
Other weighted sums are possible such as
\beq
\hat{\x}_n=\frac{1}{\sum_{r=1}^n \frac{1}{V_{unc}(\eta)_r}} \sum_{r=1}^n \frac{1}{V_{unc}(\eta)_r}\tilde{x}_{0}(\tilde{\Phi}_j, \tilde{\Phi}_k, \tilde{\Phi}_l),
\eeq
where the uncertainty volume $V_{unc}(\eta)_r$ is associated to the triplet r. There is no canonical choice for the empirical estimator, especially for the biological applications of cell navigation, because the exact mechanism for receptor flux evaluation remains unclear.
\subsection{Susceptibility of the source reconstruction: when the noise and flux amplitudes are of same order } \label{ss:clouds}
\begin{figure}
    \centering
    \includegraphics[scale=1]{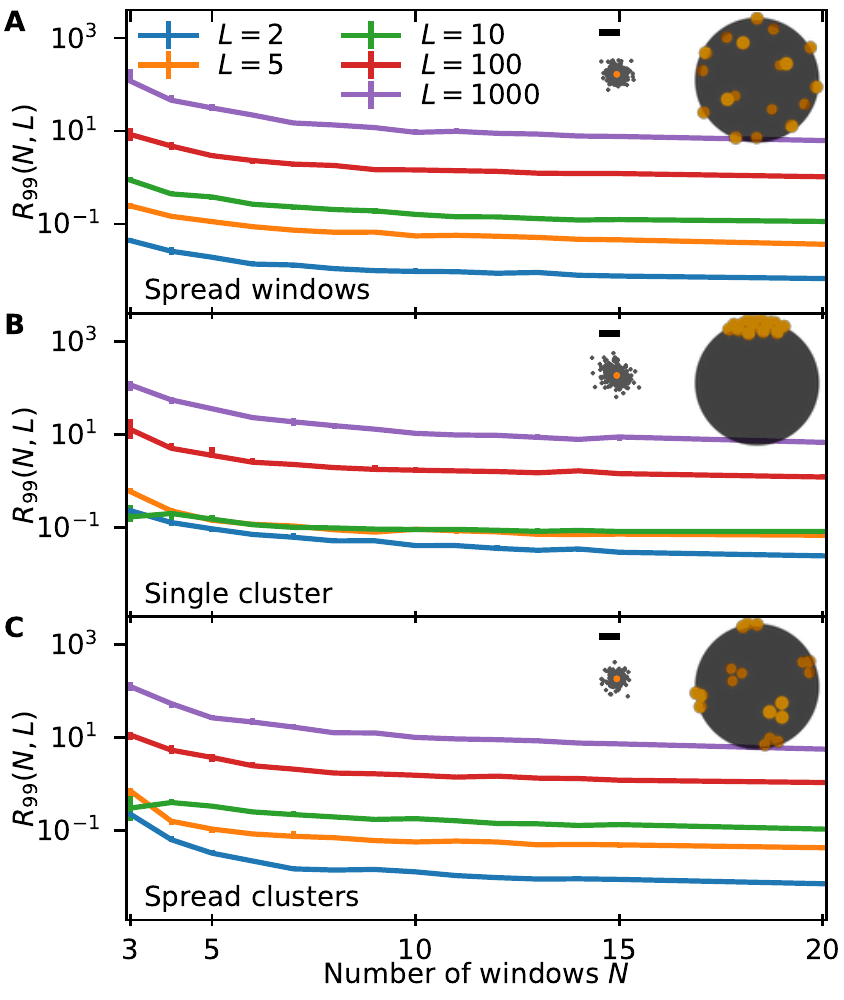}
    \caption{{\bf Uncertainty of the source location.} Deviation of the reconstructed source from its exact position {measured by the radius $R_{99}$ around $\x_0$ in which $99\%$ of recovered points fall}, when the fluxes are subject to {multinomial noise with $N_t=10^7$ trials}. The source location is computed by averaging over reconstructions from all combinations of three windows. {$R_{99}$ is calculated from} $1000$ realizations and random source positions (with $|\x_0|=L$ fixed). {\bf (A)} Windows are spread uniformly across the ball surface, {\bf (B)} a single cluster of windows, and {\bf (C)} window clusters are spread uniformly across the ball surface (at most three windows per cluster). The error bars indicate the $95\%$ confidence interval. The insets show the spread of points around the exact source position for $L=5$ and $m=5$ (scale bar corresponds to {one-tenth of the ball radius}).}
    \label{fig:figure4b}
\end{figure}

\begin{figure}
    \centering
    \includegraphics[scale=1]{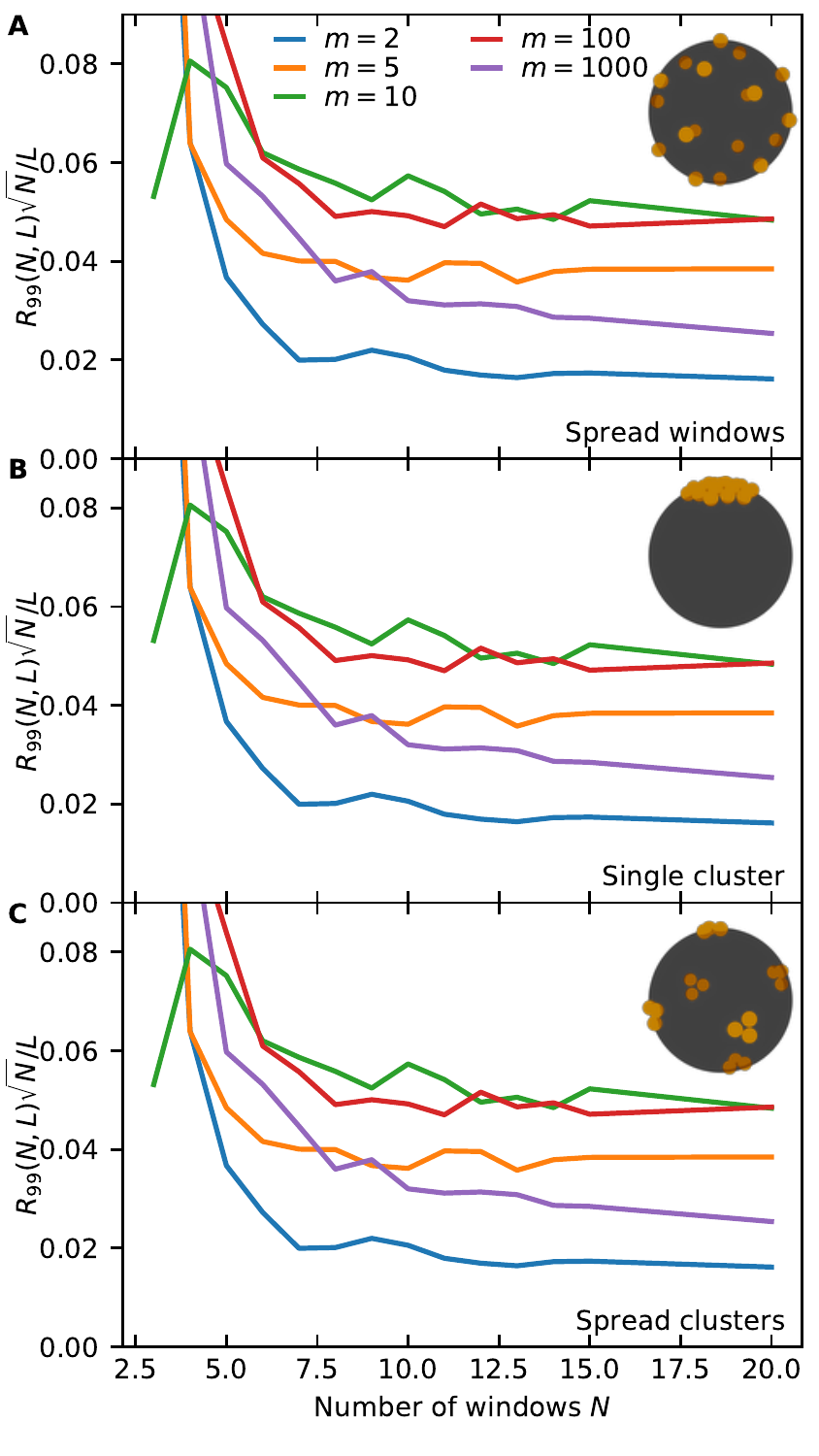}
    \caption{{\bf Dependence of $R_{99}$ on the number of windows.} (A-C) Plots of $R_{99}(N,L)\sqrt{N}/L$ showing that $R_99$ has a $1/\sqrt{m}$ behavior for $m>8$. (A) Spread windows, (B) single cluster, and (C) spread clusters.}
    \label{fig:oneovern}
\end{figure}

The Gaussian model for flux noise becomes invalid when the noise is of the same order of magnitude as the fluxes. In this case, the noisy fluxes can be redefined using a multinomial noise model~\cite{dobramysl2020PRLTria}. Then, the noisy flux coordinates are $\tilde{\Phi}_i=K_i/N_{t}$, where $i=1,\dots,N$ and $K_i\sim \mathrm{Multi}(N_t,\{\Phi_1,\dots,\Phi_N,1-\sum_{i=1}^N\Phi_i\})$ are multinomially distributed. Here, the noise level is determined via the number of trials $N_t$. For large $N_t$ the noisy fluxes $\tilde{\Phi_i}$ approach the true fluxes $\Phi_i$. As above, we have $n=N(N-1)(N-2)/6$ possible window triplets, and each triplet gives rise to a source localization $\x_{0,r}$, using the procedure described above.
Just as above, these positions are all be shifted relative to each other due to the stochastic perturbation, hence the final estimator of the source position is again $\hat{\x}_n=\sum_{r=1}^n\tilde{\x}_r/n$\\
The susceptibility of the source reconstruction to random fluxes is then measured by using the radius $R_{99}(N,L)$ centered at $\x_0$ in which $99\%$ of recovered points fall out of a given number of $q$ independent realizations of the $\tilde{\Phi}_i$ (Fig.~\ref{fig:figure4b}). \\
Windows can be arranged in one of three configurations: spread uniformly across the ball surface (A), concentrated in a single cluster (B), and window clusters spread uniformly across the surface (C)~(Figure~\ref{fig:figure4b}). The window distribution is generated by the algorithm given in appendix~\ref{sec:algo_window_distrib_ball}. The exact source position is chosen randomly on a shell with radius $L$. The susceptibility $R_{99}$ universally decreases with an increasing number of windows in all three cases (Fig.~\ref{fig:figure4b}A-C)~\cite{dobramysl2020PRLTria}. When $N\ge 9$, $R_{99}$ can be observed to roughly scale as $N^{-1/2}$ (Fig.\ref{fig:oneovern}).
\begin{figure}
\centering
\includegraphics[scale=0.6]{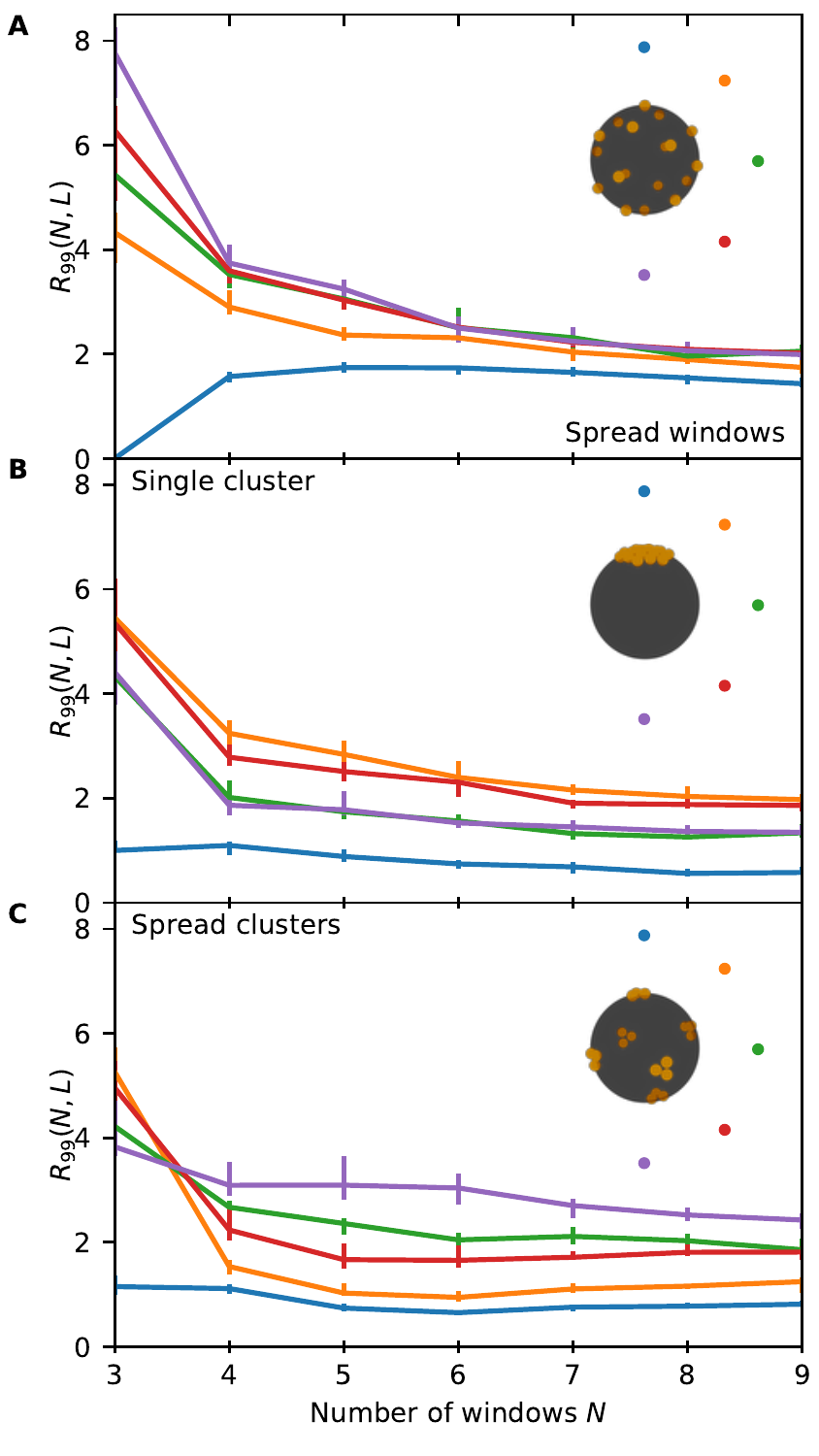}
\caption{{\bf Deviation from the individual source locations spread around the cell.} The deviation $R_99$ is shown as a function of the number of windows, with $\x_0$ varied at the indicated positions at angles $\theta=0,\pi/4,\pi/2,3\pi/4,\pi$ with the vertical axis (positions indicated in the color of the curve) and $L=|\x_0|=5$. Here, $N_t=10^4$. {\bf (A)} Windows are spread uniformly across the ball
surface, {\bf (B)} a single cluster of windows, and {\bf (C)} window clusters are spread uniformly across the ball surface (at most three windows per cluster). For better visibility, the cell is shown at double its radius.}
\label{fig:indivx0}
\end{figure}
\begin{figure}
\centering
\includegraphics[scale=0.6]{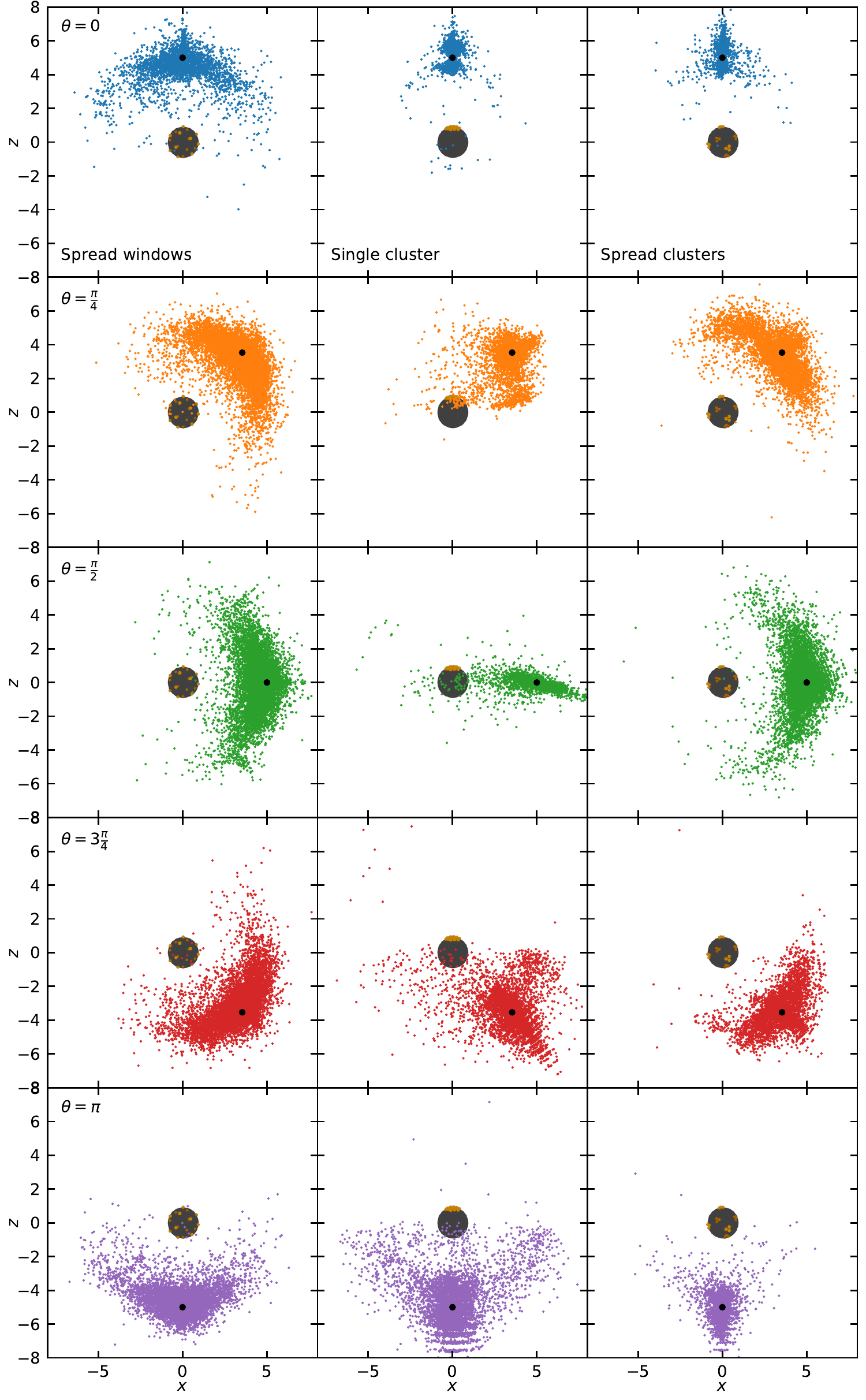}
\caption{{\bf Distribution of recovered points.} The panels show the distribution of recovered points for the different positions and window arrangements shown in Fig.~\ref{fig:indivx0} and $m=5$ windows. The colors match the positions shown in Fig.~\ref{fig:indivx0}. To make the spread visible, the number of trials in the multinomial distribution have been set to a low $N_t=10^4$.}
    \label{fig:clouds}
\end{figure}
With a fixed (non-random)source position, the spread of source localizations shows strong spatial heterogeneity, similar to the 2D case (c.f. section \ref{sec:3receptod}). This heterogeneity is most pronounced when the number of windows is low, but does persist for higher $N$ when the window distribution is non-uniform (figure~\ref{fig:indivx0}A-C). Indeed, when windows are clustered, the susceptibility to noise is highly dependent on the source position relative to the positions of the clusters. This can also be observed in figure~\ref{fig:clouds} where the clouds of localizations have roughly the same shape when the windows are uniformly distributed across the ball, but strongly depend on the source location when the windows are clustered non-uniformly.
\subsection{Effect of the distance on the sensitivity to source triangulation}
\begin{figure}
\centering
\includegraphics[scale=0.9]{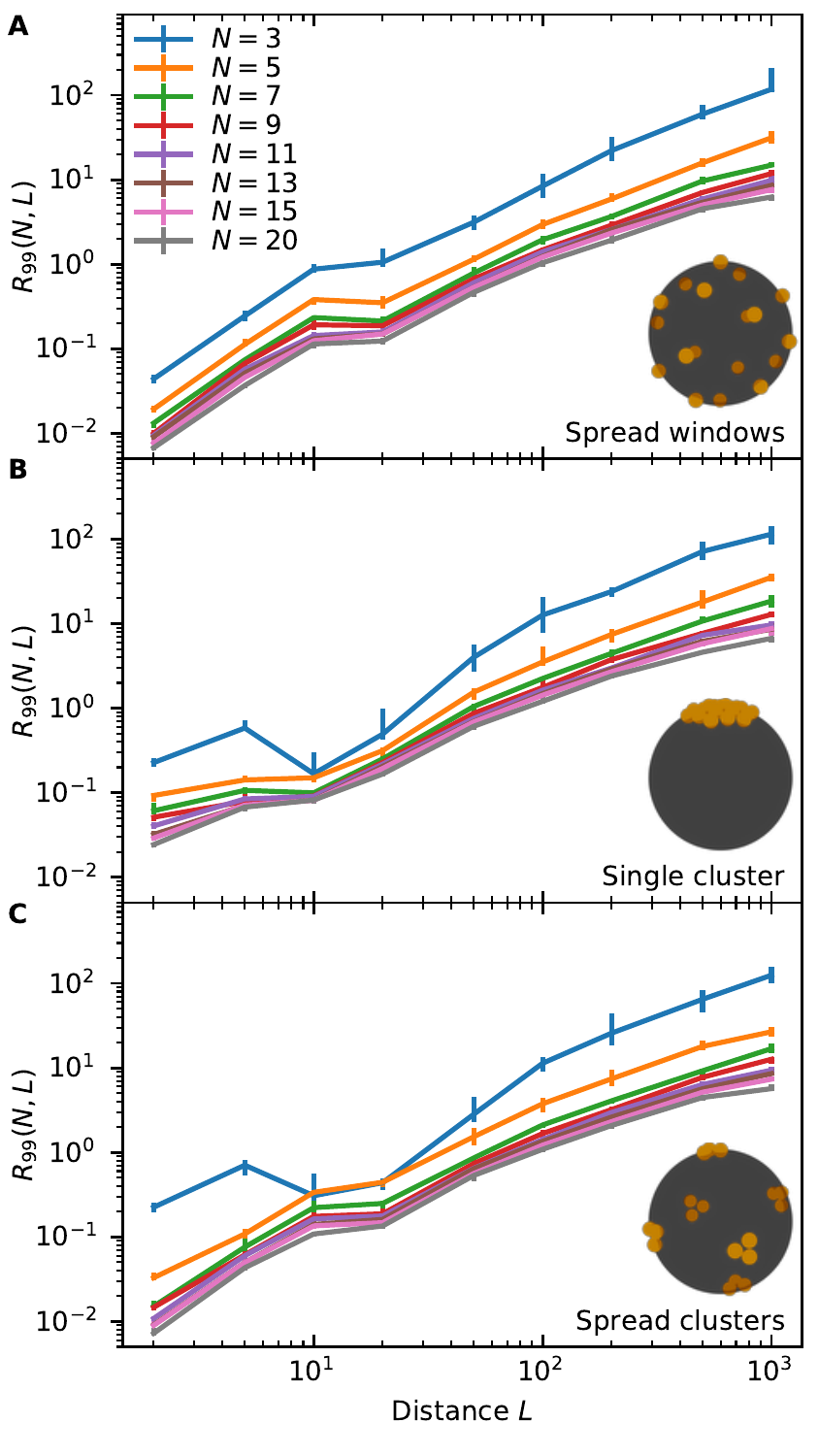}
\includegraphics[scale=0.9]{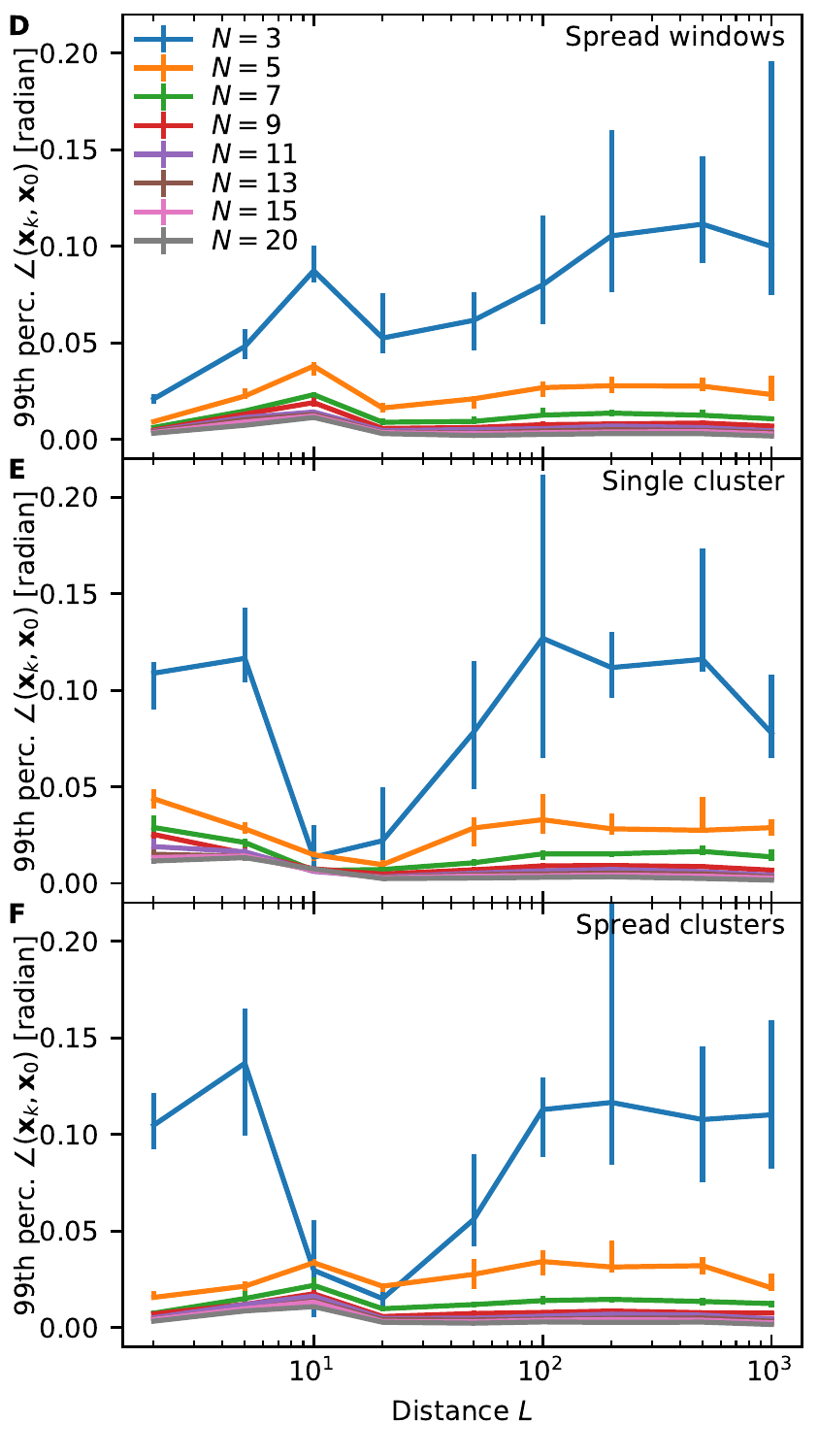}
\caption{{\bf Deviation from the exact source location.} {\bf (A-C)} Deviation of the reconstructed source from its exact position measured by the radius $R_{99}$ around $\x_0$ in which $99\%$ of recovered points fall, when the fluxes are subject to multinomial noise with $N_t=10^7$ trials. The source location is computed by averaging over reconstructions from all combinations of three windows. The deviation is averaged over {$1000$} noise realizations. The source position is randomly chosen with $|\x_0|=L$ fixed and $R_{99}$ is shown as a function of distance $L$. {\bf (D-F)} The angular spread of the recovered positions measured as the $99th$ percentile of the angle between the recovered and the exact source position. {\bf (A,D)} Windows are spread uniformly across the ball surface, {\bf (B,E)} a single cluster of windows, and {\bf (C,D)} window clusters are spread uniformly across the ball surface (at most three windows per cluster).}
 \label{fig:deviation}
\end{figure}
The statistical deviation $R_{99}$ from the exact source position increases algebraically with the distance $L$ (Fig.~\ref{fig:deviation}A-C). When windows are clustered, $R_{99}$ has a dip at $L\approx10$ (Figs.~\ref{fig:deviation}B,C). This dip coincides with a marked reduction in the angular spread at these distances compared to when windows are uniformly distributed (Figs.~\ref{fig:deviation}D-F).

\subsection{Gaussian or multinomial noise vs intersecting parallelpipeds}\label{s:noise_vs_parallelepiped}
We discussed two different measures of uncertainty: the volume of the intersection of parallelepipeds formed from linear error estimates $V_{unc}$ and a statistical measure of the spread of simulated source position localizations. The former measure rapidly decreases when the number of windows increases. However, this procedure assumes that the center of the parallelepipeds is known. Because the intersecting parallelepipeds are rotated relative to each other, the intersection of the whole ensemble decays quickly. In addition, when windows are far from each other, the error vectors perpendicular to the line connecting them decrease. Conversely, for windows that are close to each other, small differences in their measured fluxes are more susceptible to noise. Individual uncertainty volumes for the different choices of three receptor windows have aspect ratios that differ substantially from one; thus, while the individual uncertainty volumes might have comparable volumes, their intersection would have a very small volume.\\
The latter measure, $R_{99}$ rather describes a statistical spread. In contrast to the volume of uncertainty, its calculation it requires substantial computational effort. However, its advantage is its relatively straightforward interpretation. Both quantities tell of a substantial reduction in the localization error when the number of windows is increased.
%
\section{Concluding remarks - applications and principles for neuronal navigation in the brain}\label{discussion}
Here, we reviewed a combination of modeling, analysis and hybrid simulations approaches based on the theory of diffusion to ultimately reconstruct a gradient source position from fluxes to small windows. This represents a first step toward a quantitative answer to how cells, and especially neurons, can position themselves and be guided by different multiple morphogen gradients in the brain. This position finding mechanism is crucial for them to navigate to their final, proscribed location relevant to their function.\\
We presented a general computational approach to estimate the steady-state fluxes of Brownian particles to narrow windows located on a surface in two and three dimensions. We presented hybrid stochastic simulation approaches, which replace random walks between the point source and a window by mapping the source position to an imaginary surface (a half-sphere in the case of half-space and an entire sphere in the case of a ball). This is followed by a stochastic step where the Brownian trajectories are simulated in a small neighborhood of the surface. The analytical part of the method is based on computing the asymptotic solution of Laplace's equation using the Neumann-Green's function and matched asymptotics. The analytical relation between the flux expressions and the location of the source leads to a reconstruction procedure of the source from measured fluxes. In the limit of small windows, the arrival distribution of Brownian particles is Poissonian and thus the value of the steady-state flux can be estimated from averaging the duration of arrival time events.\\
In addition, this approach allows estimating how measurement fluctuations in the fluxes can be compensated by increasing the number of receptor windows. The uncertainty is represented as the volume of the Jacobian matrix for any three windows. By considering the combinatorics of any three windows out of $N$ (binomial $C^3_N$), the uncertainty can be estimated by the intersection of a large number of parallelepipeds. Finding the exact decay of the uncertainty volume with the number of windows remains an open question. In case of random fluctuation ($\eta$ is Gaussian variable in equation \ref{pertub}), the error between the source and its estimation depends on trace of the matrix $J^T J$ (see equation \ref{eq:Jacob}) and the variance of the Gaussian.\\
The present approach could be extended to the case where diffusing particles can be destroyed with a killing rate $k(\x)$. This would correspond to a (realistic) degradation of molecular cues between them being released by the source and their arrival at receptors. Mathematically, it leads to an exponential decaying diffusion profile, and thus a sharper gradient.

\subsection{Identifying the exact position of a gradient source}\label{discussion2}
\begin{figure}[http!]
\centering
\includegraphics[scale=0.9]{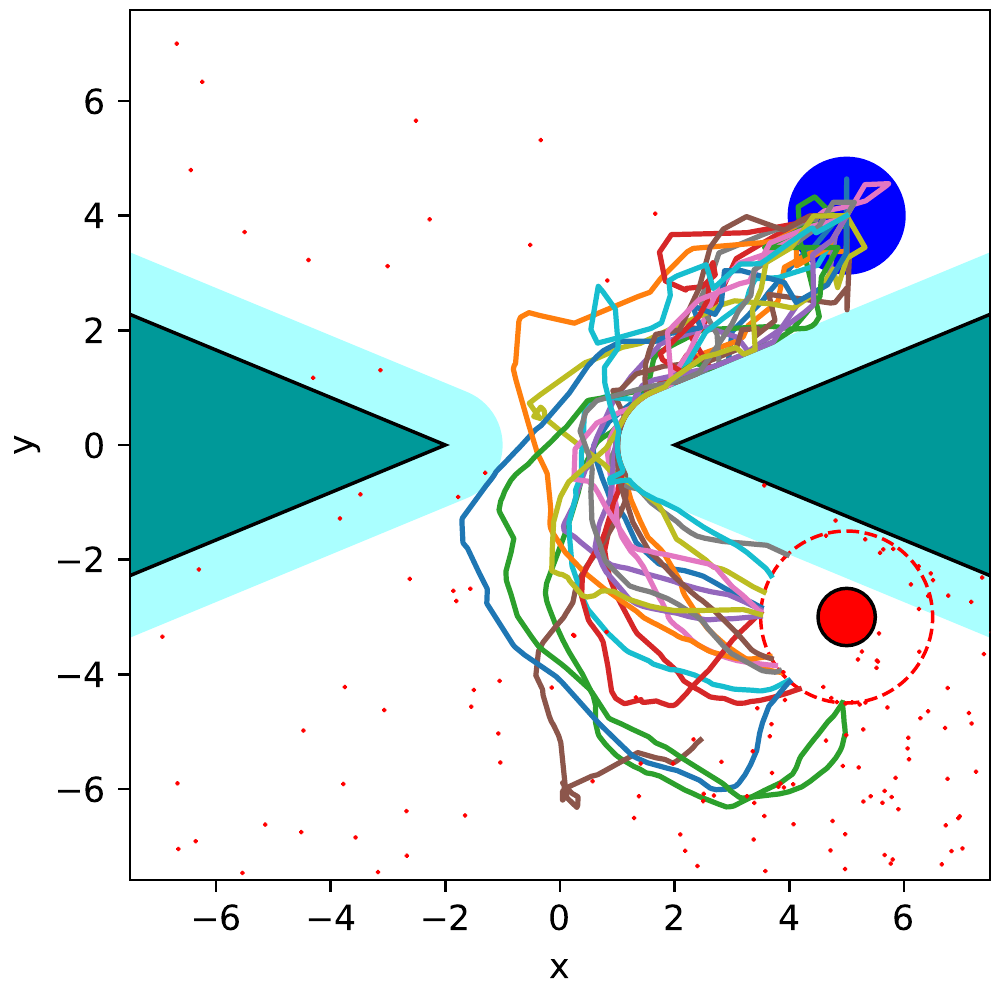}
\caption{{\bf Navigation through a slit between two wedges.} A cell of radius $R=1$ (blue circle) is guided through an opening between two reflective wedges (dark green areas). The source (red circle) emits Brownian particles with a rate $Q$ that are absorbed at $N$ receptors evenly spaced on the cell’s perimeter. Each absorption event is recorded and used to calculate the receptor’s activation level $A_i(t)=\mathrm{tanh}\left[\frac{1}{K}\left(\sum_je^{-\mu(t-\tau^{(i)}_j)}\right)\right]$ and the cell’s instantaneous velocity $\mathbf{v}(t)=v_0\frac{\sum_{i=1}^{N} A_i(t) \mathbf{x}_i}{\sum_{k=1}^{N}A_k(t)}$, where $\mu$ is the binding memory decay rate, $K$ is the receptor activation threshold, $\mathbf{x}_i$ are the receptor positions relative to the cell center and $v_0$ is the (constant) cell speed. The colored lines show 20 different realizations of the cell’s trajectory. Trajectories are terminated once the cell’s distance to the source is smaller than 1.5 cell radii (dashed red circle). The areas colored in cyan indicate excluded regions where the cell is unable to go due to its finite size. Parameters are $Q=10$, $\mu=0.5$, $K=3$, $N=12$ and $v_0=1$.}
 \label{fig:navigation}
\end{figure}

While we discussed the mathematical procedure that allows the recovery of a diffusive gradient source, the biological mechanism that allows cells to perform this feat is still elusive. Specifically, how neurons accurately identify the exact position of a gradient source remains to clarified~\cite{chedotal2010wiring,kolodkin2011mechanisms}. Even if the main molecular guidance cues have been identified and characterized, the physical mechanism that converts the external flux into a series of commands that generate the neuronal path is unclear. Possible computational rules to integrate the flow of cues to moving a cell in real time can be implemented~\cite{reingruber2014computational} as shown in Fig.~\ref{fig:navigation}.\\
The first step needs to consist of reading an external gradient field \cite{Malherbe,desponds2020mechanism}. Subsequently, this message needs to be internalized at the growth cone level to determine when to grow or to stop at a given position, a question that also remains to be understood. The present results suggest that that three receptors are sufficient to triangulate the position of the source and any additional receptors are redundant. They, however are likely required to increase the precision of the source localisation to acceptable levels at the required distances.
\subsection{Some emerging principles}
\begin{enumerate}
  \item Modeling approaches based on diffusion proposes that direction sensing can only occur for short distance (on the order of ten times the size of the cell). However, when a cell is placed in a band, not much wider than the cell itself, numerical simulations suggest that directions can be sensed over much longer distances. This is due to the very rare events of molecules that can bind to receptors located on the side of the cell facing away from the source. This situation maintains a constant asymmetry in the binding to receptors. While this case is quite particular and almost reduces to a one-dimensional scenario, it is applicable in the developing brain with neurons crossing a narrow corridor in the ventral telencephalon (Fig. \ref{fig:navigation}). In addition, the susceptibility of the source localization to random fluctuations in the fluxes is rather high, leading to source localization point clouds that are much larger than the target (Fig.~\ref{fig:clouds}). This precision loss may however be compatible with the intrinsic variability found in neuronal connectivity \cite{reingruber2014computational}. These limitations suggest that other mechanisms might be necessary for increasing the navigation reliability at large distances, such as positive or negative cooperativity, and multiple gradient sources.
  \item Obstacles could alter the distribution of a gradient and the exact effect should be studied in the future. Interestingly, obstacles not only can affect the distribution of a gradient cues but also the cell decisions in the case of conflicting gradients (i.e. chemotaxis versus topotaxis)~\cite{wondergem2019chemotaxis}.
  \item The source position is computed from a steady-state distribution of fluxes. This steady-state assumption could be further alleviated, but remains compatible with the slow velocity of neuronal growth compared to gradient formation~\cite{lander2002morphogen,kasatkin2008morphogenetic,reingruber2014computational}.
  \item The recovery of the source position requires three separate receptor windows in both two and three dimensions. This is due to recurrance of Brownian motion in two dimensions (i.e. the critical dimension for Brownian motion is $d=2$).
  \item In the models presented here, receptors are fully absorbing. However, it should not matter if receptors are only partially absorbant or if the temporal dynamics of binding and unbinding are considered. These features might influence the accuracy of the sensing, but not the general viability of the mechanism.
  \item A small number of windows is compatible with the number of receptors observed on growth-cone or neurites involved in detecting a gradient concentration, its direction to turn, its forward or retracting motion~\cite{bouzigues2007asymmetric,stettler2012engrailed}. The fast binding limit is also applicable for other cell models such bacteria~\cite{Heinrich3} or sperm~\cite{kaupp2017signaling,strunker2015physical}.
  \item Sensing gradient fluxes at receptors distributed across the cell surface is certainly the first required step, and this information is sufficient to reconstruct the location of a source. In order to understand how the cell processes this information, the secondary pathway could be modeled further: Receptors might be coupled by actin (de-)polymerization, such that a flux difference in fluxes is immediately translated into growth, retraction or turning of the growth cone into growth, retraction or turning.
  \item Are there optimal distribution of receptors? A uniform distribution or clustered receptors do not change the accuracy appreciably over large distances (Figs.~\ref{fig:deviation} and \ref{fig:clouds}). It might be interesting to extend these studies to the case of elongated elliptic domains to investigate the role of long protrusions (such as filopodia) located on the growth cone. The constant reorganization of receptors \cite{bouzigues2010mechanism} likely could be seen as a secondary pathway following the interpretation of an external gradient. Indeed, receptor activation can mediate cellular transduction that transform an external environment signal into a cellular biochemical activation cascade \cite{fain2019sensory}. When a cell compares the flux differential between one side and the other, the local transduction of the signal at the scale of a receptor must not be homogenized throughout the rest of the cell domain, and the local information about gradient direction needs to be preserved. The internally transduced signal can be carried by the concentration of second messenger or diffusing surface molecules such as calcium. Hence, receptor activations need to be localized inside the cell, leading to an asymmetrical response.
  \item If a biological cell can reliability evaluate the source location, it is conceivable that it can navigate toward a final destination that can be different from the gradient source.
\end{enumerate}
Finally, the present modeling and simulations could be apply to the case of blood vessel growth, glia cells growth during normal conditions and also during angiogenesis in gliomas.
\subsection{Developmental biology}\label{s:devbio}
Neurons in the brain require the reliable ability to compute their position in order to navigate in the Brain in three dimensions. In mice, thalamocortical axons start extending from the diencephalon into the ventral telencephalon and have to cross the striatum to reach the base of the cortex. Their growth across the striatum requires the navigation of a corridor of guidepost neurons (also called corridor cells). When corridor cells are perturbed, it prevents many thalamocortical axons from growing across the striatum. This shows that those axons actively use the signals from corridor cells to determine their position and to pass through a narrow tube. Thalamocortical axons also have the ability to measure the two parallel expression gradients of two messengers (Slit1 and Netrin-1\cite{chedotal2010wiring}) in order to be sorted correctly in the cortex computations that are still unknown.
\section*{Acknowledgements}
U.D. was supported by a Herchel Smith Postdoctoral Fellowship and acknowledges core funding by the Wellcome Trust (092096) and CRUK (C6946/A14492). D. H. 's research has received funding from the European Research Council (ERC) under the European Union’s Horizon 2020 research and innovation programme (grant agreement No 882673), Plan Cancer-INSERM Projet 19CS145-00 and ANR NEUC-0001.
\appendix
\section{Triangulation algorithm for \(N=3\) windows in 2D half-space}
\label{sec:triang_algo_2D_halfplane}
The distance between two window positions $\x_i$ and $\x_j$ is given by
$d_{ij}=|\x_i-\x_j|.$
We then define the scaled quantities
$h_{ij}=\log\frac{d_{ij}}{\eps},$
the product vector
$F=-(h_{23}[h_{12}+h_{13}-h_{23}], h_{13}[h_{12}-h_{13}+h_{23}], h_{12}[-h_{12}+h_{13}+h_{23}])^T,$
the determinant
$\Delta=(h_{12}+h_{13}+h_{23})^2-4(h_{12}h_{13}+h_{12}h_{23}+h_{13}h_{23})$
and the matrix
\beq
[b]=\begin{pmatrix}
  -2h_{23} & -h_{12}+h_{13}+h_{23} & h_{12}-h_{13}+h_{23} \\
  -h_{12}+h_{13}+h_{23} & -2h_{13} & h_{12}+h_{13}-h_{23} \\
  h_{12}-h_{13}+h_{23} & h_{12}+h_{13}-h_{23} & -2h_{12}
\end{pmatrix}.
\eeq
This gives the solution to Eq.~(1) as
\beq \label{3receptors}
\PP=\frac{1}{\Delta}(\FF-\pi\GG^T [b]),
\eeq
where $\PP=(P_1, P_2, P_3)^T$ and
\beq
\GG=\begin{pmatrix} G(\x_1,\x_0) \\ G(\x_2,\x_0) \\ G(\x_3,\x_0)
\end{pmatrix}.
\eeq
We then use Eq.~(7) for the expression of the Green's function, which yields three intersecting curves, one of which contains redundant information due to $P_1+P_2+P_3=1$. The equations are inverted numerically invert to find the intersection point $\x_0$ via the multidimensional nonlinear root finding algorithm {\it hybrj} contained in MINPACK~[21].

\section{Explicit Green's function for the semi-strip} \label{sec:green_band}
The hybrid algorithm is based on the exact expression of the Neumann-Green's function $G_{Se}$ for the semi-strip
\beq
\Omega_a=\{(x_1,x_2)\in \rR^2 |x_1>0,0<x_2<a\}
\eeq
where $a>0$. The normalized flux is the distribution of exit points \cite{Schuss4}.
We impose zero absorbing boundary condition on the boundary $\p\Omega_1=\{(0,x_2)|0<x_2<a\}$ and reflecting boundary condition on the rest of the strip $\p\Omega_2=\{(x_1,0)|x_1>0\}\cup\{(x_1,a)|x_1>0\}$. The boundary value problem is
\beq
\begin{aligned}
\label{green-semistrip_appdx}
-\Delta_{\y} G_{Se}(\x,\y) & = \delta(\x-\y), &\quad &\text{for}\;\; \x, \y\, \in \,\Omega, \\
 \frac{\p G_{Se}}{\p n_{\y}}(\x,\y) & =  0, &\quad &\text{for}\;\; \y\, \in\, \p\Omega_2,\ \x \in \,\Omega,\\
 G_{Se}(\x,\y) & =0 &\quad &\text{for}\;\; \y\, \in\, \p\Omega_1,\ \x \in \,\Omega.
\end{aligned}
\eeq
We compute $G_{Se}(\x,\y)$ by expanding in eigenfunction following the classical method of \cite{melnikov2011green} (p. 80). To start we write the following Ansatz
\beq
u(x_1,x_2)=\int_0^\infty\int_{0}^{a}G_{Se}(x_1,x_2;y_1,y_2)f(y_1,y_2)dy_1 dy_2,
\eeq
with $x_1$, $x_2$ the components of $\x$ and $y_1$, $y_2$ the components of
$\y$, which solves the inhomogeneous diffusion equation
\beq
\begin{aligned}
\label{udiffeq}
-\left(\frac{\p^2}{\p x_1^2}+\frac{\p^2}{\p x_2^2}\right) u(x_1,x_2) & = f(x_1,x_2), &\quad &\text{for}\;\; x_1>0\,,\; 0 < x_2 < a, \\
 \frac{\p u}{\p x_2}(x_1,x_2) & =  0, &\quad &\text{for}\;\; x_1>0\,,\;\; x_2=0 \; \text{and}\; x_2=a,\\
 u(0,x_2) & =0, &\quad &\text{for}\;\; 0<x_2<a.
\end{aligned}
\eeq
Because $x_2$ is bounded to between $0$ and $a$, we can write $u$ and $f$ in
terms of a Fourier series along $x_2$ \beq
\label{ufourier}
u(x_1,x_2)=\sum_{n=0}^{\infty}u_n(x_1) \cos\omega_nx_2\,,\quad\omega_n=\frac{n\pi}{a},
\eeq
\beq
\label{ffourier}
f(x_1,x_2)=\sum_{n=0}^{\infty}f_n(x_1) \cos\omega_nx_2\,,\qquad f_n(x_1)=\frac{2}{a}\int_0^af(x_1,x_2)\cos(\omega_n x_2)dx_2,
\eeq
with $\omega_n=\frac{n\pi}{a}$ By inserting this expression for $u$ into equation~(\ref{udiffeq}) we arrive at the following ODE for the $u_n$
\beq
u_n''-\omega_n^2u_n=f_n\,.
\eeq
For $n=0$, the fundamental solutions to the homogeneous equation $u_0''=0$ are $u_0^{(1)}=1$ and $u_0^{(2)}=x_1$. Therefore, the inhomogeneous problem is solved by
\beq
u_0(x_1)=\int_0^{x_1}y_1f_0(y_1)dy_1+C_1+x_1\Bigl(-\int_0^{x_1}f_0(y_1)dy_1+C_2\Bigr)=\int_0^\infty\min\{x_1,y_1\}f_0(y_1)dy_1,
\eeq
where $C_2=\int_0^{\infty}f(y_1)dy_1$ due to the boundedness condition on the solution as $x_1\to 0$ and $C_1=0$ due to the absorbing boundary at $x_1=0$.

For $n>1$, the fundamental solutions to the homogeneous system $u_n''=\omega_n^2 u_n$ are given by $u_n=\exp(\pm\omega_n x_1)$. Hence,
\beq
\begin{aligned}
u_n(x_1)&=\frac{e^{\omega_nx_1}}{2\omega_n}\Bigl(-\int_0^{x_1}e^{-\omega_ny_1}f_n(y_1)dy_1+C_1\Bigr)-\frac{e^{-\omega_nx_1}}{2\omega_n}\Bigl(\int_0^{x_1}e^{\omega_ny_1}f_n(y_1)dy_1+C_2\Bigr)
&=\frac{e^{\omega_nx_1}}{2\omega_n}\int_{x_1}^{\infty}e^{-\omega_ny_1}f_n(y_1)dy_1+\frac{e^{-\omega_nx_1}}{2\omega_n}\Bigl(\int_0^{x_1}e^{\omega_ny_1}f_n(y_1)dy_1-\int_0^\infty e^{-\omega_ny_1}f_n(y_1)dy_1\Bigr) \\
&=\frac{1}{2\omega_n}\int_{0}^{\infty}\Bigl(e^{-\omega_n|x_1-y_1|}-e^{-\omega_n(x_1+y_1)}\Bigr)f_n(y_1)dy_1
\end{aligned}
\eeq
Thus, the complete solution to equation~(\ref{udiffeq}) reads
\beq
\begin{aligned}
u&(x_1,x_2)=\int_0^\infty\Bigl(\min\{x_1,y_1\}f_0(y_1)+\frac{a}{2\pi}\sum_{n=1}^{\infty}\frac{1}{n}\bigl[e^{-\omega_n(y_1+x_1)}-e^{-\omega_n|y_1-x_1|}\bigr]f_n(y_1)\cos\omega_nx_2\Bigr)dy_1 \\
&=\int_0^a\int_0^\infty\Bigl(\frac{1}{a}\min\{x_1,y_1\}+\frac{1}{\pi}\sum_{n=1}^{\infty}\frac{1}{n}\bigl[e^{-\omega_n(y_1+x_1)}-e^{-\omega_n|y_1-x_1|}\bigr]\cos\omega_nx_2\cos\omega_ny_2\Bigr)f(y_1,y_2)dy_1dy_2\\
&=\int_0^a\int_0^\infty\Bigl(\frac{1}{a}\min\{x_1,y_1\}+\\
&\qquad+\frac{1}{2\pi}\sum_{n=1}^{\infty}\frac{1}{n}\bigl[e^{-\omega_n(y_1+x_1)}-e^{-\omega_n|y_1-x_1|}\bigr]\bigl[\cos\omega_n(x_2-y_2)+\cos\omega_n(x_2+y_2)\bigr]\Bigr)f(y_1,y_2)dy_1dy_2,
\end{aligned}
\eeq
where we inserted the fourier coefficients for $f$ from
equation~(\ref{ffourier}). By inspection, we arrive at the expression for the Green's function
\beq
G_{Se}(x_1,x_2;y_1,y_2)=\frac{1}{a}\min\{x_1,y_1\}+\frac{1}{2\pi}\sum_{n=1}^{\infty}\frac{1}{n}\bigl[e^{-\omega_n(y_1+x_1)}-e^{-\omega_n|y_1-x_1|}\bigr]\bigl[\cos\omega_n(x_2-y_2)+\cos\omega_n(x_2+y_2)\bigr].
\eeq
Using the identity~(see \cite{melnikov2011green} p. 84)
\beq
\sum_{n=1}^{\infty}\frac{q^{n}}{n}\cos n\phi=-\frac{1}{2}\ln(1-2q\cos\phi+q^2),
\eeq
we can further simplify to get
\beq
\label{greensfct-semistrip}
\begin{aligned}
  G_{Se}(x_1,x_2;y_1,y_2)=-\frac{1}{4\pi}\bigl[&\ln(1-2e^{-\omega|x_1-y_1|}\cos\omega(x_2+y_2)+e^{-2\omega|x_1-y_1|}) \\
  &+\ln(1-2e^{-\omega|x_1-y_1|}\cos\omega(x_2-y_2)+e^{-2\omega|x_1-y_1|}) \\
  &-\ln(1-2e^{-\omega(x_1+y_1)}\cos\omega(x_2+y_2)+e^{-2\omega(x_1+y_1)}) \\
  &-\ln(1-2e^{-\omega(x_1+y_1)}\cos\omega(x_2-y_2)+e^{-2\omega(x_1+y_1)}) \\
  &+\frac{4\pi}{a}\min\{x_1,y_1\}\bigr]\,.
\end{aligned}
\eeq
with $\omega=\pi/(2a)$. The exit
probability distribution is again given by the flux through the $\p\Omega_1$
boundary
\begin{equation}
  \label{eq:strip-y-distribution_appdix}
  p_{ex}(x_2;y_1,y_2)=\frac{\p G_{Se}}{\p y_1}\Bigl|_{y_1=0}=\frac{\sinh\omega y_1}{2a}\Bigl[\frac{1}{\cosh\omega y_1-\cos\omega(x_2+y_2)}+\frac{1}{\cosh\omega y_1-\cos\omega(x_2-y_2)}\Bigr]\,.
\end{equation}

\section{Exact Neumann-Green's function for the ball}\label{appendix1}
The Neumann's function  $\mathcal{\tilde{N}}(\x,\x_0)$ is the solution of Laplace's equation
\beq
\Delta \mathcal{\tilde{N}}(\x,\x_0)&=&-\delta(\x-\x_0) \hbox{ for } \x \in \mathbb{R}^3 \nonumber \\
\frac{\p \mathcal{\tilde{N}}}{\p n} (\x,\x_0) &=& 0 \hbox{ for } \x \in S_a=\p B_a.
\eeq
where $S_a$ is the sphere of the three-dimensional ball $B_a$ and the source point $\x_0 \in \rR^3-B_a$. The analytical expression of the Neumann function~\cite{lagache2017extended} is
\beq \label{Neumann}
\mathcal{\tilde{N}}(\x,\x_0)&=& \ds\frac{1}{4\pi |\x-\x_0|}+\frac{a}{4\pi |\x_0||x-\ds \frac{a^2 \x_0}{|\x_0|^2}|}\nonumber \\&+&
\ds{\frac{1}{4\pi a }\log\left( \ds\frac{\ds\frac{|\x_0||\x|}{a^2}\left(1-\cos(\theta)\right)}{\ds 1-\frac{|\x_0||\x|}{a^2} \cos(\theta)+\left(1+\left(\frac{|\x_0||\x|}{a^2}\right)^2-2\frac{|\x_0||\x|}{a^2} \cos(\theta)\right)^{\frac{1}{2}}}\right)}. \label{N}
\eeq
When $\x$ and $\x_0$ are on the sphere $S_a$, $|\x_0|=|\x|=a$, we obtain the expression
\beq
\mathcal{\tilde{N}}(\x,\x_0)&=&\frac{1}{2\pi |\x-\x_0|}+ \frac{1}{4\pi a }\log\left(\frac{|\x-\x_0|}{2a+|\x-\x_0|}\right). \label{Neumann2}
\eeq
The far field expansion for $|\x|\gg 1 $ is given by
\beq \label{decay}
\mathcal{\tilde{N}}(\x,\x_0)&\approx&  \frac{1}{4\pi |\x|} +\frac{3 \x .\x_0}{8\pi |\x|^3} +O(\frac{1}{|\x|^3})\\
\mathcal{\nabla \tilde{N}}(\x,\x_0)&\approx& O(\frac{1}{|\x|^2})
\eeq
We provide now the derivation of the Neumann's function $\mathcal{N}(\x_i,\x_j)$ using a two parameter expansion in $a$ and $\eps$.  For $\x$ and $\x_0$ in the neighborhood of the sphere $S_a$, we expand the Neumann function $\mathcal{N}(\x,\x_0)$
\beq
\mathcal{N}(\x,\x_0)=\mathcal{\tilde{N}}(\x,\x_0)+O(1),
\eeq
where
$\mathcal{\tilde{N}}(\x,\x_0)$ is solution of the Laplace equation
\beqq
\Delta \mathcal{\tilde{N}}(\x,\x_0)&=&-\delta(\x-\x_0) \hbox{ for } \x \in \mathbb{R}^3 \nonumber \\
\frac{\p \mathcal{\tilde{N}}}{\p n} (\x,\x_0) &=& 0 \hbox{ for } \x \in S_a.
\eeqq
To compute the $\log$-term, we first decompose $\mathcal{\tilde{N}}(\x,\x_0)=\frac{1}{4\pi |\x-\x_0|}+\Phi(\x,\x_0)$ where $\Phi$ is solution of the system:
\beq\label{equadiff}
\Delta \Phi(\x,\x_0)&=&0, \hbox{ for } \x \in \mathbb{R}^3 \nonumber \\
\frac{\p \Phi}{\p n} (\x,\x_0) &=& -\frac{\p}{\p n}\left(\frac{1}{4\pi  |\x-\x_0|}\right), \hbox{ for } \x \in S_a.
\eeq
To solve equation (\ref{equadiff}), we choose a coordinate system for which the point source $\x=\x_0$ is on the positive $z-$axis. Since $a \Phi=0$ and $\Phi$ is axisymmetric, it has the series expansion
\beq
\Phi(\x,\x_0)=\sum_{n=0}^{\infty} b_n(|\x_0|) \frac{P_n(\cos(\theta))}{|\x|^{n+1}},
\eeq
where $P_n$ are the Legendre polynomials of integer $n$, $\theta$ is the angle between $\x$ and the north pole and $b_n(|\x_0|)$ are coefficients, determined from boundary condition \ref{equadiff}.\\
For $\x \in S_a$ and $\rho=|\x|$,
\beq\label{bound1}
\frac{\p \Phi}{\p n} (\x,\x_0)=\frac{\p \Phi}{\p \rho} (\rho=a)=-\sum_{n=0}^{\infty}\frac{\left(n+1\right) b_n(|\x_0|)}{a^{n+2}} P_n(\cos(\theta).
\eeq
For $|\x|<|\x_0|$, we have the expansion
\beq
\frac{1}{4\pi |\x-\x_0|}=\frac{1}{4\pi }\sum_{n=0}^{\infty}\frac{|\x|^n}{|\x_0|^{n+1}}P_n\left(\cos \left(\theta\right)\right),\label{gen}
\eeq
which leads to the boundary condition:
\beq\label{bound2}
-\frac{\p}{\p \rho}\left(\frac{1}{4\pi  |\x-\x_0|}\right)\left(\rho=a\right)=-\frac{1}{4\pi }\sum_{n=0}^{\infty}\frac{n a^{n-1}}{|\x_0|^{n+1}}P_n\left(\cos \left(\theta\right)\right).
\eeq
Injecting relations (\ref{bound1}-\ref{bound2}) into the boundary condition (\ref{equadiff}), we obtain that for all $n \geq 0$:
\beq \label{b_n}
b_n(|\x_0|)=\frac{1}{4\pi } \frac{n a^{2n+1}}{(n+1)|\x_0|^{n+1}}.
\eeq
The Neumann function $\mathcal{\tilde{N}}(\x,\x_0)$ is then given by:
\beq
\mathcal{\tilde{N}}(\x,\x_0)=\frac{1}{4\pi |\x-\x_0|}+\frac{1}{4\pi } \sum_{n=0}^{\infty} \frac{n a^{2n+1}}{(n+1)|\x|^{n+1}|\x_0|^{n+1}} P_n(\cos(\theta)),
\eeq
that we rewrite
\beq\label{expa}
\mathcal{\tilde{N}}(\x,\x_0)=\frac{1}{4\pi |\x-\x_0|}+\frac{1}{4\pi } \sum_{n=0}^{\infty} \left(\frac{a^{2n+1}}{|\x|^{n+1}|\x_0|^{n+1}}-\frac{a^{2n+1}}{(n+1) |\x|^{n+1}|\x_0|^{n+1}}\right) P_n(\cos(\theta)).
\eeq
Using expansion (\ref{gen}), we have for the first term of (\ref{expa}),
\beq
\frac{1}{4\pi } \sum_{n=0}^{\infty} \frac{a^{2n+1}}{|\x|^{n+1}|\x_0|^{n+1}} P_n(\cos(\theta))=\frac{a}{4\pi |\x_0||x-\frac{a^2 \x_0}{|\x_0|^2}| }.
\eeq
To compute the second term $I(\rho)=-\sum_{n=0}^{\infty}\frac{a^{2n+1}}{ (n+1) \rho^{n+1}|\x_0|^{n+1}} P_n(\cos(\theta))$, we differentiate
\beq
I'(\rho)=\sum_{n=0}^{\infty}\frac{a^{2n+1}}{ \rho^{n+2}|\x_0|^{n+1}} P_n(\cos(\theta))=\frac{a}{\rho |\x_0||x-\frac{a^2 \x_0}{|\x_0|^2}|},
\eeq
that is
\beq
I'(\rho)=\frac{1}{\rho a \left(1+\frac{|\x_0|^2 \rho^2}{a^4}-2\frac{|\x_0|\rho}{a^2} \cos(\theta)\right)^{\frac{1}{2}}}.
\eeq
Because $\lim_{\rho \to \infty} l(\rho)=0$, we have:
\beq
l(\rho)=-\int_{\rho}^{\infty} I'(s)ds= -\int_{\rho}^{\infty} \frac{ds}{s a \left(1+\frac{|\x_0|^2 s^2}{a^4}-2\frac{|\x_0| s}{a^2} \cos(\theta)\right)^{\frac{1}{2}}}.
\eeq
Thus,
\beq
\ds{l(\rho)=\frac{1}{a}\log\left(\frac{\frac{|\x_0|\rho}{a^2}\left(1-\cos(\theta)\right)}{1-\frac{|\x_0|\rho}{a^2} \cos(\theta)+\left(1+\left(\frac{|\x_0|\rho}{a^2}\right)^2-2\frac{|\x_0|\rho}{a^2} \cos(\theta)\right)^{\frac{1}{2}}}\right)}. \label{I5}
\eeq
Finally, we obtain the Neumann function $\mathcal{\tilde{N}}(\x,\x_0)$ and the exact expression  with respect to the inner ball of radius $a$:
\beqq
\mathcal{\tilde{N}}(\x,\x_0)&=&\frac{1}{4\pi |\x-\x_0|}+\frac{a}{4\pi D |\x_0||x-\frac{a^2 \x_0}{|\x_0|^2}|}\nonumber \\&+&
\ds{\frac{1}{4\pi a }\log\left(\frac{\frac{|\x_0||\x|}{a^2}\left(1-\cos(\theta)\right)}{1-\frac{|\x_0||\x|}{a^2} \cos(\theta)+\left(1+\left(\frac{|\x_0||\x|}{a^2}\right)^2-2\frac{|\x_0||\x|}{a^2} \cos(\theta)\right)^{\frac{1}{2}}}\right)}. \label{N1}
\eeqq
When $\x$ and $\x_0$ are on the sphere $S_a$, $|\x_0|=|\x|=a$, we have
\beqq
\mathcal{\tilde{N}}(\x,\x_0)&=&\frac{1}{2\pi |\x-\x_0|}+ \frac{1}{4\pi a }\log\left(\frac{|\x-\x_0|}{2a+|\x-\x_0|}\right). \label{fi2}
\eeqq
\section{Algorithm to generate distribution of windows on the ball surface}
\label{sec:algo_window_distrib_ball}
To compute a uniform, single cluster, and uniform distribution of clusters arrangement of windows  on the surface of a all (cases A, B and C respectively in subsection \ref{ss:clouds}), the following Monte Carlo algorithm can be used:
\begin{enumerate}
\item Initialize the first window angles to $\phi_1=0$, $\theta_1=0$. Set the window counter to $c=2$.
\item Initialize the next window angles $\phi_c=0$, $\theta_c=\pi$.
\item Initialize the step counter $j=1$. Set $E_0=-\infty$.
\item Calculate the window positions $\x_i=(\sin\phi_i\sin\theta_i,\cos\phi_i\sin\theta_i,\cos\theta_i)$, $i=1,\dots,c-1$.
\item Generate candidate angles $\tilde{\phi}_c=\phi_c+\Phi_{j,c}$ and $\tilde{\theta}_c=\theta_c+\Theta_{j,c}$ where $\Phi_{j,c},\Theta_{j,c}\sim N(0,\rho)$. Wrap $\tilde{\phi}_c$ and $\tilde{\theta}_c$ such that they are proper continuous coordinates on the unit sphere.
\item Calculate the candidate position $\y_c=(\sin\tilde{\phi}_c\sin\tilde{\theta}_c,\cos\tilde{\phi}_c\sin\tilde{\theta}_c,\cos\tilde{\theta}_c)$. Evaluate the total energy as $E_j=\sum_{i=1}^{c-1}V(|\x_i-\y_c|, \y_c)_{i,c}$ and calculate $\Delta E=E_j-E_{j-1}$.
\item If $\Delta E<0$, or $u<\exp(-\Delta E/T)$ where $u$ a random variable uniform in $[0, 1]$, accept the step and set $\phi_c=\tilde{\phi}_c$, $\theta_c=\tilde{\theta}_c$. Otherwise, set $E_{j}=E_{j-1}$ and discard the candidate position.
\item Increment the step counter $j$. If $j=1000$, go to step 2. Otherwise go to step 3.
\end{enumerate}
The potential $V(r)$ is different for each case:
\begin{itemize}
\item Case A, uniformly spread windows:
  \beq
  V_A(r,\x_c)_{i,c}=-r
  \eeq
\item Case B, single cluster:
  \beq
  V_B(r,\x_c)_{i,c}=4\left[\left(\frac{1}{5r}\right)^{12}-\left(\frac{1}{5r}\right)^6\right] - \x_c\cdot\mathbf{e}_z
  \eeq
\item Case C, spread out clusters of three windows each:
  \beq
  V_C(r,\x_c)_{i,c}=\begin{cases}
    V_A(r,\x_c)_{i,c} & i < c - \mathrm{mod}(c, 3) \\
    V_B(r,\x_c)_{i,c} & \mathrm{else}.
  \end{cases}
  \eeq
  This ensures that a new window position is repulsed by any existing window not in the current cluster of $3$, and attracted to any window already in the current cluster.
\end{itemize}
\normalem
\bibliographystyle{ieeetr}
\bibliography{PRLCellsensingbiblio3,RMPbiblio4newN2,biblioPierre,bibs_ud}
\end{document}